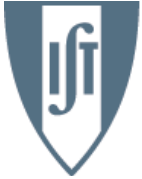

# Mnemonical Body Shortcuts

## Gestural Interface for Mobile Devices

**Ricardo João Silveira Santos Gamboa**

Dissertação para obtenção do Grau de Mestre em

**Engenharia Informática e Computadores**

**Outubro de 2007**

# Acknowledgments

First of all, I would like Tiago Guerreiro for the amazing support he has given to this project since day one. Whenever I had any doubt, needing any new ideas or looking for a renewed direction on my project, Tiago was always present to give me the right guidance. I would like to express my gratitude to my adviser Professor Joaquim A. Jorge for his knowledge, constantly innovative ideas and the way he always motivated me to keep up the pace of the work.

I also thank people at PLUX *Engenharia de Biosensores*, for their total support on the work material. Hugo Gamboa, thank you for all the technical guidelines and for being one major inspiration not only in my academic path but also in my life.

To all my colleagues at IST-Taguspark, thanks for your constant friendship that helped through the development of this work.

To my family, my mother Maria, my father Antonio and my sister Mafalda, thanks for the unconditional dedication and patience during the most difficult phases. To my sister Mafalda, my special appreciation for all those good works done in the context of this document. Also, very special thanks to Ana Rita and my cousins Alice, Diana and Carolina for all that happiness you bring to my life.

Finally, I would like to thank Maria for motivating me in difficult times, be patient in all those phases when I couldn't be present and for your inspiring strength and affection.

# Agradecimentos

>vi

# Abstract

Mobile devices' user interfaces are still quite similar to traditional interfaces offered by desktop computers, but those can be highly problematic when used in a mobile context. Human gesture recognition in mobile interaction appears as an important area to provide suitable *on-the-move* usability. We present a body space based approach to improve mobile device interaction and mobile performance, which we named as Mnemonical Body Shortcuts. The human body is presented as a rich repository of meaningful relations which are always available to interact with. These body-based gestures allow the user to naturally interact with mobile devices with no movement limitations. Preliminary studies using Radio Frequency Identification (RFID) technology were performed, validating Mnemonical Body Shortcuts as an appropriate new mobile interaction mechanism. Following those studies, we developed inertial sensing prototypes using an accelerometer, ending in the construction and user testing of a gestural interface for mobile devices capable of properly recognizing Mnemonical Body Shortcuts and also providing suitable user control mechanisms and audio, visual and haptic feedback.

**Keywords:** Gestures, Mnemonics, Shortcuts, RFID, Accelerometer, Mobile, Feedback.

# Resumo

Existem algumas semelhanças entre as actuais interfaces oferecidas por computadores pessoais e as interfaces de dispositivos móveis, mas a existência dessas semelhanças pode ser muito problemática quando esses dispositivos são usados num contexto móvel. Assim, o reconhecimento de gestos para interacção móvel surge como uma importante área para promover a usabilidade dos dispositivos para utilizadores em movimento. A nossa proposta é a de utilizar interacção gestual e o espaço corporal para melhorar a interacção em movimento, que nomeámos como *Mnemonical Body Shortcuts*. O corpo humano é apresentado com um vasto repositório de significados, com os quais podemos interagir a qualquer momento, e os gestos baseados em partes do corpo possibilitam uma interacção natural com dispositivos móveis sem qualquer limitação de movimento. Foram realizados estudos preliminares usando identificação por radiofrequência (RFID), onde este conceito foi validado. Seguindo este estudo, nós desenvolvemos protótipos baseados num sensor inercial (acelerómetro), que resultaram numa interface gestual para dispositivos móveis capaz de reconhecer eficazmente os atalhos corporais mas também dar aos utilizadores mecanismos de controlo e feedback auditivo, visual e de vibração.

**Palavras-Chave:** Gestos, Mnemónicas, Atalhos, RFID, Acelerómetro, Mobilidade, *Feedback*.

x

# Index

# List of Figures

# List of Tables

# Acronyms

| | |
|---|---|
| **API** | Application Programming Interface |
| **DVD** | Digital Video Disk |
| **EMG** | Electromyography |
| **GPS** | Global Positioning System |
| **GUI** | Graphical User Interface |
| **HCI** | Human-Computer Interaction |
| **IST** | *Instituto Superior Técnico* |
| **kNN** | k-Nearest-Neighbours |
| **MEMS** | Micro-Electro-Mechanical System |
| **MP3** | Mpeg-Layer 3 (Audio Compression) |
| **PDA** | Personal Digital Assistant |
| **RF** | Radio Frequency |
| **RFID** | Radio Frequency Identification |
| **SMS** | Short Message Service |
| **TCP** | Transmission Control Protocol |
| **WLAN** | Wireless Local Area Network |
| **2D** | Two Dimensions |

# 1

# Introduction

Over the last few decades, we have been witnesses to an extraordinary development on mobile technology. Not so long ago, computers were meant to be used only in static environments. However, communication development, component miniaturization and a general education on the use of computers dictated the emergence and success of portable computational devices. In their genesis, those mobile devices generally had an awkward design, large size and only a couple of simple functionalities besides standard communication. But the state of development rapidly changed and actually we can use small mobile devices with colourful screens, stylish designs and featuring an extensive list of functionalities such as digital camera, calendar, video and mp3 player or web browser. These multi-task devices are still under a significant development and constant mutation in available functionalities, communication facilities, design or user interfaces. This dissertation will focus on user interfaces area, studying and developing a new method of interaction. In this particular area of interest, mobile devices have adopted a button-based interaction featuring visual display and extensive menus, in some aspects copying and adapting user interfaces developed for desktop computers. Over the years there were not many breakthrough innovations interfaces for mobile devices, excluding the usage of touch screens and voice recognition. It is crucial for the development of better user interfaces the research on innovative interaction methods that can enhance usability. It is also important to study the main limitations that characterize mobile interaction. A typical user wants to interact with the mobile device in variable conditions: noisy environments, light variations, while moving or even in emergency situations. A successful user interface for mobile devices has to be usable in all those conditions and also surpass the input/output and processing limitations inherent to a mobile device. This dissertation converges on the study of gesture-based interaction with mobile devices. Gestures are a natural and expressive way of communicating, providing also a suitable interface for Human-Computer Interaction. Gestures may be recognized using diverse methods (e.g.: Vision, Touch Screens, Inertial Sensing, Radio Frequency Identification and Electromyography) and are applicable in various areas, including our area of interest, Mobile Interaction. The remainder of this chapter describes the problem and our proposed approach. It also overviews the present work, enumerates the main contributions and publications and finishes with the dissertation outline.

## 1.1. Problem

Due to their limited size, existence of many different platforms and possible interaction in multiple scenarios, mobile devices have a set of limitations closely related to the development of a suitable user interface. The study on these limitations is mainly based on empirical research, and is described on many different works. Forman and Zahorjan [6] are generally referenced as pioneers on the description of such limitations, but many authors have followed the same trail, such as Kristoffersen and Ljundberg [8], Landaya and Kauffman [3] and Brewster [4]. These works characterize mobile devices as having a modest screen, a small amount of little-sized buttons and limited processing capabilities. Besides, interaction is not only desktop-based but should also be appropriate for different light conditions, human motion and social environments. These constrains are known for years; however, existing interfaces have reproduced some features of desktop computers. As Stephen Brewster stated [4], it is "clear that taking the desktop interface and implementing it on a mobile device does not work well; other methods must be investigated to make mobile interfaces more usable". In truth, recent mobile interfaces do not take in account some interaction issues:

- While visual attention on desktop computers can always be given, that does not happen while interacting in mobile devices in various conditions, when the user has to choose between the mobile device and other main task. Oulasvirta *et al* [5] demonstrated that "Continuous attention to the mobile device fragmented and broke down to bursts of just 4 to 8 seconds, and attention to the mobile device had to be interrupted by glancing the environment up to 8 times during a subtask of waiting a Web page to be loaded." Users have to control the environment with this frequency because "attentional resources must remain with the main task for safety reasons" [30].

- Most mobile devices do not provide direct selection capabilities, which leads to the creation of multiple menus and difficulties when a specific task is wanted. Touch screens support direct selection but represent a two-handed and visually demanding interaction.

- When using desktop computers, there is an eminent need to control multiple tasks at the same time, but mobile devices are used to make one task simultaneously with other activities in the physical world. This characteristic implies the growing importance of how fast one can reach the applications and move to second plan the ability to manage and access different ones.

The imitation of Graphical User Interfaces (GUI) present in desktop-based computers resulted on a mobile interface slow to use, visually demanding and requiring high workload from users. They spend much of interaction time trying to reach the chosen applications rather than using them. Given the existence of a core of applications that are constantly used, one solution is the creation of appropriate shortcuts to ease access to the most used functionalities. Some solutions were developed and applied in commercial devices, namely key shortcuts and voice recognition. Key shortcuts are the most used ones, yet they fail on long-term usage because they do not usually provide any auxiliary memorization about which application is in which key shortcut. This fact leads people to forget the functions of each key and return to the slow and visually demanding menu selection. Regarding voice shortcuts, there are some unresolved issues that compromise their performance: low recognition rates, especially in noisy environments; low acceptance on a public usage; voice commands do not provide much privacy because they are too revealing of the task to perform.

Since actual interaction with mobile devices does not provide users with the most appropriate tools, interface developers should study new mechanisms that enable a faster and less visually demanding experience, an easy access to main applications and also an easy integration across platforms while taking in account all the limitations of a mobile device and the new possibilities that arise from their use in Human Computer Interaction (HCI) area.

## 1.2. Proposed Approach

In order to provide mobile devices with a more appropriate interface, our approach will focus on the creation of gesture-based shortcuts. Gestures are one of the most important means of communication between humans, and one could say more: they were certainly one of the first. It is remarkable that they surpass speech since they are rather international. When people do not share the same language, gestures are usually a very effective resource. They have proven capabilities in diverse areas of HCI, providing a more natural way to interact with diverse applications, but they are an unexplored method in commercial mobile devices. Mobile interaction with gestures had not such an early start mainly caused by the difficulties on detecting gestures with mobile devices, since the main method for gesture recognition is based on external cameras that are not usable in a mobile context. In chapter 2 the main methods for gesture recognition with mobile devices will be discussed.

This work proposed approach is based not only in the interaction capabilities of gestures but also in the extended meaning that gestures have when combined with body parts, inspired by Ängeslevä *et al* work [64]. In regular communication between humans, gestures are often combined with body hints to empathize an idea (i.e. sincerely apologising with a hand over the heart, asking the time with a touch on the wrist or asking someone to be quiet with a finger on the mouth, examples in Figure 1.1).

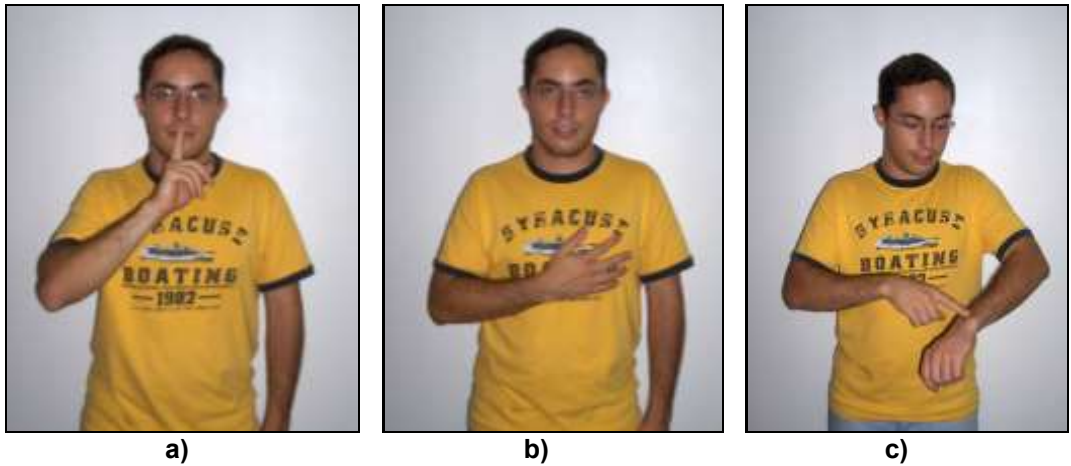

**Figure 1.1 – Gestures and Body Parts emphasize communication a) Silence b) Emotion c) Time**

Using the undeniable capacities of gestures and the possibility of joining them with the rich significance of the different body parts is possible to create strong associations between them and provide a new interaction modality for mobile devices. In this dissertation, those associations are referenced as Mnemonical Body Shortcuts. There are multiple potential Mnemonical Body Shortcuts, depending on the different mnemonics that each user may want to choose, but these are some possible examples:

- An approximation to the ears opens the music player.
- A gesture to the heart calls a beloved person.
- A gesture towards the wrist triggers the clock or time information.
- A movement to the head shows the contact list.

This approach cannot be only based on gesture recognition and shortcut triggering, but it also takes in account the importance of an appropriate user-control mechanisms and feedback such as audio, vibrational or visual feedback to fully complete the interaction. We intend to enhance Ängeslevä *et al* work proving the concept of body-based gestural interaction with user tests but also developing and exploring different approaches to achieve appropriate gesture recognition and user interface.

## 1.3. Overview of Present Work

Following our proposed approach, we defined as first priority the research of related work in the area of gestural interaction with mobile devices, and five different main technological approaches were found. Those approaches are the usage of Radio Frequency Identification (RFID), Accelerometers, Cameras, Touch Screens and Electromyography (EMG), but specific gestural recognition in mobile devices is also possible with Capacitance Sensing and Laser Beams. The main works in each area were studied and compared in the most important aspects such as the type of recognizable gestures, detection complexity, accuracy, additional hardware and implementation cost. The complete research is present in the chapter 2 of this dissertation. With the knowledge of what has already been done in the area, the next step was to perform a task analysis. The objective of the task analysis was to get the actual panorama of user interaction habits with mobile devices, with special interest in the utilization of any type of shortcuts and the efficiency when triggering the most used applications. Results demonstrated that key shortcuts are frequently used while users are still reluctant to use voice shortcuts. However, user observation when triggering the most used applications and call the most used contacts lead to the conclusion that users end up using menu interaction even when there are available key shortcuts. Mobile Interaction is still keystroke consuming and mainly based on menu Interaction, and key shortcuts, even popular, still lack a good efficiency rate.

With the objective of creating a first prototype that recognizes Mnemonical Body Shortcuts, we invested our efforts on the creation of a RFID prototype. RFID technology provides direct point recognition using tags on clothes and a RFID reader in a mobile device, what makes this technology the best candidate for a first prototype. A RFID prototype would be able not only to prove its own feasibility but also to compare the concept of Mnemonical Body Shortcuts against the most popular shortcut method, key shortcuts. The developed application permitted proper tag reading, discarding multiple tag detection and keeping a log about the information of each tag read, which enables a correct analysis of the efficiency and identification of tag sequence. User tests were made to measure the accuracy of the prototype but also to test the validity of Mnemonical Body Shortcuts approach. The RFID prototype achieved good recognition results, however, it also lead to the conclusion that it might generate too many unintended recognitions (false positives) and some discomfort with the usage of RFID tags on clothes. But, altogether, the concept of Mnemonical Body Shortcuts was a success. Users were able to make meaningful associations between body parts and applications, and we also found many different clusters of associations, reflecting that this method is capable to support personal associations but also some more transversely used. The memorization aspect was also tested against key shortcuts, with clarifying results supporting the high remembrance rate we expected to achieve with our approach.

Although RFID proved to be a suitable method to be used under specific restrictions, we suspect that it would not have a good acceptance for a daily basis interaction: users would have to stick RFID tags on clothes everyday to have an all-time available interface. Since accelerometers are being introduced in mobile devices, and provide the needed tools to gesture recognition as will be proved in chapter 2 and 3, they were chosen to be used for the new prototypes. In a first prototype, we developed an algorithm featuring the calculation and analysis of both position and angle variations throughout the movement towards a body part, which permits the definition of some default recognizable gestures that are able to used if knowing the height of the user. The second prototype was based on an implementation of a feature-based algorithm. This algorithm extracts 12 different features of the signal, such as the maximum and minimum peak and the final value of the signal on the three axes and on the movement amplitude. After feature extraction, gestures are classified using K-Nearest-Neighbours for default gestures and Naive Bayes classifier when we introduce user training to recognize personalized gestures. Both prototypes were pre-evaluated to select which would give a better accuracy. The second prototype was more accurate in all contexts, and it was used to create a final prototype, joining gesture recognition, haptic, audio and visual feedback, and a user interface to control shortcut triggering. Usability tests on this final prototype were made in diverse scenarios, using both personalized and pre-defined gestures (named "default gestures"), under different mobility settings. Interesting results were achieved, not only regarding recognition rate but also in terms of feedback, user-control mechanisms and user acceptance of the system. With the final prototype, we were able to create a suitable system to provide Mnemonical Body Shortcuts while standing or moving, giving users the needed tools to feel comfortable when using the system and accomplishing the usability goals we defined in the start of our work.

## 1.4. Contributions

This work is expected to develop a renewed view on the interaction with mobile devices through the research of different methods on gesture recognition applied on this specific area, but also proposing a concept that was already referenced but poorly validated, developed and tested. The outcome of our research can be divided in six main points:

- **Survey on gestural interaction with mobile devices**

We studied different methodologies and works that enable gestural recognition. Diverse themes have been surveyed regarding gestural input in HCI, but none focused on the different capabilities and limitations that are present while using mobile devices. This survey provides an overview on this area and may serve as starting point for other investigation works that focus on the creation of gestural interfaces for mobile devices.

- **Characterization of actual mobile usage**

We performed a task analysis on the subject of mobile interaction and shortcut usage with mobile devices. This task analysis was developed to clarify the actual panorama on the subject. General information about actual tasks was gathered, such as diary frequency of use, most used applications and most used contacts. We also identified some issues on current usage of shortcuts in mobile devices. Voice recognition is not used due to its inexistence on some mobile devices, but mainly because of users discomfort about its recognition rate and low intimacy, while key-shortcuts have an extremely low support for memorization of personalized shortcuts. This characterization reports that users' interaction with mobile devices is still keystroke-consuming in many areas, which proves its inefficiency and urgency for finding new interface solutions.

- **Guidelines for development of gestural interfaces for mobile devices**

Based on the analysis on the different works that enable gestural recognition and also in the characterization of actual mobile usage, we were able to define and describe a set of guidelines that should be accomplished in order to develop a gestural interface appropriate to mobile devices: Support shortcut memorization; Give appropriate feedback; High recognition rate; Grant social and user acceptance; Allow mobile interaction. In fact, most of the works in this area seem to accomplish a couple of the guidelines we propose, but lack in fulfilling all of them. If some of these guidelines are not followed, many issues can emerge during user tests on the mobile gestural interface.

- **Creation of a signal visualization platform**

A signal visualization platform was developed. This platform was extremely helpful to analyze and categorize accelerometer signal, take snapshots of specific signals to insert in reports and also analyse the results from diverse user tests. Besides, it is independent of the source of the signal. The platform can receive the signal from any sensor and use its tools to help not only in the visualization process, but also in the first steps necessary to the construction of a pattern recognition algorithm.

- **The validation of the concept of Mnemonical Body Shortcuts**

We were able to validate the concept of Mnemonical Body Shortcuts as a suitable method of shortcut interaction. Using a RFID-based prototype it was possible to test the concept in two main areas: users ease to create relations between body parts and applications and the remembrance rate that can be achieved when using this approach. In fact, these tests confirmed body-related gestures as an efficient and meaningful method to create shortcuts. The remembrance capabilities of this approach were compared with key shortcuts, achieving much higher remembrance rate in long-term memory, thus surpassing the main issue present on key shortcuts.

- **Mnemonical Body Shortcuts complete gestural interface**

A gestural interface for mobile devices was constructed, based on the guidelines we defined for this kind of user interface. It was based in a feature-based algorithm, specifically developed for the recognition of personalized and default Mnemonical Body Shortcuts, but also in the development of a user interface with appropriate feedback and user-control mechanisms. We also pioneered the development of a new type of probability-based haptic feedback and a *Multichoice* mechanism to switch between applications both stored on the same body part or to trigger the second or third recognized application. The final prototype for Mnemonical Body Shortcuts, using feature-based recognition, achieved an accuracy of 89.5% when using personalized gestures and 92.5% for default gestures, but when using feedback and user-control mechanisms, errors were around 3%. These results demonstrate the success of our development approach in both gesture recognition and user-control mechanisms.

## 1.5. Publications

The work present on this dissertation yielded three publications in International Conferences and one in a National Conference. They are listed on chronological order of publication:

1. Ricardo Gamboa, Tiago Guerreiro, Joaquim Jorge , *Mnemonical Body Shortcuts*. Proceedings of the 7th International Workshop on Gesture in Human-Computer Interaction and Simulation, Lisboa, Portugal, 05/2007

2. Ricardo Gamboa, Tiago Guerreiro, Hugo Gamboa, Joaquim Jorge, *Mobile Interaction Based On Human Gesture Analysis*. Proceedings of ISHF 2007 - International Symposium on Measurement, Analysis and Modeling of Human Functions, Cascais, Portugal, 06/2007

3. Ricardo Gamboa, Tiago Guerreiro, Joaquim Jorge , *Mnemonical Gesture-based Mobile Interaction*- HCII 2007 – Proceedings of the 12th International Conference on Human-Computer Interaction, Beijing, China, 07/2007

4. Ricardo Gamboa, Vasco Costa, Joaquim Jorge, *Multimodal Presentations on Multi-Projection Displays,* Proceedings of the 15º Encontro Português de Computação Gráfica (EPCG) , Oeiras, Portugal, 08/2007

5. Ricardo Gamboa, Tiago Guerreiro, Joaquim Jorge, *Mnemonical Body Shortcuts***,** Springer Lecture Notes on Computer Science issue of GW2007, Submitted for publication.

6. Ricardo Gamboa, Tiago Guerreiro, Joaquim Jorge, *Mnemonical Body Shortcuts: Body Space Gesture Recognition,* 13ª Conferencia Portuguesa de Reconhecimento de Padrões, 10/2007, Submitted for publication.

7. Ricardo Gamboa, Tiago Guerreiro, Joaquim Jorge, *Mnemonical Body Shortcuts: Gestural Interface for Mobile Devices,* International Conference on Intelligent User Interfaces (ACM IUI 2008), Submitted for publication.

## 1.6. Dissertation Outline

The remainder of the dissertation is compound by five chapters:

In Chapter 2 we present our research on different methods that provide gestural interaction with mobile devices, namely Radio Frequency Identification, Accelerometers, Touch Screens, Mobile Cameras and Electromyography.

Chapter 3 is divided on three main parts. It starts with the description of a task analysis performed to clarify the characteristics of current interaction with mobile devices and also how users interact with the shortcuts that are currently available. In this first part, we identify the main problems of mobile interaction resulting of user observation and the most important design guidelines and usability goals to produce a suitable gestural mobile interface. In the second part of the chapter, our approach based on Mnemonical Body Shortcuts is explained in detail, and we present some user scenarios and also the description of the RFID prototype development and evaluation, because it mainly served to validate the body-space gestures concept. Finally, the third part consists on an overview on the most important aspects of gestural recognition using accelerometers, serving as introduction for the next chapter.

In chapter 4 we describe in detail the two accelerometer-based approaches that were developed in this work, using the theoretical basis referenced in the final of chapter 3. The first implementation consists on a convergence of both position calculation and rotation measurement to identify the gesture, while a second prototype uses feature-extraction and classifiers to the same purpose. Furthermore, the final prototype is also described, focusing not only on the recognition algorithm but also in the technological solution, feedback and user-control implementation

Prototype evaluation is reported in the Chapter 5. Since the RFID prototype evaluation is already referenced in chapter 3, it is only focused on the results of the pre-evaluations on both approaches using accelerometers and in the final usability studies.

Finally, Chapter 6 concludes this dissertation, with an overall discussion of the benefits and limitations of the prototypes and conclusions regarding the whole investigation. We also present some suggestions for future work in the area.

# 2
# Related Work

Mobile Interaction gained vast importance in the actuality. However, mobile devices interfaces are still mainly graphical-based, creating difficult situations for a constantly changing and *on-the-move* context. A gestural interface, with its natural characteristics and expressivity, is able to fill the lack of consistency of those interfaces and provide the user with a suitable and fast mobile access. In the latest years, many different works have already enhanced mobile devices with technologies able to recognize gestures without mobility constraints. This chapter will overview the state-of-the-art regarding the usage of gestures in mobile devices such as cell phones or wearable devices. The different approaches will be divided accordingly to the used technologies. A technological comparing was chosen because the main objective of this research is finding and comparing the diverse options to be implemented within the context of our work on a mobile device. The validation and implementation of such a concept has to use a technology with specific characteristics, such as recognizing gestures towards body parts and have an accessible cost. In this study, we found 5 distinct gesture-recognition tools: RFID, Cameras, Accelerometers, Electromyography and Touch Screens. Because it is important to find the most versatile technology to solve the current problem, each one will be analyzed keeping in context some vital characteristics:

- **Variety of possible gestures**

The range of possible gestures performable while holding a mobile device is innumerable. Different technologies can be characterized by the set of gestures each one is able to accurately recognize.

- **Implementation Cost**

We are aware that research is not fruitful when detached form reality. A low implementation cost is essential to, in a near future, start a large-scale deployment of gestural interaction software and hardware on commercial mobile devices.

- **Social Acceptance**

Gestural interaction with computers in public environments is still uncommon, so a major concern about the approach to be used is its possible low acceptance when used in public. This characteristic is connected with gesture variety but also with the needed additional hardware.

- **Recognition Complexity**

A high recognition complexity using a certain technology is not an immediate reason to not use it. The knowledge about the development cost of each approach gives the possibility to adapt the chosen approach with time limitations or different research objectives. Besides, the approach has to be sufficiently lightweight to be implemented on a mobile device.

- **Self-containable hardware**

While some technologies need additional hardware outside the mobile device to provide gestural recognition, others only rely on self-containable hardware. A self-containable approach provides usage comfort and might also be more easily implemented on commercial devices.

- **Accessibility**

A gestural-based interaction has the advantage to be independent of visual or audio senses, or even be performed when there are limited motion capabilities. A technology that provides gestural recognition for the majority of population but also for people with disabilities is truly valuable.

- **Accuracy**

A critical issue on gestural recognition is the impact that a low accuracy might have on user acceptance and future usage. Users would not accept a technology that triggers shortcuts without any user command, known as false positives. In fact, there are some approaches that may create false positives easier than others, and special actions have to be studied for those situations.

- **Further contributions**

Gestural interaction technologies might also be useful to other interesting activities in mobile contexts, which would give more arguments to implement it on mobile devices

## 2.1. RFID

Radio Frequency Identification Technology (RFID) is one of the main areas of interest in wireless identification, resultant of years of investigation on Electromagnetism. It is based on three basic components:

**RFID tag**: a small sized chip which has an antenna emitting radio frequency (RF) waves and stores data, usually an unique identifier. They can be either batteryless (passive), getting power to transmit data through inductive coupling from the RFID reader, or active when they have an energy source. (Figure 2.1)

**RFID Reader**: used to interrogate the RFID tag using radio waves to obtain its data. The reader exists in various sizes and forms, in big boxes to small integrated readers in mobile phones.

**Host Computer**: A RFID host can be a PC, a PDA or any other computer with processing and communication capabilities. The host wirelessly receives the data from the RFID reader and filters it, discarding multiple readings.

The first RFID concept application appears in the last century during the World War II [1][2]. Nowadays, a myriad of applications use RFID with different objectives: animal tracking, personal access, collecting tolls or product identification on retail shops. These different applications exist due to some important characteristics: tags have different forms, high durability and reusability. Besides, tag reading does not require line of sight and it is possible to read multiple tags at the same time. These characteristics also allowed some investigators to think RFID as a suitable tool to be used with mobile devices and detect gestures.

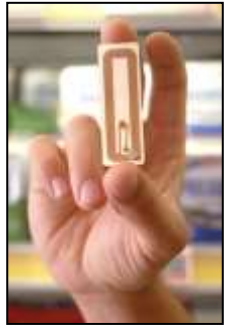

**Figure 2.1 – One kind of RFID tag**

### 2.1.1. RFID as a Mobile Gestural Input Technology

In HCI, RFID commonly serves as a bridge between the real and digital world. This characteristic is especially useful outside static environments, because computer terminals and wireless communications are not always available. Wearable devices and cell phones have been adapted with RFID readers to extract information from tags present in the "real world". This component is applied in two main mobility areas:

**Location Awareness**

Due to its various forms, RFID tags can be placed on the floor, walls or objects. Diverse applications use tags to deliver location information to a wearable device, in areas such as augmented reality [12] or blind people location systems [14].

**Object Interaction**

RFID tagged objects and RFID readers in wearable devices and cell phones are the basis for many approaches that use objects to interact with. Those approaches embrace diverse areas like augmented reality, human-activity detection, information retrieval and mobile interaction. For instance, Want *et al* [9] attached RFID tags on objects in an effort to bridge "*physical and virtual worlds with electronic tags*". Some possible interactions with a laptop computer were implemented: RFID was used to tag physical documents; Personal cards had embedded tags, allowing direct e-mail messaging; Furniture with RFID tags, serving as context information for personal computers. Gloves are used in [16] to show a web page based on the URL referenced by a tagged object, RFID bracelets allow detection of human activity [17] or download information about CD's or DVD's [21] and there is also a lot of research on the usage of cell phones combined with RFID for communication with interactive spaces [23] or provide mobile payment and book availability [24].

From all the last examples, two types of interaction with the tags can be spotted – implicit and explicit. While implicit interaction uses RFID readers to read tags without order of the user, the implicit approach only works when the user demands it. In the last case, the user has to make a gesture with the tag reader towards the tag to make some action. These gestures can be simple point recognitions or multiple point-to-point gestures when various tags are recognized in sequence. Physical tagged objects ensure interaction with wearable and mobile devices. However, these objects are not always present, and users need to interact with the devices many times during the day, in the most various places and situations. The solution to use RFID to support a mobile gestural interface is the embodiment of RFID tags – if tags are used in the body (attached to clothes, wallets, bracelets, etc), they are always accessible. Headon and Coulouris [20] (Figure 2.2) have one of the few works that does not rely in physical objects to interact with the wearable device, providing a truly *on-the-move* interaction with all sorts of applications. They created a wristband to read RFID tags attached to some devices or worn on the body. When the wristband reads a tagged device, it communicates with a PDA and changes the context of the application. The application can be controlled with gestures, using a grid of 2x6 tags on the shirt of the user. For example, the user can touch a digital camera and then take a self-photo without temporization. However, the display of the RFID tags lacks association, and because tags can be used for many applications, the action of each one could turn to be confusing. Besides, the issue of false positives in tag reading was not discussed and user tests only measured the accuracy of tag reading.

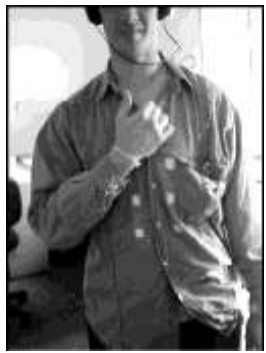

**Figure 2.2 – Headon and Coulouris Work [20]**

### 2.1.2. RFID-Based Mobile Gestures Evaluation

Gestural interfaces using mobile devices and RFID are only based on object interaction, and there is not a reference work where to extract the main characteristics. Since Headon and Coulouris work is also applicable to other mobile devices, we will use it as reference in this evaluation.

**Variety of possible gestures**

For a truly RFID-based mobile interaction, tags have to be placed on clothes or personal objects. For that reason, the set of possible gestures is limited to single or multiple point recognition, but in those classes the possible gestures are many. For instance, one single-point gesture is a gesture with the mobile device towards a tagged wallet and a multiple-point gesture could combine this first point recognition with other point on clothes were the user wears a tag.

**Implementation Cost**

Some major companies have plans to install RFID readers on their mobile devices [7], but that seems to be a long-term plan, with no guaranteed success. To use this approach, users would have to afford an external RFID reader compatible with their mobile device. RFID tags should also be bought but they are generally inexpensive.

**Social Acceptance**

There are no social acceptance issues on the usage of a RFID-based system because RFID tags can be worn invisibly. The set of gestures is large on its diversity, so it is possible to choose socially acceptable gestures.

**Recognition Complexity**

From the point of view of the programmer, gesture recognition is straightforward. To recognize a gesture using RFID, the hardware delivers the tag ID or the information stored in it, and multiple readings have to be discarded. The host computer can simply perform some action based on the tag information.

**Self-containable hardware**

One of the main problems of this approach is the mandatory use of RFID tags on clothes. Even knowing they could be used invisibly, it may still be a great discomfort for the user to place them everyday and also replicate their usage on different objects and clothes. Although tags have to be used outside of the mobile device, RFID readers can actually be implemented on mobile devices.

**Accuracy**

False positives rate is high since a simple unintentional approximation with the device to a tag would trigger an action. A switch button is recommended to solve this issue and also reduce energy consumption.

**Further contributions**

The existence of an RFID reader on mobile devices can become, in future, an important link to get information about different interactive spaces and products.

## 2.2. Accelerometers

An accelerometer (Figure 2.3) is an electromechanical inertial sensor device that measures its own acceleration, or the acceleration of an object equipped with it. Examples of other inertial sensors are gyroscopes (rotation) or altimeters (altitude). The acceleration might be static (gravity acceleration) or dynamic, which refers to the acquired acceleration of the body. The acceleration measurement can give other variables like the gravity, object vibration, velocity, space position, revolutions per minute or angle.

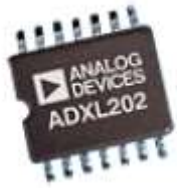

**Figure 2.3 – ADXL 202 Accelerometer**

### 2.2.1. Accelerometer as a Mobile Gestural Input Technology

When accelerometers started to be produced with MEMS (Microelectromechanical Systems) technology they became suitable and practical for usage in mobile devices – their size, weight, power consumption and cost was reduced like never before. MEMS technology consists on the integration of mechanical and electronic elements in the same chip using micro fabrication techniques. Accelerometer's motion sensing capabilities and self-operability can transfer useful information to mobile computers. This captured information can be split in explicit and implicit data. The implicit data is the motion that accelerometer captures without user knowledge. Explicit data exists when the user wants to interact with the wearable device and does some gesture expecting some retrieval and feedback of it. Accelerometers also have some problems – the captured data has some noise (like gravity and velocity error when integrating acceleration), which difficult the pattern detection algorithms and there is a clear lack of guidelines when it comes to analyse accelerometer data. The main works in both implicit and explicit gestures will be presented, focusing on explicit gestures because they provide intentional interaction with computers.

### 2.2.2. Implicit Gestures

Accelerometers are one of the most used sensors to provide motion context to computational devices, permitting body-motion information retrieval to perform gait [35] and posture [36] analysis. Some of these works use multiple accelerometers in different body parts, but researchers concluded that even one or two accelerometers can give a high percentage of recognition in many activities [52], enabling its usage on mobile devices with a single accelerometer. One example of a motion sensing device is the "eWatch" [44], used to determine if the user is "sitting, standing, walking, ascending stairs, descending stair, and running". Some PDA's and cell phones are also equipped with accelerometers to identify different individuals [48] or adapt the screen layout while moving based on gait analysis [49].

### 2.2.3. Explicit Gestures

Explicit gestures are the type of gestures that have a higher value when interacting with a computer. They can have a specific meaning and if a gesture is detected by a computer an appropriate response can be given. Explicit gesture recognition using accelerometers is a commonly used method to interact with mobile technologies, and has been subject to many studies. Because explicit gestures are studied in great depth, it is important to be objective and restrict the research in two main areas – hand and arm gestures. Those gestures are natural and expressive; they play an important role on non-verbal communication and are often used in explicit interaction. The importance of this kind of gestures is clear when people cannot verbalize and use sign language. Furthermore, the majority of actions people do in daily life are based on their hands. Mobile devices such as cell phones can be held and hand-moved, making them a suitable platform for hand and arm gesture recognition.

**Hand Gestures**

The typical accelerometer-based manipulative movements in handheld devices are the vibrational, tap and tilt input.

In "Muscle Tremor as an Input Mechanism" [50], Strachan *et al* prototyped a PDA with an accelerometer that was able to detect vibration. A simple game was developed, where the user can inflate a balloon by squeezing the PDA. It was proved that squeeze gestures are recognizable and can be an added feature to mobile phones with motion sensing. Squeezing the phone should be useful, for example, to accept/reject a call or trigger a gesture movement. Tap input on handheld devices is also based on vibration. Small tap gestures generate vibration in a certain direction. In [65] a PDA user can silence the phone with three taps on it and can swap between applications with lateral taps. In [66] a simple ball game was made to demonstrate haptic input and output potentialities.

Finally, tilt input is based on the movement of the device to left, right, up and down. These movements can be easily done when holding the device, and can create specific metaphors to interact with applications. The first use of accelerometers with tilt detection was made by Rekimoto [67], using the movement to select an item in pie menus, scroll documents and maps. Harrison *et al* [68] used tilt to navigate through sequential lists, while Hinkley *et al* [69] firstly suggested an automatic screen orientation for PDA's, adapting the screen information both in vertical and horizontal display. "Rock 'n' Scroll" [70] is other important work in this area, because it presented a "clutch button", that was not present in other works. This button is important to minimize the occurrence of undesired tilt interaction. Interesting applications have also been made in text-entry methods. In TiltType [72] a new text-entry method for very small devices is proposed, based on 4 buttons on a watch-type device and tilt in 8 directions. Different characters are selected by the combination of selected buttons and tilt direction. TiltText [71] is focused on mobile devices and permits the user to stroke any button in the pad and then select the correspondent character in 4 directions, with a double tilt to the upper case.

**Arm Gestures**

When equipped with an accelerometer, mobile devices can be arm-moved in the tri-dimensional space and perform recognizable gestures. The role of the accelerometer is to capture the acceleration signal and deliver it to proper software to recognize the gestures. The possible types of gestures are immense, because the human arm has a high degree of freedom when comparing with the hand. This variety and the possibility to create gestures that are natural but also unusual in the daily life gives the arm gestures high importance in computer interaction. Since a single accelerometer is sufficient to capture this data, the research in arm gestures converged in handheld devices.

Some inertial sensing handheld devices are used to recognize gestures and communicate and control external devices. Two similar approaches were made in order to control multiple devices in ubiquitous environments with simple gestures, the "XWand" [63] and the "BlueWand" [62]. There are other applications that provide control of specific devices. In [57], a handheld ("eMote") is able to control a compact stereo. Gestures include skipping music with a throwing gesture to the front, turn the volume up and down with vertical tilt and turn the music off when the control is upside down. A group of researchers has made important developments on this area. In their first work [60] they addressed the problem of gesture recognition with accelerometers and proposed an algorithm based on Hidden Markov Models. Later, they prototyped a DVD controller [59] and enhanced the algorithm with noise data to make it able to perform with less training repetitions by the user. Finally, the system was used to control a design environment with gestural input and proper user tests were carried [58].

Accelerometric data can be used to control applications within the device itself. One of the main works in gesture analysis with an accelerometer was made by Benbasat *et al* [56]. In this article, the authors studied the algorithmic problem of arm gesture recognition in a PDA but the implementation, even if successful, lacks on the description of functionalities and user testing. Body Mnemonics [64] is a project that provides a real application for gesture recognition. This work, which served for us as a main reference for our approach, makes preliminary studies in the association of gestures with parts of the body and the possibility to trigger applications with meaningful connections with those parts. Some associations are online banking information in the back pocket or GPS information in the feet. A technical report was later done with an approach similar to Benbasat *et al*. A commercial mobile phone with an integrated tri-axial accelerometer (Samsung SCHS130) was recently used by Choi *et al* "BeatBox" [55] (Figure 2.4) to create a gestural application later used in the cell phone. This application permits the user to draw numbers in the air with the cell phone to trigger a call to the correspondent number in the phone book. They also made it possible to navigate in a mp3 player with left to right (and inverse) gestures, delete a message lifting the phone twice, used a shaking detection algorithm to create musical applications ("beat box" and "electronic orgel") and created new games, such as rolling a dice when shaking the cell phone.

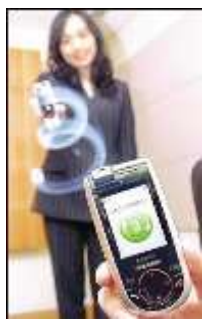

**Figure 2.4 – Samsung SCH-S130**

### 2.2.4. Accelerometer-Based Mobile Gestures Evaluation

There are many different approaches using accelerometers, but they share the following interaction characteristics:

**Variety of possible gestures.** Using an accelerometer is possible to recognize a great variety of both hand and arm gestures. The limits to this recognition are only on the different characteristics of the developed algorithms.

**Implementation Cost.** This technology could be used to recognize gestures when mobile devices begin to be equipped with accelerometers, and in that case, the implementation cost is low. It is also possible to buy external adapters with accelerometers, but they are expensive and very uncommon.

**Social Acceptance.** Socially accepted gestures can be selected within the large set of possible gestures to be performed.

**Recognition Complexity.** The recognition complexity varies with the gestures to be detected. However, this process is usually long, and pattern recognition algorithms have to implement different solutions such as Hidden Markov Models or Feature Extraction and Classification, and many others. A development using accelerometers has to be carefully planned and tested, especially for high-complexity gestures.

**Self-containable hardware.** Accelerometers can be integrated on mobile devices, making this technology totally self-containable.

**Accuracy.** Handheld devices are manipulated during the day, and if the device was always recognizing gestures many false positives would be triggered. Most works using accelerometers rely on the usage of an action button to turn on and off the gesture recognition

**Further contributions.** If an accelerometer is present on mobile devices, there are much new functionalities that can be added, and this is a field in great development yet. As described in this chapter, implicit gestures can be recognized, gait analysis may be performed, and accelerometers will certainly be used to get stable camera pictures. Furthermore, they can also serve as a biometric security system [50].

## 2.3. Cameras

Vision-based methods to detect body movement have been largely used in HCI, as they are considered one of the most natural man-machine interfaces. When applying image analysis algorithms, the movement of the human body can be detected, recognized and serve as an interaction method with a computer. However, these algorithms have a high level of complexity. Various techniques are chosen, using either monocular systems or multiple cameras. Detailed detection algorithms usually need markers on the body to recognize movements, but other systems also track human motion without any body-mounted markers. The diversity of algorithms varies with the environment, type of recognized gestures, dynamic of gestures and many other characteristics. In the last years the interest in vision-based tracking of human motion is growing and recognition of facial expressions, lip motion, hand and arm gestures and full body activity are the main study areas.

### 2.3.1. Wearable Cameras

The usage of vision based methods to recognize gestures while moving has been proved to have severe limitations. It is not possible to have static cameras used in most studies because users are free to move. One of the solutions is the selection of wearable cameras. These devices are truly mobile and may be situated in positions that only permit image retrieval out of the body-frame of the user, making it impossible to analyse the full motion of the body. However, hands are easily viewable and current works on this area focus on hand gesture recognition, which may have different applications. As example, in Augmented Reality, hand gestures are used as a pointing and selecting mechanism to interact with the system. "HIT-Wear" [28] uses the hand as a menu and each finger as a different selection. The user has a head mounted display that shows each menu name in the end of the fingers, and may select the menus touching the corresponding finger with the free hand. Finger pointing can also select menus that are presented in head mounted displays. Kurata *et al* developed the Hand Mouse [26] that allows a user with a head worn display to click on visual tags such as a visual soft keyboard. Other gesture-based approaches based on hand movements and wearable cameras feature text interaction in real world [25], sign language [22] and handwriting recognition [19].

Wearable cameras as a gesture recognition sensor are particularly useful interaction method with augmented reality systems. The problem for this kind of interaction is the needed hardware, because wearable cameras depend on caps, hats, glasses, head mounted or shoulder devices.

### 2.3.2. Mobile Cameras

Mobile phones and PDA's are now equipped high definition digital cameras to take photographs and capture video. Those cameras can be used also as a sensor and be a useful tool in gestural interaction with mobile devices, without needing the usage of wearable cameras. One of the most usual appliances of mobile cameras as a sensor is the identification of visual tags. The user can aim and click on a visual tag with their phone and trigger different applications. This approach is similar to RFID, but requires line of sight, more workload by the user (aim and click) and this type of tags are not wearable. One of the pilot works in visual tag recognition is Rohs visual codes [18], with a detailed description of the detection algorithm, but other applications followed by many researchers [15] [29]. Other options were studied were the phone can be moved to interact with external or internal applications when analysing the running optical flow of the camera images. These possible gestures are the movement, rotation and tilting of the phone which makes it feasible interactions such as scrolling, zooming or pointing. Some of these studies use visual tags [51] or simple geometric forms [47] to detect motion, thus dependant on their constant existence. Rohs work is also important in this area because it provided the idea of "movement detection algorithm that solely relies on image data obtained from the camera", using a feature based tracking algorithm. This algorithm provides the same cited applications, but can be used when pointing to almost any background. Some recent work by Haro *et al* [46] enhanced Rohs algorithms to scroll documents, navigate and zoom in maps and provide new game interaction possibilities. The problems of this type of interaction appear with different light conditions, because too bright or dark environments do not provide sufficient significant features to be analysed. Algorithms may usually fail when the camera is filming other objects in motion, and there are actually no references to arm gesture recognition on this area.

### 2.3.3. Camera-Based Mobile Gestures Evaluation

The following considerations are made in the context of the utilization of mobile devices cameras to recognize gestures using the optical flow of the camera.

**Variety of possible gestures.** This technology only enables the recognition of a limited set of gestures we characterize as "hand gestures" such as small left/right or up/down movements, rotation and tilting.

**Implementation Cost.** Since mobile devices are already equipped with cameras, and the tendency is to integrate more powerful ones, the implementation cost is inexistent.

**Social Acceptance.** The set of possible gestures is mainly very subtle, and users would only need current mobile devices to perform gestures. These facts allow a possible good social acceptance of those gestures.

**Recognition Complexity**. Algorithmic complexity is high, but this area has been studied for many years, and different approaches can be experienced and tested. However, it is important to adapt those approaches to the processing limitations of a mobile device.

**Self-containable hardware.** The camera is the only sensor used and is self-contained on the mobile device

**Accuracy.** This technology should only be used through an activation method, on certain applications or whenever the camera is active. Otherwise, false positives and unintended actions would be triggered most of the time.

**Further contributions.** The usage of cameras for gesture recognition would probably raise some interest on mobile camera usage outside of its custom functionalities, namely the introduction of visual tag reading software.

## 2.4. EMG

Electromyography (EMG) refers to the measurement of muscle tension and has been used in HCI in diverse areas, with special interest in accessibility to physically disabled individuals. To apply this technique the user has to place differential wet electrodes on the skin or use dry electrodes in elastic bands to measure muscle activity. These electrodes are able to capture minimal electric pulses generated by any muscle contraction and then send the information to a desktop or wearable computer by cables or wireless transmission, where the signal is processed.

### 2.4.1. EMG as a Mobile Gestural Input Technology

The received signal of muscle activity can be processed, using simple threshold detections or more complex pattern recognition systems, enabling gesture recognition. EMG can be used to detect a great variety of gestures because each one involves different muscles: as an example, with electrodes placed on the forearm the movement of each finger can be detected with accuracy. Because muscle tension is individual, gesture detection algorithms need user training. When trained, this technique is able to detect gestures that have no motion involved (isometric), providing a subtle interaction with the computer.

The most of the work using EMG as a gesture interface was done in static environments. There are, however, a small amount of interesting applications which propose an EMG interface to use while mobile and interact with mobile devices. Fistre and Tanaka [45] proposed an interface to control a portable music device based on EMG gesture recognition. The prototype is based on two EMG electrodes in the forearm and can detect 6 hand gestures related to the basic functions of a music controller, but no user testing was made. Forearm electrodes were also used by Wheeler and Jorgensen [53] when developing a system to detect simulated joystick and keyboard gestures but the intention of applying it to mobile and wearable devices was not achieved. More recently, Costanza *et al* [54] successfully prototyped and tested an interface for mobile devices. It is based on three electrodes in the bicep and wireless transmission to a portable computer to analyze the signal (Figure 2.5). The user is able to interact with the computer by reacting to specific cues, where an affirmative response is the contraction of the muscle and a negative ignoring the cue. Formal usability experiments were realized in [39].

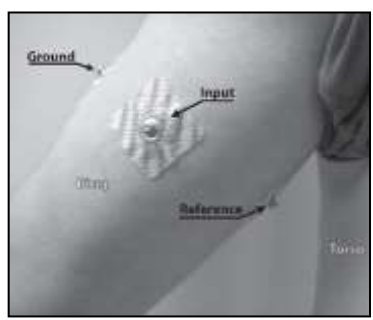

**Figure 2.5 – From Costanza *et al* [54], electrode position in the upper arm.**

### 2.4.2. EMG-Based Mobile Gestures Evaluation

The following evaluation is based on the potentialities and limitations of an approach based on Costanza *et al* work.

**Variety of possible gestures.** There is a large set of possible gestures to be made, because the same electrodes are able to detect subtle or larger gestures, and these electrodes can also be used in different muscular body parts. Besides, gestures are not made holding an object, making this interaction hands-free.

**Implementation Cost.** We consider the cost of this approach to be high. Wet or dry electrodes, cables, wireless emission devices and receivers have to exist to install an electromyographic system. These components are not common and they will not be part of standard mobile devices

**Social Acceptance.** Using an EMG-based system it is possible to recognize subtle gestures that would be hardly noticeable, thus having a comfortable usage and good social acceptance. Electrodes can be placed without being noticed.

**Recognition Complexity.** Gesture recognition using EMG is well studied and it is possible to implement in mobile devices. However, if trying to recognize many different gestures, more time-consuming implementations using pattern recognition have to be considered.

**Self-containable hardware.** One of the main issues is the need of dry or wet electrodes (preferably dry for user comfort) and the possible existence of cables or wireless modules to receive electromyography signal. This amount of additional hardware would probably make this technology hard to accept by most users.

**Accessibility.** Electromyography is generally used to bring HCI to people with some disabilities. If using EMG, mobile devices would be controllable by people with severe motion injuries.

**Accuracy.** Continuous interaction is not possible because the possible chosen movements may interfere with normal activities of the user, and would create too much false positives. The solution is to perform gestures under specified contexts such as reaction and response to audio/visual cues, but this problem largely reduces interaction possibilities.

## 2.5. Touch Screens

Pressure sensitive surfaces are commonly used in some mobile devices, usually integrated with their screens. The technique behind touch-enabled screens is based on a panel that covers the viewable area of the normal screen. This panel generally has an electric current changed by the touch, which allows the determination of the screen's touch location. There are essentially 4 main techniques for touch sensing devices [43] – capacitive sensing, infrared detection, resistive membrane and surface acoustic wave detection.

### 2.5.1. Touch Screens as a Mobile Gestural Input Technology

Touch screens are usually present in wearable computing as a graphical pointing interface in devices like watches [41] or vest-integrated wearables [40]. Besides, PDA's and other commercial mobile devices already have touch screens, mostly to provide stylus interaction. They are used as a technique to provide a pointing paradigm to select icons or menus, but they are also appropriate to create visually undemanding 2D gestures [33] such as characters or simple strokes using a finger or a stylus. One of the most well-known gestural input applications in PDA devices is the calligraphic recognition software "Graffiti" [34]. A lot of research has been done in this area, and there are actually some free source software packages that provide handwriting and stroke recognition for desktop and mobile devices.

One of the most relevant works in this area was made by Pirhonen *et al* [30]. They designed and implemented the "Touch-Player", a music player controlled with simple metaphorical gestures and with non-speech audio feedback, using an *iPAQ* portable device. The device was placed on the belt with the screen facing outside, and the user was able to control music when *on-the-move* with simple finger gestures, like a sweep right-left to the next track, or a simple tap to play and pause. The feedback was given by pre-recorded stereo sound samples played through headphones. Mobility studies demonstrated that Touch-Player is a usable and low-workload interface for interaction *on-the-move* with the device, in contrast with a visually-demanding media player. Brewster *et al* [33] also focused on metaphorical gestures to interact with wearable devices, using 12 directional single and multi-strokes, featuring also alphanumeric and geometric symbols, audio feedback and double-tapping on the screen to abort gestures. Kostakos and O'Neill [32] have a similar work, studying single and multiple direction strokes as a form of input in wearable devices, testing it in a wearable computer with touch screen and pen-based input. They state stroke gesture recognition method as "usable on small devices with limited processing capabilities and small input areas".

All these works do not use gesture potential to provide shortcuts to applications in mobile devices. In this area, Friedlander *et al* [31] suggested an eyes-free type of menu to be used in limited screen devices. It is based on a ring of options (which can be selected by directional strokes) and audio feedback when the user moves across a menu item. Even this technique had some good results it has a clear lack of metaphors when it comes to the interaction, making it hard to retrieve where a specific application is in the concentric rings.

### 2.5.2. Touch Screen Mobile Gestures Evaluation

The following general evaluation of touch-screen gestures takes into consideration an interaction method based on Pirhonen *et al* work, where users can perform finger gestures when mobile with a touch-screen device placed on the belt.

**Variety of possible gestures.** It is only possible to recognize 2D gestures, such as numbers, characters, geometrical forms or simple drawings, which reduces the possible metaphors.

**Implementation Cost.** There is no cost because many mobile devices, especially PDA's, are already equipped with touch screens.

**Social Acceptance.** Gestures recognized using this technology might be considered very subtle, while standing, sitting or moving. Since gestures are almost not perceived, and there is no special visible hardware to be used, this is a suitable method to be used in any social environment.

**Recognition Complexity.** There are already PDA's able to recognize finger-drawn characters. There are no software issues, and some open source libraries could also be used.

**Self-containable hardware.** Besides the mobile device with touch screen, users need also a belt to place it, but those are affordable and easy to find.

**Accessibility.** A touch-screen based interaction only needs hand movements to be performed, and does not use any buttons. This characteristic makes this method suitable for motion disabled individuals that still maintain hand and some arm mobility.

**Accuracy.** Users must be careful when using the touch screen, and avoid hand contact with it when mobile, but we believe this is not a significant issue, since it can be resolved with a "hold" button.

## 2.6. Discussion

In previous sections we reported the most important related work in the top 5 different areas of gestural mobile interaction. However, some other interesting works and technologies were not mentioned. For example, capacitance sensing may be used on clothes to detect the proximity and position of a conductive object (a hand, for example) [61] and infra-red laser beams serve an in-ear wearable device to recognize moving fingers [38]. However, in the present section we will not mention these works, and will focus only on RFID, Accelerometers, Touch Screens, EMG and Mobile Cameras.

### 2.6.1. Comparative Results

Table 2.1 summarizes the conclusions made during the analysis on the 5 different technologies.

| | Possible Gestures | Implementation Cost | Social Acceptance | Recognition Complexity | Self-Containable Hardware | Enhanced Accessibility | Accuracy |
|---|---|---|---|---|---|---|---|
| RFID | Single or multiple point-to-point recognition | High | Good | Low | No | No | Action Button needed |
| Accelerometers | 3D Hand and Arm Gestures | Medium | Good | High | Yes | No | Action Button needed |
| Cameras | Rotation, Tilting, Left/Right or Up/down | Low | Good | High | Yes | No | Action Button needed |
| Touch Screens | 2D Gestures | Low | Very Good | Medium | Yes | Yes | No |
| EMG | Muscle Contractions | High | Very Good | Medium | No | Yes | Yes Limits Interaction |

**Table 2.1 – Comparative results on Technologies Characteristics**

Each one was compared on the previously defined characteristics, and some proved to be more distinguishing between methods. One of those characteristics is the possible set of gestures. In this field, accelerometers are the most complete technology, enabling a wide variety of recognizable gestures. Using touch screens, we also have many possibilities, but only on a 2D plan. RFID tags and Cameras recognize simpler gestures, namely point-to–point gestures and small hand movements respectively. EMG is the most limited sensor, because it only recognized small-scale gestures such as muscle contractions. In terms of implementation cost, there are options that are more accessible, such as using cameras or touch screens, while RFID and accelerometer may soon exist in commercial mobile devices. The recognition complexity on technologies such as RFID, EMG and Touch Screens was considered simpler than when using inertial sensing or vision-based algorithms. Finally, one of the most important characteristics is the hypothesis of having a recognition only based on sensors present within the mobile device. In that context, Accelerometers, Cameras and Touch Screens provide that type of interaction, while EMG needs surface electrodes and RFID the existence of tags on clothes to interact with. One final consideration on the usage of an action button to provide a flawless gestural interaction: since muscle contractions or arm and hand gestures with mobile devices are common during the day, a full-time recognition would lead to too many false positives.

### 2.6.2. Objective-Driven Technology

Each of the technologies has, as proven in the different works described during this chapter, characteristics that make them especially useful under different circumstances and to reach different goals. When trying to implement a mobile gestural interface, developers face various difficulties and objectives, and the technology to be used has to solve problems and accomplish objectives. We defined some common objectives and selected the most suitable technologies for each one. For other goals, one would have to consider the evaluations we have made for each technology and select the most appropriate one.

**Rapidly test a gestural interaction technique.** Gestural interaction with computers is still under development and facing all the inertia that exists when trying to create new forms of interaction. Thus, sometimes developers face the need to test these new options before they are fully developed. Regarding mobile gestural interfaces, there are technologies that provide easy gesture recognition and enable a faster testing on a large set of gestures. For 3D gestures, RFID permits the recognition of point gestures within the body-frame without needing any complex algorithmic solution. Besides, Touch Screens gestures are also a good choice because they are based on available open-source algorithms to recognize 2D gestures and simple gestures can be recognized with EMG with signal thresholds and a correct placement of electrodes. If trying to make preliminary tests on a mobile gestural interaction method, those are the best choices.

**Recognize Arm Gestures.** If the gestures to be recognized are mainly arm gestures moving the mobile device on the 3D space, the most powerful and versatile choice is the usage of accelerometers. This sensor has many interesting applications on mobile devices, even outside the scope of gesture recognition, so it is possible that most mobile devices will be equipped with one accelerometer in the future. With an accelerometer, is possible to detect simple gestures to more complex ones, using diverse algorithmic approaches that can be found through literature on this area.

**Recognize Hand Gestures.** Accelerometer-based algorithms are also capable of recognize hand gestures with a mobile device, such as tilting or side movements, but if those gestures are the main focus of a work, we believe that using mobile devices cameras is a more correct approach. A work using cameras aims to a wider range of mobile devices, while maintaining a good set of recognizable hand gestures. For 2D finger gestures, the common approach always uses touch screens to interact with, but cameras and laser beams [37] might also be used for similar purposes.

**Provide a Subtle interaction.** If the set of gestures to be detected is not intended to be perceived by others, there are two major interaction choices. In one hand, EMG provides a subtle interaction and can be used with various muscles. However, the interaction is limited to response gestures because of the false positives issue. On other hand, it is also possible to use touch screens to make 2D gestures in a subtle manner.

**Enhance current mobile devices.** It is possible to create useful gestural interaction capabilities using sensors already available in the major part of mobile devices and PDA's. Those sensors are the Cameras and Touch Screens. If a development group is interested on an easy and large-scale deployment of gestural interaction techniques for mobile devices, these technologies should be used.

**Use gestural interaction to provide more accessibility.** In general, gestural interaction without visual feedback is already an accessibility improvement for blind people, but this is valid for all technologies. One of the most interesting technologies to enhance interaction with mobile devices for motion-disabled individuals is EMG, since subtle muscle contractions serves as interaction modality, and these users would not probably reject the usage of some electrodes in order to permit interaction with mobile devices.

## 2.7. Conclusion

In this chapter, we performed a research on the available technologies that are able to enhance mobile devices with gestural recognition capabilities. With the research, we found a main block of five different technologies that have been used to this purpose through the latest years: Radio Frequency Identification, Cameras, Accelerometers, Touch Screens and Electromyography. Each one of these technologies provides different advantages and has to be chosen accordingly to interaction objectives. In general, EMG and Touch screens may be considered a more limited technology for a gestural recognition purpose, because they only recognize a reduced set of gestures, but those gestures are very subtle and may be performed by disabled people. Cameras are useful to recognize hand gestures for specific applications and RFID, even using extra hardware and only recognizing point-to-point gestures, are a good solution to test certain gestural interfaces. Finally, accelerometers are considered the most powerful and versatile technology on this area, as they are now starting to be introduced on commercial mobile devices. With accelerometers, it is possible to detect a wide range of arm and hand gestures, as well as important contextual information. Developers should carefully select the most adequate approach and adapt it correctly to provide mobile devices with a suitable gesture-based interaction to be used under different circumstances.

# 3

# Mnemonical Body Shortcuts

The research on technologies that support gestural recognition with mobile devices was the first step to realize that is technically possible the creation of our own approach to this subject: the recognition of the Mnemonical Body Shortcuts. We define Mnemonical Body Shortcuts as gestures made using a mobile device towards different body parts, resulting on the triggering of applications within the mobile device that are culturally or personally associated with that specific body part. Such an interface will ease shortcut remembrance while maintaining the natural aspects of a gestural-based interaction and the advantages of using gestures while *on-the-move*. Through the following sections we will introduce in detail this approach. In section 3.1 we will begin with an overview on the preliminary studies we have made on the current panorama of mobile devices usage. Following those studies, we present the found problems and the main design guidelines for a successful gesture based system. In section 3.2, we take an in-depth look to the Mnemonical Body Shortcuts method we intend to implement. We also describe some possible use scenarios of our gestural interaction method and present an experimental validation of the concept using a RFID-based prototype. Finally, section 3.3 concludes this section with an analysis on gesture recognition using inertial sensing (accelerometers), namely its characteristics, potential and a description of the most used algorithms. This final section gives the essential background to a better understanding of the accelerometer-based prototypes descriptions made in section 4.

## 3.1. Preliminary Studies

Our work has a main objective: the creation of a new tool that enhances the interaction with mobile devices within the various environments where they can be used. The start point of such work has to begin with a correct evaluation of the current panorama on the usage of mobile devices, with special focus on the creation and triggering of shortcuts. This evaluation will guide to an analysis of the existing issues on the area and the main goals to be achieved when designing a gestural interface. If those goals are kept in mind during development, user difficulties when interacting with mobile devices may be correctly surpassed.

### 3.1.1. User Observation

In order to capture the actual panorama considering shortcuts in mobile devices, 20 individuals were interviewed and observed (Appendix A). The task analysis consisted on a first part with questions about current habits on mobile phone interaction, to know the type of mobile device users have, which are the most used applications, frequency of use and finally if and how users interact with both key and voice shortcuts. In a second part, users were asked to reach the most common applications and contacts. They performed those actions while observed in a controlled environment, and the numbers of buttons pressed for each action was registered.

First part results present some already expected conclusions: the majority of users have classic mobile devices instead of PDA's and use them more than 10 times per day, usually to make calls, send SMS, consult the contact list, agenda, clock, set the alarm clock and take and visualize photos. The average number of the most used contacts was set on 6 (most used contacts are contacts that users call at least one time per week). Results on shortcut usage reported that 75% of the interviewed uses key shortcuts, while none used voice shortcuts. When asked about why they do not use voice shortcuts, three main reasons were presented: they are not available in their mobile device; they used it but the recognition rate was low in many situations; finally, usage limitations under diverse social environments. It was clear that the remaining analysis had to be focused on the current habits on key shortcut interaction. We concluded that an average of 5 programmed key shortcuts is used, and 93% of the users execute them on a daily basis. When asked about memorization issues on their key shortcuts, we observed that users with more programmed shortcuts reported more difficulties, and they stated that because of that difficulty they generally only use a couple of shortcuts.

In observation, results show that users need an average of 4 keystrokes to access the 3 most personally common applications and 5 keystrokes to call the 3 most used contacts. In fact, users were more likely to choose menu selection when prompted on the applications and contacts they defined as the most used rather than using key shortcuts available in the majority of the cases. The usage of menu selection was reflected in a larger number of keystrokes and task errors, resulting on a slow selection of the wanted task. In Appendix B we present the answers to the 11 Task Analysis questions that followed the results present in Appendix A.

### 3.1.2. Problems

With the user observation we performed, some issues on current mobile interaction were found. Firstly, it is clear that voice recognition is still not used by a general audience. The most important reasons are not only the inexistence of this modality on many mobile devices but also the difficulties that this method presents when used on noisy or socially busy environments. Voice recognition on mobile devices still has a long way to be able to perform well on complex environments, but there are some social issues inherent to this kind of interaction that can not be surpassed. In the context of key shortcuts, the most important problem seemed to be their memorization and efficiency. The memorization problem was referred by users with many programmed shortcuts, but this issue will be further tested as described in section 3.2.5. User observation showed that, even when asked to perform actions that should be rapidly repeated using available key shortcuts, users spent a large number of key presses and often returned to the classical menu operations to reach the intended application. A conclusion to be made based on these results is that mobile interaction, even most of the times based on repeated actions, is still slow, keystroke consuming and does not give full appropriate support to rapidly reach those actions

### 3.1.3. Design Guidelines

After user observation, we concluded that mobile interaction still has diverse issues that need to be addressed in order to provide a more usable mobile interaction. Using that knowledge together with the main issues that are also present or referenced in the literature on this area, we can list a set of design guidelines for a mobile gestural interface that we intend to accomplish:

**Support Shortcut Memorization**

The expressivity of a gesture is a powerful tool to create metaphors and mnemonics when interacting with computers. This tool is often given a small use because the same gestures are used in most of the works. Generally, simple directional strokes, tilt or characters or are performed with gestures to interact with the device. Those gestures, although useful, are not practical to a wider range of applications and do not always provide a correct creation of memory aids to associate gestures with actions. There are two options to make them more natural to the user. The first one is trying to create mnemonics to static operations on a device, such as twisting the hand simulating key unlock which could unlock the cell phone, point it to the sky to retrieve meteorological information or associate gestures with body positions as it was did in Ängeslevä *et al* project [64]. One other approach is the personalization of the gestures. If users could make their personal gestures, they would apply their memories, personal information and subjective thoughts making gestures meaningful to them. A good approach should be using some default gestures with well defined mnemonics, suggest some others but let the user to choose its own gestures to interact with the mobile device.

**Give appropriate Feedback**

The major part of current gestural input applications have researched in the implementation problem, technical solution, algorithmic difficulties or possible gestures to be used. However, there is a main problem that is forgotten in a general way – users need suitable feedback. Gestural input is normally intended to achieve non-visual interaction, making visual feedback less important. However, when screens are used, it should be possible to retrieve visual feedback as an auxiliary method. Prototypes that implemented some feedback generally use it in audio format to advise the users about the state of the gesture detection. The origin of the audio is not usually studied. It can be performed by the device, but it is not appropriate for specific social environments. Earpieces are a good solution, especially when users are already listening to music, but they would hardly be used only as a feedback channel. The type of sounds can be single beeps, multiple beeps or continuous increasing/decreasing sound. When gestures are performed with the mobile device in hand, vibrational feedback also has a great potential to inform user with subtleness and give the gestural action more haptic sense. Works in gesture input interaction should analyse which is the most appropriate feedback to the diverse range of applications, and test that different approaches with users, focusing on audio and haptic feedback. Applications should also give users the needed tools to control the confirming/cancelling each gestural action, namely by using a small time window for confirmation purposes.

**High Recognition Rate**

A good gestural interface has to be supported on an excellent recognition rate. As we have noticed in user observation, the lack of a good recognition rate of voice shortcuts is sufficient for users to drop its usage and go back to a button-based interaction that they are accustomed and intend as reliable. Users have to be confident on the system and know that it will respond as it is supposed to.

**Grant social and user acceptance**

There have been some commercial applications using gestural input, but the fact is that they are not common and seem to have some problems when entering the market. There is a full hand of mobile devices with gesture recognition but they are not a success (many were discontinued). The main problem seems to be the user acceptance on this novel method and also the social implications that actions based on gestures can have. User acceptance seems to be the minor problem, because it should only take some more practice to make users more interested in this natural technique. Social acceptance is an issue that has to be carefully analysed. Some people might be constringed to make gestures in a public area because it is usually not accepted if it does not come along with a clear action or speech. These social constraints may limit the use of gestures when interacting with mobile and wearable devices. One possible solution to this problem exists in a new area of thinking in HCI, the Aesthetic interaction [27][22]. The idea is simple: if aesthetics in GUI are generally given so much importance, this concept should be also be used in other areas in HCI, taking in account not the graphical aesthetics but the aesthetics of the interaction. The two referenced follow the same guideline and are a sample of the growing interest in this area applied to gestural input. In Peterson *et al* work [27] it is suggested that interfaces should be designed considering social and cultural background, link the mind and body and take in account the instrumentality of the interaction. If some of these issues are carefully studied when designing gestural interaction, it is possible that their success will arise and become a generally accepted technology. Besides, the chance of having personalized gestures also opens the door to an interface with better user acceptance.

**Allow Mobile Interaction**

When developing an interface for mobile devices, we should not forget that they are intended to be used while standing or sitting, but also while moving. When mobile, visual attention has to be focused on other main task, especially for safety reasons. For example, it is not unusual to see regular users of mobile devices sending text messages without looking to the mobile device, freeing their eyes to perform other actions at the same time. A desirable gestural interaction platform has to provide sufficient tools to be used in different mobility settings so its potential of providing a rapid and natural access to the main functions of the device can be fully explored. One of the most important features to achieve a suitable mobile interaction is also the existence of a suitable and personalized feedback, compatible with the variant environments where the system can be used.

### 3.1.4. Usability Goals

After user observation and the definition of the design guidelines for a mobile gestural interface, we were able to specify a set of usability goals we intend to accomplish in our shortcut-based gestural interface:

**Shortcut triggering error inferior to 10%**

In the design guidelines we reinforced the importance of an appropriate feedback, user-control mechanisms and a good recognition rate. All these characteristics should work together to make a robust gestural interface, capable of triggering gestural shortcuts correctly in the most part of the situations. Since this objective is based on many different variables, we defined a minimal value that should enable users to have confidence that this system would trigger the intended shortcut: in at least in 90% of the situations it should trigger the correct application, supported by a good gesture recognizer but also by a suitable user-control and feedback mechanisms.

**Users should need, in average, less than 4 clicks to trigger an application**

During user observation, we found that users needed an average of 4 keystrokes to access the 3 most common applications and 5 keystrokes to call the 3 most wanted contacts. Even knowing that our approach is gesture based, users will still need to click buttons to, for example, trigger the gesture recognizer. With our approach, we intend to reduce the observed average number of 4 clicks to trigger an application.

**Time to trigger an application should be less than 5 seconds**

When using shortcuts, it is extremely important to rapidly recognize users' intention. We did not registered the average time that users take to trigger an application with voice or key shortcuts, but with our approach we should be able to trigger shortcuts within a similar time frame of those two approaches. We defined as acceptable an average time less than 5 seconds to trigger a shortcut.

**Unintentional stops during shortcut triggering should not happen in more than 5% of the gestures**

One of the main advantages of using a gestural-based shortcut system is the possibility of using it while mobile, thus we intend that no more than 5% of the gestures take users to stop their movement to correct any error or doubt that has as its origin the gestural shortcuts user interface. This objective is crucial to allow a correct mobile interaction, as defined in the Usability Guidelines.

**Gesture memorization errors should not happen in more than 10% of the gestures**

When triggering a key shortcut, is essential to have a well-defined map for where is each of the applications in the keypad. The same happens to a gestural interface, and in our opinion the system should be supported in a method that enables users to correctly remember how to perform their gestures to reach a certain application. Memorization errors regarding long-term memory should not occur in more than 10% of the cases.

**At least 50% of the tested users should be willing to use the system.**

All these usability objectives have a relation to the design guidelines we defined, and an essential role in the final result of the interface. However, they may all be accomplished, but if tested users did not respond positively to the system, it would be hardly successful. Due to that consideration, we determine that a minimum of 50% of the users that experiment our system should be willing to use it in the future, when it would be available in commercial mobile devices.

## 3.2. Body Space Interaction

Based on our empirical knowledge about mobile interaction and the observation described in 3.1, we propose a new interaction method for mobile devices, capable of changing the way mobile devices are used. This new method is based both in the powerful characteristics of gestures when used in HCI but also in the body as a rich repository of different meanings and personal associations.

### 3.2.1. Gestures in Human Computer Interaction

Gestures are a communication method commonly used by humans. The first picture that may appear in our minds when referencing gestural communication is the sign language, mainly used by deaf individuals, but gestures are constantly used by non-disabled people. We use gestures by maintaining different body postures, facial expressions or making gestures with our hands and arms. Furthermore, when people with different languages meet, their communication has to be based on gestures. This fact shows one of the most important characteristics of gestures: they are a universal form of communication, and people have intrinsically recorded the meaning of many gestures that are valid all over the world.

Interaction system developers had, in the last decades, noticed the importance that gestures can also have in both directions of the communication between computers and humans. Nowadays, it is possible to find different gestural interaction platforms to control computers, either based on camera recognition, inertial sensing, touch screens and many other methodologies. The second communication link is also being studied and applied, enabling the simulation of human gestures on virtual environments agents or robots, thus creating more a more realistic interaction. Following the course of development, it is inevitable that gestures are studied to be an interaction modality also in the recent field of mobile computing, transporting all the advantages of a gestural interaction to this new area.

### 3.2.2. Human Body: A Rich Meaningful Space

The human body is a set of diverse parts where each one plays a different role in what we call "life". The hands are our work tool, our brain the space where all the decisions are made, the mouth and tongue essential not only to feed us but also in the way that we communicate with each other. Our body is a space densely rich on functions, and because of that fact it is possible to think those diverse parts as a symbol for emotions and actions. For example, the heart, one of the most important organs in our bodies, is often related with emotional feelings, our hands are related with physical work and our shoulder with our intention to comfort someone in hard times. The human body is full of this type of associations, which are typically transversal to many societies, but can also be very personal and intimate.

### 3.2.3. Mobile Body-related Mnemonical Gestures

One idea emerge from the last sections: we have, at the reach of our hands, meaningful tools that are often used to communicate. Those tools are the gestures we can perform but also our body as a strong meaningful space. In fact, gestures are often combined with body hints to emphasize an idea. There are many examples of the relation between gestures and body parts: someone reaches his heart to apologise, touches his mouth to ask for silence or puts a hand in the head when he forgot something. When we combine gestures with body parts, a whole new set of possible relations and meanings appear, and they can be universal like those described, but they can also be very personal. (Figure 3.1)

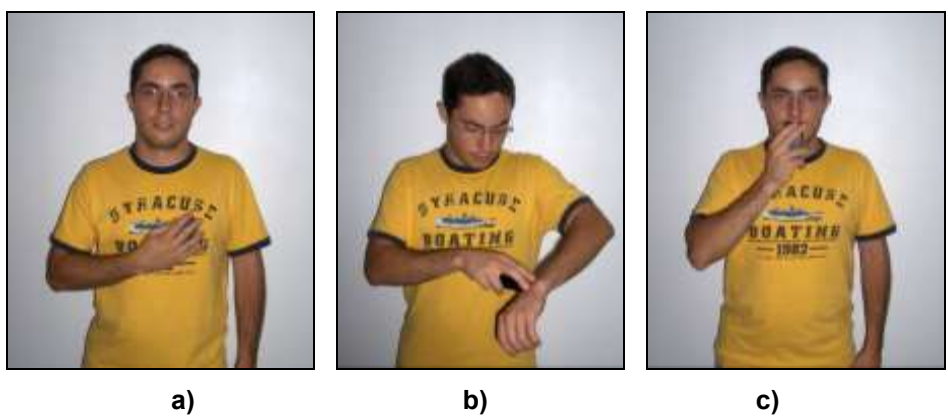

a)     b)     c)

**Figure 3.1 – Mnemonical Body Shortcuts a) Gesture to the chest; b) Gesture to the wrist; c) Gesture to the mouth**

A gestural interaction with mobile devices might be created using a free set of gestures, with any direct relation with the task that is intended to perform, but such approach would fall in the same memorization issues that exist while using key shortcuts. We are convict that cooperation between the possibility of making gestures with a mobile device and the ability to direct it to a body part may create diverse strong mnemonical cues that could be easily remembered and performed by users, in distinct mobility conditions. It is easier to remember how we perform a gesture towards a body part than gestures that are performed with any type of mnemonical aid. We will, from now on, reference these gestures as Mnemonical Body Shortcuts. This gestural concept for mobile devices is not new. We were inspired by Ängeslevä *et al* [64] work, where they presented the theoretical background and preliminary studies on the possibility to associate gestures with parts of the body and trigger applications on a mobile device using those body space mnemonics. This study, dated from 2004, also justifies that users have the needed accuracy to make the gestures towards body parts and they can create multiple meaningful associations between the applications existent on mobile devices and the body space. However, the work stagnated on these considerations and on a low-detail description of a non-tested algorithmic approach to recognize those gestures, and left much work to be done. It is important to validate the concept of Mnemonical Body Shortcuts with users, test if this method really enhances memorization, when compared with the most used type of shortcut interaction. Besides, a full implementation has to be developed, described and tested in detail, with proper feedback, supporting gesture personalization with a high level of recognition rate. We believe in the potential of this approach and want to close the blanks left open by the previous work, refreshing the idea and preparing it to be used on commercial mobile devices.

### 3.2.4. Use Scenarios

There are some possible use scenarios for a system based on the Mnemonical Body Shortcuts concept:

**First Scenario**

*After two weeks of practice, Peter was already familiarized with its new mobile device with motion sensing capabilities and the new functionalities it has available. When he bought the mobile device, he started to define a set of gestures to be used as shortcuts to some existent functions in the device. Later that day, Peter had scheduled a party in a nearby friend's home. When on the party, and since his girlfriend was very late, Peter grabbed the phone and made a gesture towards his hand, what automatically accepted the gesture with a small vibration and opened the SMS editor in the mobile device. It was the first time that Peter made gestures with the mobile device in public, but he felt comfortable as no one seemed to notice it. Because his girlfriend did not respond to the SMS he sent, he tried to phone her, making a gesture to the heart. She answered and told him she was arriving at that moment. The rest of the night was truly amazing and he even had time to show his friends how he is capable to interact with gestures with his new cell phone!*

**Second scenario**

*While walking on Marquês de Pombal metro station in Lisbon, Alice suddenly felt bored, especially because she remembered the long way she still had to accomplish. She plugged in the earphones on her cell phone and made a gesture with the cell phone to the ears, which launched the music player. She heard the sentence "Music Player" on her earphones and knew that her gesture was well recognized and now she could hear her favourite music. When leaving the station, she noticed a really funny street clown that was doing an elephant-balloon figure just outside the metro entrance. She rapidly took her mobile device, approximated it to her eye to turn on the camera and took a picture of that moment, to show later to her parents when finally she arrives home.*

### 3.2.5. RFID Experimental Evaluation

Before we have started a full implementation of the Mnemonical Body Shortcuts, we decided to perform an evaluation of the concept and test if it would perform as expected. Accordingly to the conclusions present on the related work of this thesis, the best technologies to provide a fast implementation of gestural interaction for mobile devices are RFID, EMG and Touch Screens. Touch screens are clearly not able to reproduce body-space gestures, while EMG can recognize contractions on different body parts but our intention is to perform gestures with a handheld device and not all body parts have voluntarily contractible muscles. RFID appears as an excellent choice because it is possible to stick RFID tags on clothes and read them using a RFID reader in the mobile device, simulating gestures towards different body parts. This methodology gave us the possibility to make a preliminary test on Mnemonical Body Shortcuts with only a few weeks of development around the RFID prototype. This prototype was developed using a Pocket Loox 720 with a compact flash ACG RF PC Handheld RFID reader. (Figure 3.2 a)) In terms of software, the system was able to discard multiple tag reading and keep the log about all the tag readings during evaluation.

With the RFID-based prototype we were able to simulate the association of body parts (through sticker tags) with any given mobile device shortcut (i.e. an application or a call to a certain contact). The prototype was evaluated with 20 users in a controlled environment (Appendix C for protocol and results). In the first stage of the evaluation users were asked to select the five most frequently tasks effectuated with their mobile phones and associate them both with a body part and a mobile device key (in their own mobile device). Considering body shortcuts, it is interesting to notice that 89%, out of 18 users, related message writing with the hand, 88%, out of 17 users, related making a call to their ear or mouth and 91%, out of 11 users, related their contacts to their chest, among other meaningful relations (Table 3.1). An hour later, users were asked to access the previously selected applications, following both approaches (body and key shortcuts). For each of the approaches they were prompted randomly 20 times (5 for each application). Although several users selected already used key/application relations, 50% (10 users) made at least one error, with an average of 9% errors/user. Considering body shortcuts, only 15% (3 users) made a mistake with an average of 0.8% errors/user. Results were still very favourable for Mnemonical Body Shortcuts one week later, with an error rate of 22% for key shortcuts and 6% for the gestural interaction Results showed that, even against some established key shortcuts, gestural mnemonics had better results and may surpass the problem of low memorization of key shortcuts, providing also a wide range of possible associations, when compared with the physical limit of mobile devices keys. With these results, it is possible to state that Mnemonical Body Shortcuts concept accomplishes one of the usability goals we defined, because we observed that users remember gestural shortcuts using their long-term memory in 94% of the gestures, thus achieving the result of less than 10% memorization errors. These results were also a main motivator to follow this approach and find a solution that does not have the inconveniences of using RFID tags on clothes to perform gestures with a mobile device.

|  | Mouth | Hand | Chest | Head | Wrist | Eye | Finger | Ear |
|---|---|---|---|---|---|---|---|---|
| SMS |  | 10 | 1 |  |  |  | 6 |  |
| Call | 3 |  |  | 1 |  |  |  | 12 |
| Contacts |  | 3 | 5 | 2 |  |  |  | 1 |
| Clock |  |  |  |  | 10 | 1 |  |  |
| Photos |  |  |  | 2 |  | 8 |  |  |
| Calculator |  | 3 |  |  |  |  |  |  |
| Mp3 |  |  |  |  |  |  |  | 2 |
| Agenda |  | 1 | 3 | 1 |  |  |  |  |
| Alarm-clock |  |  |  | 2 | 2 | 2 |  | 3 |

**Table 3.1 – Most common associations gesture-application**

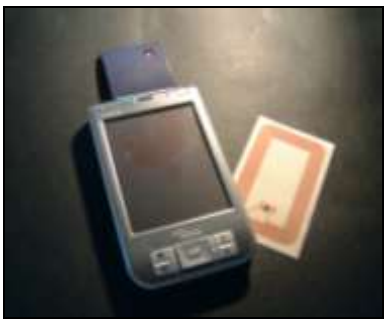 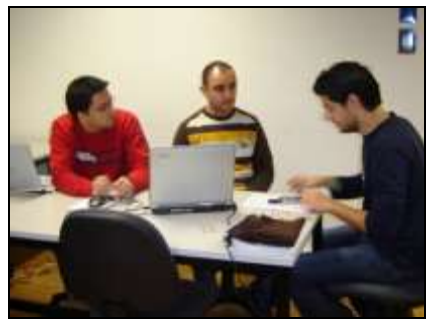 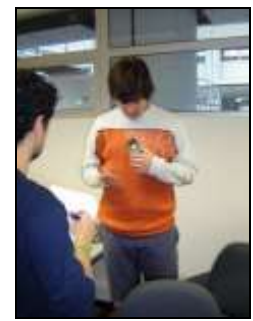

a)         b)         c)

**Figure 3.2 – a) Pocket PC with RFID reader and a RFID tag; b) and c) User tests**

## 3.3. Inertial Sensing

In this section we will introduce the area of Inertial Sensing, which refers to the measurement of both rotational and translational movement of an object, using an accelerometer. Firstly we will briefly describe how this sensor captures the acceleration and delivers the raw data. Then a brief description about the basics of gestural detection using an accelerometer will be made (Figure 3.3). We will also focus on the pre-processing of raw data and on the different data transformation methodologies and classification algorithms that are applied to the pre-processed data to correctly recognize different sets of gestures.

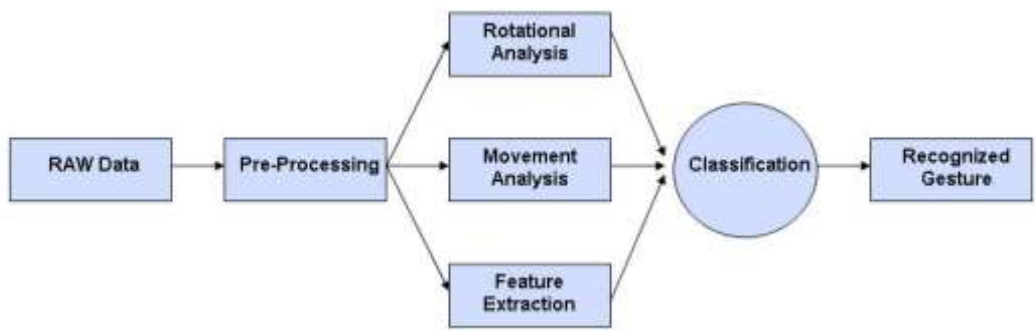

**Figure 3.3 – Possible steps for gesture recognition with accelerometer data**

### 3.3.1. Accelerometer

An accelerometer is classified as an Inertial Sensor because it is able to measure acceleration, possibly in multiple axes. It measures not only the dynamic acceleration (the acceleration provoked by a movement of the object attached to the accelerometer) but it also measures static acceleration (gravity force present in each axis). Nowadays, the majority of accelerometers are of the pendulous type, consisting of a proof mass attached to the rest of the sensor by a spring joint. When an accelerometer suffers some translation, the proof mass and the angle of the spring tends to be displaced for the original position. (Figure 3.4) These deformations are transformed to the output signal of the sensor. Actual accelerometers use MEMS technology, which reduced the cost and size of the sensor, allowing integration with microelectronics and micromechanics. Other technologies are still used in a smaller scale, like piezoelectric or capacitance-based accelerometers. The shift to MEMS technology enabled the usage of accelerometers in many industries: for example, they are now used to trigger airbags when an abnormal acceleration is detected. They are also being adapted to mobile devices, and one of the most mediatised devices, *Apple*'s *Iphone,* already has an accelerometer.

Accelerometers can be either analogical or digital, whether they output a continuous signal or a sampled signal. They can have a different amount of axis, depending on its purposes. For 3-dimensional space accelerations, it is needed a 3-axis accelerometer, but 1- and 2-axis are commonly used to other applications. Accelerometers with 3 or more axis can be mounted using more accelerometers together. This sensor can detect variable accelerations, measured in g (earth gravity, +/- 9,81 ms$^{-2}$), ranging from small amounts to hundreds of gs. Bandwidth is also variable, referring to the number of times a signal is read for second. For slow movements, 50Hz bandwidth is enough, but to detect faster accelerations a larger bandwidth is required.

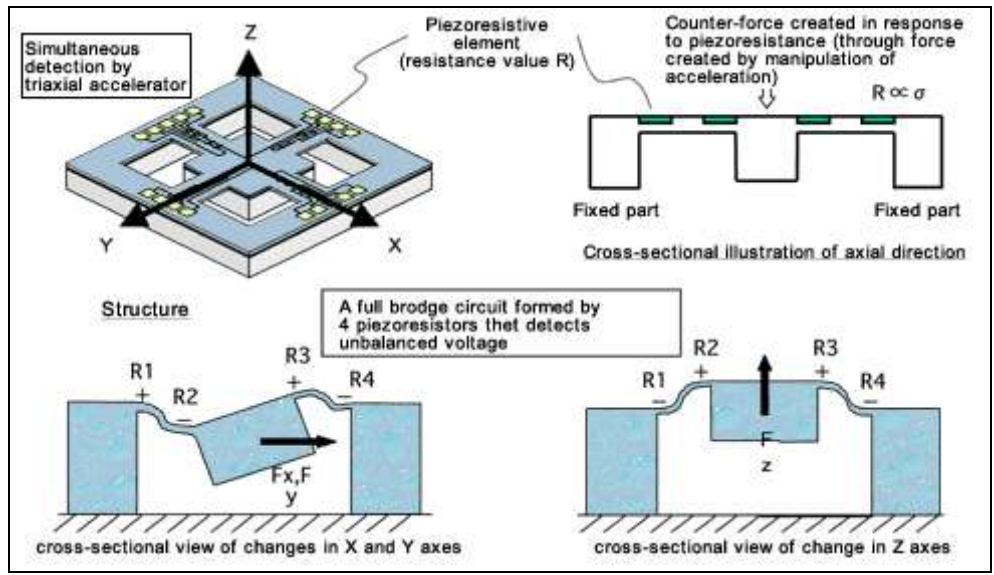

**Figure 3.4 – MEMS Accelerometer Functional Diagram [13]**

### 3.3.2.    Pre-Processing

The raw signal of an accelerometer has to be calibrated, because different accelerometers generally output different values. This calibration might be done using reference values. If the accelerometer is stable on a horizontal position, we know that the value on the perpendicular axis reflects the gravity acceleration of 9.8 ms$^{-2}$. If all gravity values in each axis are known, the rest of the values can be recalculated and the output of the accelerometer be given in ms$^{-2}$. After a well performed calibration, the same acceleration signal when the accelerometer is horizontally stable has to be near 0 ms$^{-2}$ on two axes and near 9.8 ms$^{-2}$ in the orthogonal axis.

The calibrated signal of the accelerometer is already a useful tool to be analyzed by many algorithms, but it is essential that more data pre-processing is done to clear the noise that is characteristic of this sensor. The accelerometer captures much noise because of its great sensitivity to movement, and the appliance of smooth filters such as a Moving Average may clear it. This algorithm applies a variable sized window starting in each point of the signal to generate a new signal based on the calculation of the mean value of the multiple windows. A larger window will reflect a more accentuated smooth, while a smaller window might have little effect on the final result. This approach can be easily implemented but it has some problems, especially in the start and end of the signal, because there is no signal before and after to calculate the mean value. Values on the centre of each window should be given more value than those values on the extremities. These issues are handled by algorithms that use

Gaussian-like functions to give more weight to values on the middle of the window (i.e. Gaussian, Hanning, Hamming or Blackman windows). Besides, it is possible to extend both the end and start of the signal with the inverted signal to have a better mean calculation on those areas. In Figure 3.5 we present the results of applying the smooth filter based on the Hanning window with two different windows (50 and 500)

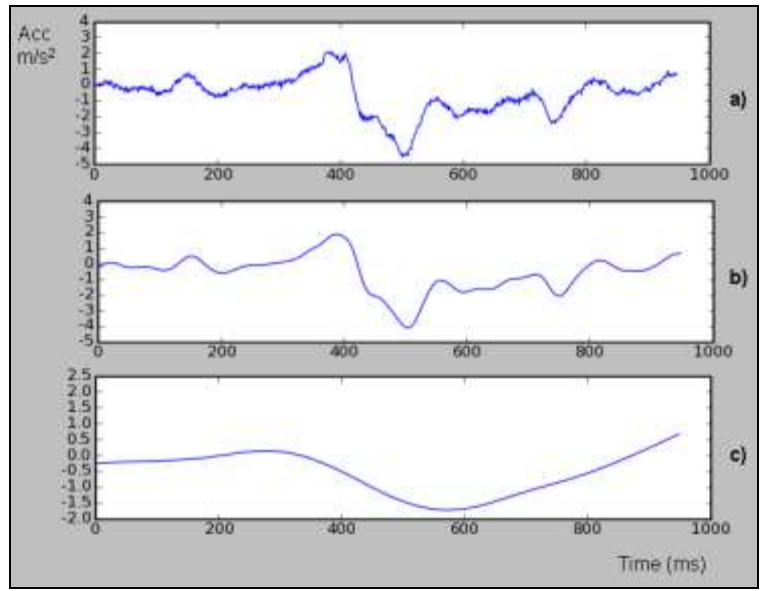

**Figure 3.5 – Signal Smooth using Hanning Window   a) Raw signal  b) Smooth with window = 50 c) Smooth with window = 500**

### 3.3.3. Rotational Analysis

Accelerometers are often used to make rotation measurements, substituting gyroscopes that are still an expensive inertial sensor. This is possible when considering that the object will not suffer dynamic movement, and all the acceleration detected is the gravity acceleration. We know if there is no dynamic acceleration if the amplitude of the signal (formula 3.2) in the three axes is calculated and it remains lower than 10 ms$^{-2}$. In those conditions, we are able to calculate the rotation of the accelerometer on one axis using the formula 3.1. This formula is applicable to all axes only when no dynamic motion is present. We also have to be aware that accelerometers have a limited degree of freedom when recognizing rotations. When a rotation is performed between two axes that do not have gravity to pass between them, it is not possible to recognize the rotation using this methodology. Furthermore, the rotation measurement using an accelerometer is considered to be less sensitive when an axis is turned away from the perpendicular, as they loose resolution. Finally, to provide a good rotation measurement to support gesture detection such as tilt interaction, it is important that the signal is centred in the beginning of the movement. This is done using the first values recorded by the signal and subtracting it from the gathered values during the whole movement. In this way, we start the rotation measurements always from the point where the movement started.

$$xrot = \arcsin(ax/g) \quad \text{(3.1)}$$

$$amp = \sqrt{x^2 + y^2 + z^2} \quad \text{(3.2)}$$

### 3.3.4. Movement Analysis

One of the utilities of having an accelerometer is the possibility to, given its acceleration data, calculate how it behaves during time, firstly calculating its velocity and then the position variation. In this case, contrary to what we have described in the last section, the only acceleration that interests us is the dynamic acceleration, since is this specific acceleration that provokes movement variation. Because of this fact, the first step to make movement measurements using an accelerometer is to centre the signal as described in rotation measurement (Figure 3.6 a)), but in this case we do not expect gravity to appear in any axis. If that happens, the gravity acceleration is considered as dynamic acceleration, creating a significant error margin on the final result. Using a gravity-free acceleration signal, we must perform a double integration (Formula 3.5). With the first integration (Formula 3.3, Figure 3.6 c)), we get the velocity variation during time in each axis. Velocities near 0 ms$^{-1}$ have to be considered 0 because that residual value would have a large effect on position results if accumulated during time. From the second integration (Formula 3.4, Figure 3.6 d)) results the position variation during time, in meters. This approach, even with some error brought by the double integration and the small rectifications we have to do on velocity, is able to calculate distances especially when the accelerometer is moved on the top of a table or in other controlled movements with minimal rotation

$$v(t) = v0 + \int_0^t a\,dt \quad \text{(3.3)} \qquad r(t) = r0 + \int_0^t v\,dt \quad \text{(3.4)} \qquad a(t) = \frac{dv}{dt} = \frac{d^2 r}{dt} \quad \text{(3.5)}$$

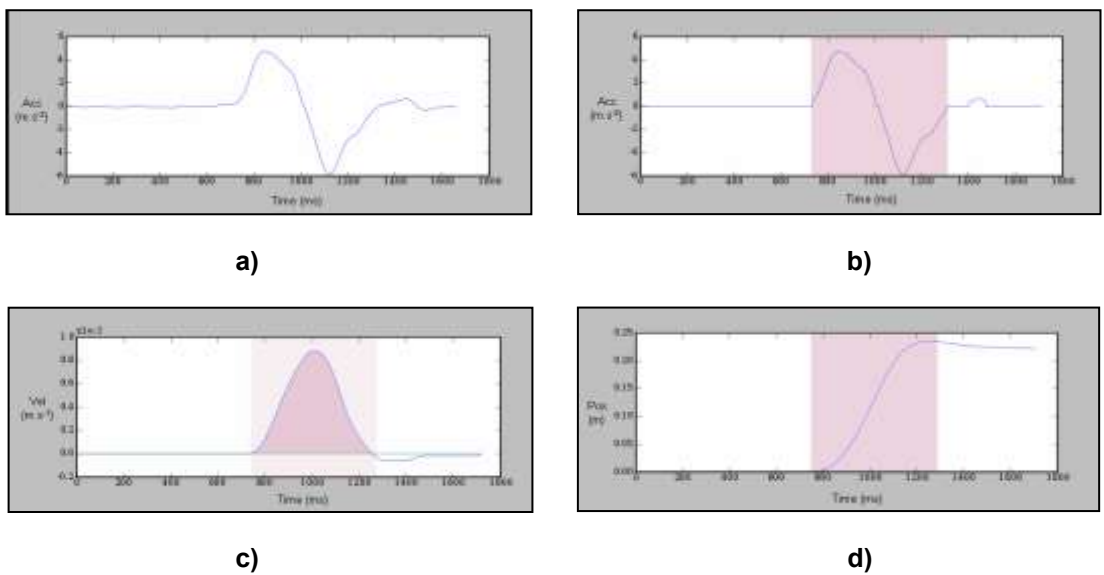

**Figure 3.6 – Movement Analysis Evolution a) Raw Centred Signal b) Filtered c) Velocity d) Position**

### 3.3.5. Pattern Recognition

As presented in the last subsection, both rotation and translational movements can be calculated from acceleration signals, but there are many issues when trying to use them altogether. For that reason, it is still very difficult to have an absolute measurement of a 3 dimensional position over time when freely manipulating an object based on the signals provided by a 3 axis accelerometer. It is only possible to gather the absolute position with some outside support, such as using cameras on infra-red communication. If that position was known, many approaches could be applied to recognize gestures, such as those applied in vision-based systems with markers. When using only an accelerometer, developers can use the relative position and acceleration of the device.

There are many alternative solutions to recognize gestures using the accelerometric data and detect different movement patterns. In general, pattern recognition systems intend to collect the raw data, categorize a pattern and perform an action based on that recognition. The process typically has 3 phases:

#### 3.3.5.1. Collect and Pre-Process Raw Data

The data is collected by sensors and then pre-processed in order to transform the raw data into more useful information (some examples for inertial sensing were given in 3.3.2).

#### 3.3.5.2. Feature Extraction

It is possible to extract measurements and properties that are able to differentiate between patterns, and they are named as features. Patterns are usually represented in an n-dimensional Euclidean space by its different features and are organized in different groupings named classes. A common issue in this phase is to select features with more information in terms of separation of the classes. The best choice is obviously the selection of those with more discriminating power between classes, and several techniques of automatic feature selection methods are available [11].

#### 3.3.5.3. Classification

The classifier module is responsible to categorize a pattern, using the set of extracted and selected features. Firstly, classifiers pass through a learning process, where sample patterns are used to give the classifier sufficient data to train the data models and then make the decisions. The training process might be supervised, when the sample patters are already classified, or unsupervised when the training set has to be organized in different clusters to detect possible data agglomerates. There are many different classifiers that can be applied to the task, namely K-Nearest-Neighbours, Neural Networks or probabilistic classifiers such as Naive Bayes or Hidden Markov Models (based on Bayesian Networks). Since the K-Nearest-Neighbours and Naive Bayes Classifiers will be referenced in the fourth chapter, we will make a brief description of both.

**K-Nearest-Neighbours Classifier**

In a K-Nearest-Neighbours (kNN) classifier, a pattern is classified based on the nearest neighbours present on the feature space formed by the training set (Figure 3.7) using some distance metric. A neighbour is more or less near the pattern accordingly to the distance function used, that might be i.e.: Euclidean or Manhattan distance (among others). The most common classifying method is to classify the sample in the class with more training samples in the group of *k* neighbours, but it is also possible to give different weights based on the distance or ranking of proximity [73]. The parameter k may be chosen taking into account the size of the training set but there are some heuristic techniques to define the value. A high k value may embrace too many unintended values while a low one may leave out too many neighbours that would help a correct classification. This classifier tends to have better results when a large training set is available, but it is computationally heavy for large training sets and may be incorrectly influenced by bad chosen features.

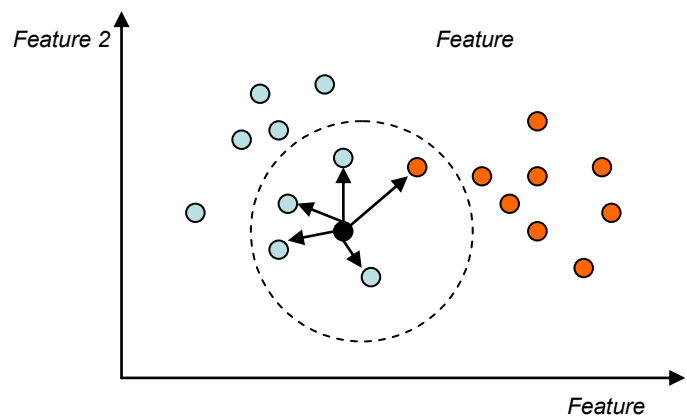

**Figure 3.7 – kNN example. The blue/red circles represent the training set and the black circle is the test object. In this case, it would be classified as blue for a parameter k of 5.**

**Naive Bayes Classifier**

A Naive Bayes classifier [42] is a probabilistic model based on the Bayesian theorem (Formula 3.6). It is based on 2 main characteristics, the prior probability and the likelihood. The prior probability is calculated based on the number of patterns for each class: if a class is present in a larger number on the training set, it is possible to *a priori* define that class as more likely to happen than other with a minor training set. The likelihood is computed by assuming a data model and using the training set to estimate the probability density function. Both prior probability and likelihood are used to generate the final classification decision, producing the posterior probability from the application of the Bayes theorem. The pattern is recognized as belonging to the class with the major posterior probability. Even simple, this classifier is known to have a better performance than many complex classifiers, and is especially useful for a large feature space with a small training set.

$$P(A|B) = \frac{P(A)P(B|A)}{P(B)} \quad (3.6)$$

# 4

# Inertial Sensing Prototypes

The research present on chapter 2 was concluded with the consideration that accelerometers are, in the actual panorama, the most powerful technology to provide mobile devices with an efficient and versatile gestural interface. RFID emerged from the same chapter as a technology suitable for a rapid development and testing. Following that conclusion, we developed a RFID prototype described in section 3.2.5, with the objective of testing the concept of Mnemonical Body Shortcuts. As we already expected, the use of an RFID prototype, even with a high recognition rate and being extremely appropriate for an implementation of Mnemonical Body Shortcuts, was not appropriate for a full-scale deployment, mainly because users rejected the possibility of using RFID tags on clothes on a daily basis. It was clear that we had to take another perspective on the implementation, and accelerometers appeared as the best choice. Technically, there was no clear limitation regarding the possible gestures to be recognized, so it is theoretically possible to, using accelerometers, recognize gestures towards different body parts.

In this chapter we will describe the implementation of the diverse prototypes we have developed using accelerometers. Firstly, in section 4.1 we reference the material we used throughout the work, while section 4.2 discusses the preliminary tasks we have performed to develop a better understanding of accelerometer usage. Sections 4.3 and 4.4 describe, respectively, the first and second inertial sensing prototype. The first prototype was based on the combination of both movement and rotational analysis while the second prototype uses feature-based classification. The final prototype, reported in section 4.5, uses the prototype with better recognition performance as basis and completes the Mnemonical Body Shortcuts interface with new user-control facilities and also visual, audio and haptic feedback.

## 4.1. Working Material

The ideal development set one could have to develop a gestural interface for mobile devices would be the existence of a cell phone or PDA with an integrated accelerometer and an open API. In the actual market, we know two models with these characteristics, *Apple*'s *Iphone* and *Nokia 5500*. The *Iphone* was not released at the start of our work, and both *Apple* and *Nokia* models are still expensive. Excluding those two mobile devices, we had two options available. The first one was to use the *Nintendo Wii* remote controller, but this option, even appropriate for an algorithmic development, was not very strong because it would not enable us to simulate the interaction with a mobile device. The second choice, and this is the one we followed, was to use a Wireless Bluetooth transmitter with a cable-linked accelerometer. This device (a 14x8x3 cm lightweight box) named Bioplux4 and an accelerometer were temporarily given to our group during 6 months by the enterprise Bioplux Systems. Using this system, we were able to capture data from gestures with a mobile device if the accelerometer was correctly placed on its back, simulating the existence of an internal sensor. However, user tests have to be performed while carrying the Bioplux4 device. Even with this limitation, this was still a more appropriate technological approach when compared with the *Wii* remote controller.

### 4.1.1. Bioplux4 Device

The Bioplux4 device (Figure 4.1) is able to collect data from four analog channels and delivers it wirelessly through Bluetooth connection to a range up to 100m, with a sampling rate of 1000Hz. This system already had available libraries written in C and C++ that allowed Bluetooth communication between the device and a computer. Since we intended to program in C# (an object-oriented programming language developed by Microsoft) in both mobile and non-mobile computers, we had to port this library to C# and also to the Compact Framework, in order to receive the accelerometer signal on a PDA.

### 4.1.2. Bioplux Accelerometer

The Bioplux accelerometer is, in essence, an ADXL 330 MEMS accelerometer, linked by a teflon cord to three analog channels (one for each axis) of the Bioplux4 device. This exact type of accelerometer is used in the *Wii* Remote, so we were convinced that this accelerometer was a secure choice. This accelerometer is triaxial, 10000g shock resistant and can measure +/- 3g of acceleration. Since we use the raw signal of this accelerometer, our work could be easily reproduced using other accelerometer of this type if a correct calibration is performed.

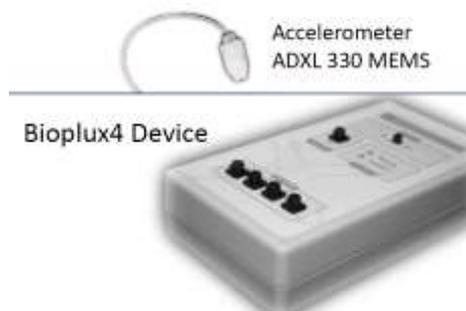

**Figure 4.1 – Bioplux4 Device and Accelerometer ADXL 330 MEMS**

## 4.2. Preliminary Tasks

In the beginning of our studies on how we should use the accelerometer data to detect gestures, we made some small applications that handled the motion signal to detect different mobility contextual data, and also an application to detect tilt gestures in two directions. In those cases, we were able to use some of the concepts we discussed in section 3.3, such as pre-processing and rotational analysis. In section 4.2.1 we will briefly introduce our analysis on contextual data and in 4.2.2 an also brief description and analysis of the tilt algorithm.

### 4.2.1. Context Analysis

A mobile device with motion sensing capabilities has the ability to capture implicit data about the user movement, providing different functionalities in each situation. When we finally had an accelerometer available, we started to create simple applications to find different movement characteristics. We used thresholds to detect the presence of movement (when the user is walking or running) and a lower threshold to detect if the user is holding the device. When both thresholds are not passed, it indicates that the device is stable. Those characteristics can be useful to decrease energy consumption of the device or chose different user profiles for different motion characteristics. In Figure 4.2 we present an example of an accelerometer signal (amplitude) for different movement stages that are recognizable using an accelerometer (Stopped, Picked, Holding, Walking and Running). We also developed some work on fall detection, because this kind of contextual data is important provide faster medical care to the elder. Our algorithm is based on two main characteristics when people fall: an unusually large acceleration magnitude and an unexpected final angle of the device. When both characteristics are combined, it is possible to recognize falls and have a low false-positive rate. However, since this was not the focus of our work, we did not further developed or tested this approach. These first studies on contextual data worked as a tutorial for the next phases of development.

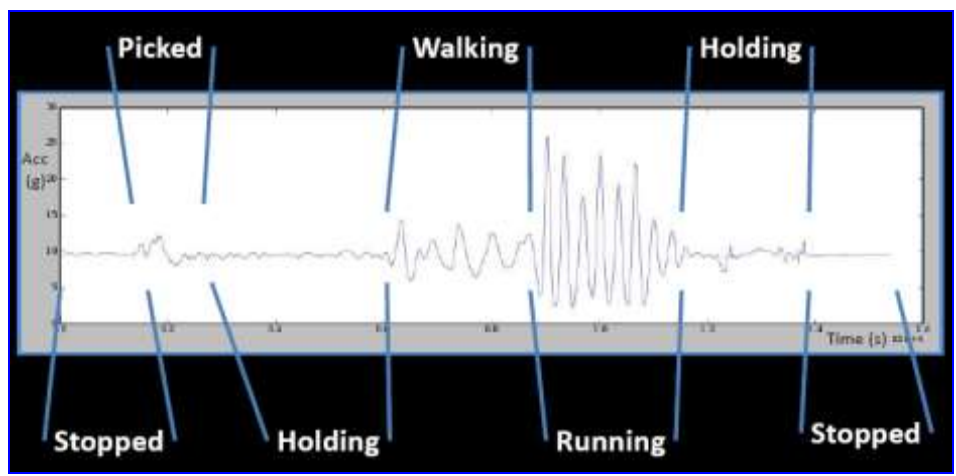

**Figure 4.2 – Contextual analysis**

### 4.2.2. Tilt Gestures

We also find interesting the possibility of detecting tilt movements, because it would enable, if paired with our Mnemonical Body Shortcuts, the existence of an interface for mobile devices with tilt navigation and a shortcut method only based on gestures. This concept is very strong, because in the future we could imagine mobile devices without any buttons, only controllable by gestures. Tilt recognition using accelerometers is documented in many articles, but we did not have any description of an algorithmic approach to this type of recognition. Because this should be a good way to use rotational analysis to detect gestures, we decided to construct and test a tilt recognition application (Figure 4.3). The tilt detection was achieved with four distinct phases:

1) Pre-processing of the signal (calibration and adjustment)

2) Calculation of degree variance in all axis and joining y and z axis to have a better sensitivity in the up/down axis

3) Threshold the signal to find possible candidates to tilt

4) Finally, comparing recognized tilts of phase three to know if there are some in the same time frame. In those cases, the tilt with less amplitude is discarded.

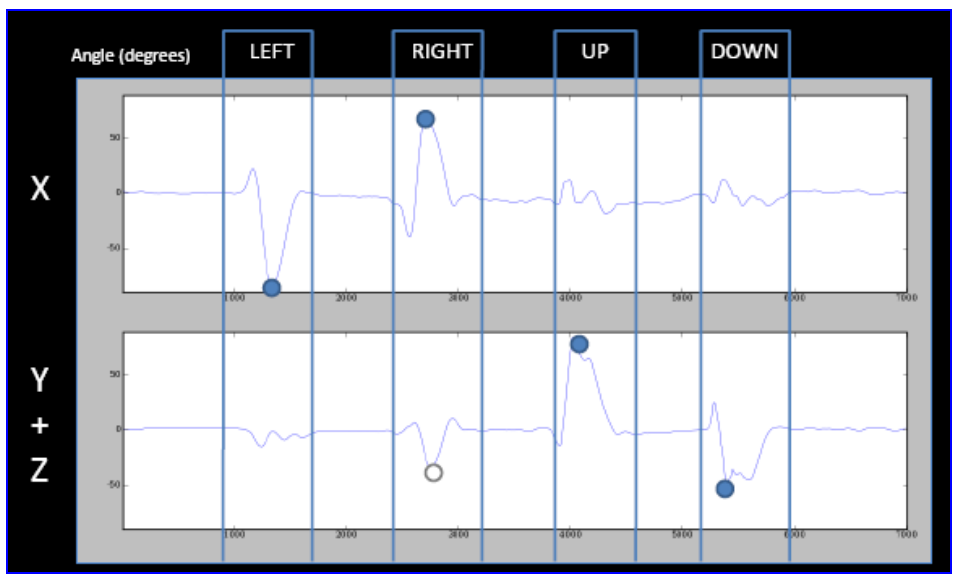

**Figure 4.3 – Tilt, final phase of recognition. Discarding one "down" tilt because of its lower amplitude.**

The tilt algorithm was a tested with 20 users along with the tests of the first prototype. The final recognition rate of this algorithm was set on 86%, which is truly acceptable. It would be possible to fine-tune this algorithm to achieve higher results, but it was not the main focus of our work, and this development already gave us some basic knowledge to face the most important phase of our work – the recognition of Mnemonical Body Shortcuts and the construction of its user interface.

## 4.3. Position-Based prototype

As first approach to the recognition of Mnemonical Body Shortcuts, we decided to explore the possibilities of dislocation measurement offered by an accelerometer. Since our intention was to detect the movement towards body parts, we choose to map de dislocation of the mobile device, calculating the distance between an initial and fixed point (the chest, with the screen facing it) and a final variable point (Figure 4.4).

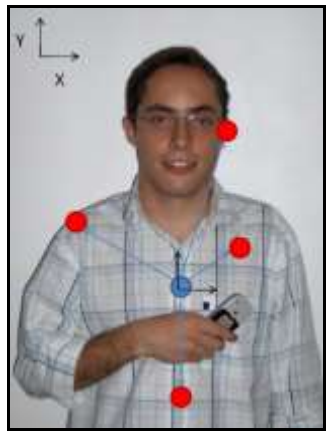

**Figure 4.4 – Mapping the dislocation of the mobile device on the 2D plan**

Furthermore, it is important to detect user's intention to trigger a gesture, and this is done using an "action button". This action button had to be pressed during the whole gesture, from the chest to the intended body part. This prototype was firstly developed for the Pocket Loox 720, but we did not use it in the test phase because it was too hard to handle. The solution was to stick the accelerometer in the back of other mobile device, and the action button role was made using the "Enter" key on a laptop device.

The algorithmic approach has 4 main phases (Figure 4.5): pre-processing, movement analysis, rotational analysis and classification.

### 4.3.1. Pre-Processing

The pre-processing phase is compound by three basic data modifiers, essential to a proper analysis. Firstly, the signal is calibrated, in order to receive correct values in $ms^{-2}$. Secondly, we define the starting rotation of the device using the signal adjustment. This adjustment is crucial for both movement and rotational analysis, because we assure an initial acceleration of 0 $ms^{-2}$ in all three axes. Finally, we apply a moving average filter to remove most of the noise of the signal, which is important to a correct detection of the start and end of the gesture in the movement analysis.

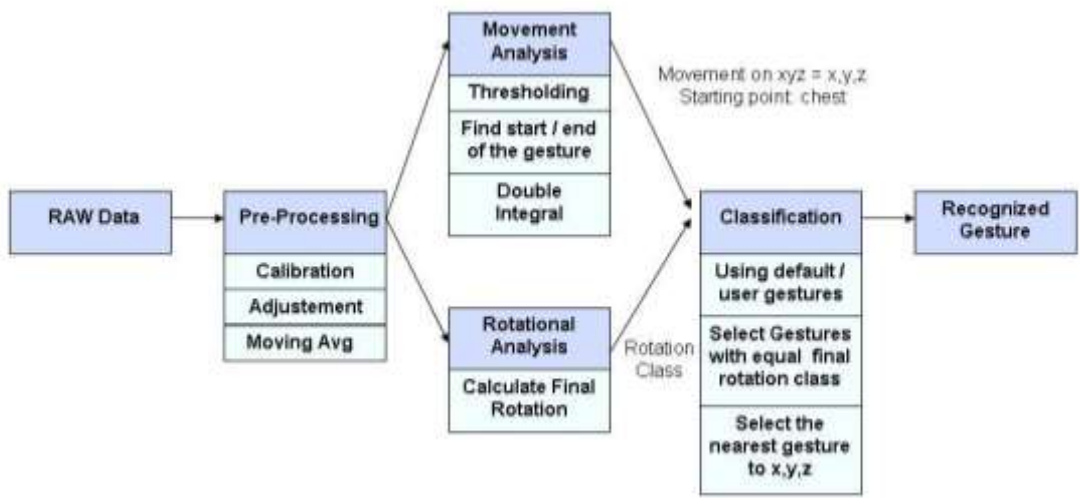

**Figure 4.5 – Position-Based prototype diagram**

### 4.3.2. Movement Analysis

After the pre-processing we have made, it is already possible to apply a double integral and reach a value in meters for the dislocation in each axis. However, even filtered, the signal still has some minor acceleration variations before and after the actual gesture that are caused by the vibration of the hand, and when we use the first integral to the acceleration we get a residual incremental velocity. When we apply the integral to that incremental velocity, a major error margin is created when, in fact, the mobile device was still stopped. To eliminate this error, we developed a method to calculate the exact start and end of the gesture. The residual acceleration before and after is eliminated using a threshold of 0.4 $ms^{-2}$, but the problem sometimes persists when the value surpasses the threshold. To solve that problem, we find the maximum and minimum peak of the signal and calculate the probable end and start of the gesture, eased by the threshold calculation. With this approach, we are able to apply the double integral only to the area that we assure as part of the gesture, thus minimizing the error margin.

### 4.3.3. Rotation Analysis

In this prototype, the rotation analysis was introduced in order to surpass the limitation of the movement calculations when users make gestures towards a body part with a pronounced rotation of the device. We performed the rotation measurement on the last received signal values of each gesture, which gave us the rotation relative with the point we fixed as starting point of the gesture (the chest). Each gesture was classified under one of the 6 rotation classes based on its final rotation (Figure 4.6). These 6 rotation classes were defined because they represent the six rotational degrees of freedom that are detectable using an accelerometer (rotations made on the axis aligned with the gravity acceleration cannot be detected). As suggested in the figures, each class is represented by a 90º sector of the rotation circle, and the class 1 represents the class where minimal rotation was made.

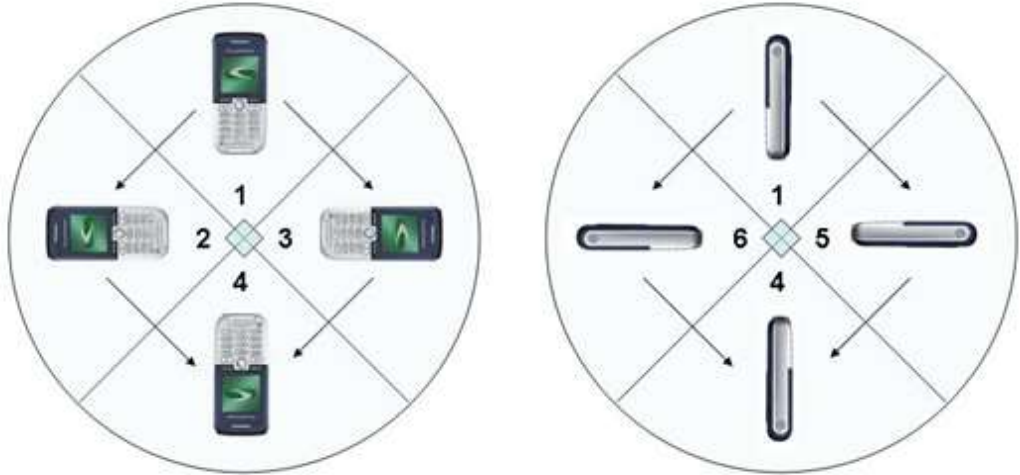

**Figure 4.6 – Six different rotational classes**

### 4.3.4. Classification

At this stage of the algorithm, we have the two essential measurements needed to recognize a gesture: the xyz point given by the double integral on the acceleration and the final rotation class of the gesture. However, we still need a set of labelled examples to be able to classify the incoming movement and rotation in a specific gesture class. We decided to follow to two different approaches to this problem:

**Recognize default gestures**

For this prototype, our intention was to recognize a set of 10 different gestures: Mouth, Chest, Navel, Shoulder, Neck, Ear, Head, Leg, Wrist and Eye. To recognize these gestures we had to specify different position measurements and rotational classes and label them as a body part. In order to have those values, we used a model to perform each gesture and retrieve the position and rotational measurements. However, if we only used those values, they would turn to be extremely inadequate for a lower or higher individual. To surpass this limitation, we followed a generally acceptable hypothesis that considers the human body as tall as 8 heads: the second head is down to the nipples, the third to the navel, the fourth to the genitals area, the sixth to the knees and finally the eight head ends in the feet (Figure 4.7, using the Vitruvian Man as reference). When given the height of the user, the points we got through the model can be proportionally recalculated if following the eight-head approach.

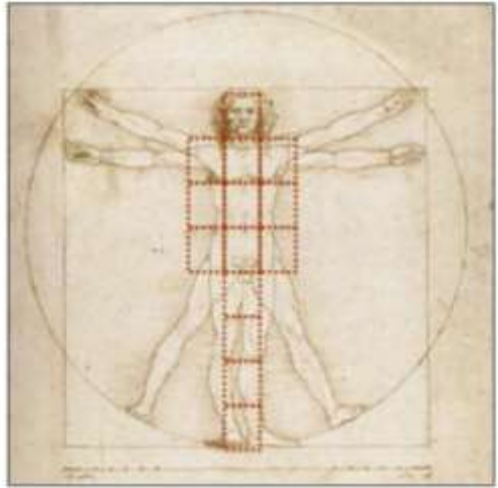

**Figure 4.7 – Eight-Head approach. The body height is usually represented as the sum of 8 heads.**

With the new default gestures, more adapted to the height of the user, we have available a position and rotation measurement labelled for each of the 10 gestures. When a new gesture is detected and passes all the processing already mentioned, we also have available its position and rotational class. In order to classify this gesture, we calculate the distance of the new gesture position to the different default points, but only in the same rotational class, e.g.: the gestures towards mouth, chest, elbow and navel are in the group 1 of rotational classes (Figure 4.6), while gestures toward the eye or the wrist where in class 2 and gestures to the leg and neck in class 3. If there is no gesture present in the same class as the one detected, the gesture is not recognized. Finally, the classification algorithm returns the labelled interface related with the nearest point.

**Recognize personalized gestures**

This second approach uses a variable training set of the user. Each training gesture retrieves a position point and a gesture class. When all training gestures are performed, we calculate the central point from all the available points and we set as rotation class the most frequent rotation class in the test set. The final recognition is made in a similar way as described for default gestures, with the calculation of the nearest training gesture point within the same rotational class of the gesture to be recognized.

## 4.4. Feature-Based prototype

The second approach we have made towards a correct recognition of Mnemonical Body Shortcuts was focused on a feature-based approach using pattern recognition techniques. This prototype was produced to recognize offline gestures. The main objective was to develop a prototype to be compared, in terms of recognition rate, with the position-based prototype. The description of the evaluation procedures and tests is present on section 5.2. This offline prototype follows the same basic phases present on the first prototype: data is pre-processed, then transformed (in this case by a feature extraction algorithm) and finally is classified. In this prototype, there are two classification methodologies, one based on a kNN classifier and other based on the Naive Bayes classifier. We used these classifiers because their characteristics were more adapted to the problem in hands, and they were both available in the open-source software we decided to work with. We will continue this chapter with the description of all phases but firstly we will introduce ORANGE, the software package we used to provide the classification tools. In Figure 4.8 we present the diagram for this prototype. The signal acquiring and pre-processing is present on a Pre-Processing module developed in C#. A Feature Extraction module was implemented in Python (a high-level programming language) and ORANGE widgets were used to construct the Classification module.

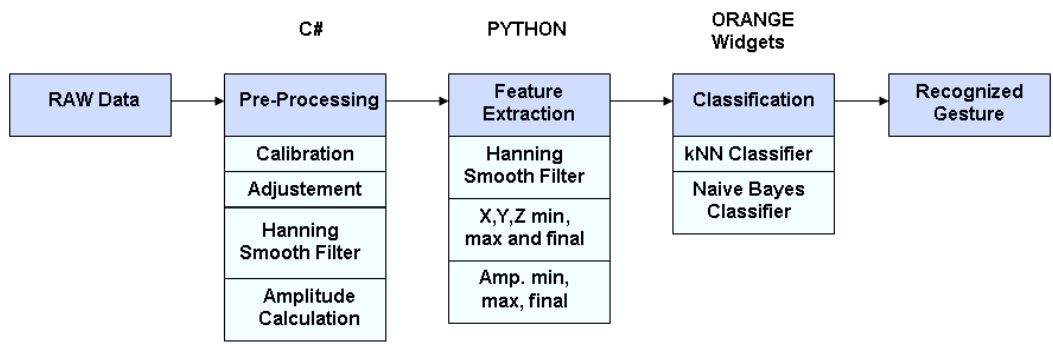

**Figure 4.8 – Feature-Based prototype diagram**

### 4.4.1. ORANGE Data Mining Software

ORANGE (Figure 4.9) is a component-based open source data mining software, developed by a group of researchers in the Faculty of Computer and Information Science of the University of Ljubljana in Slovenia [10]. This software has components to import, visualize, process, model and explore data sets with different techniques. It also supports predictive modelling tools such as classification trees, naive Bayesian classifier, K-Nearest-Neighbours, majority classifier, support vector machines, logistic regression and rule-based classifiers. Although developed in C++, their components can be accessed and modified through python scripts or also through visual programming with a graphical user interface objects named Orange Widgets.

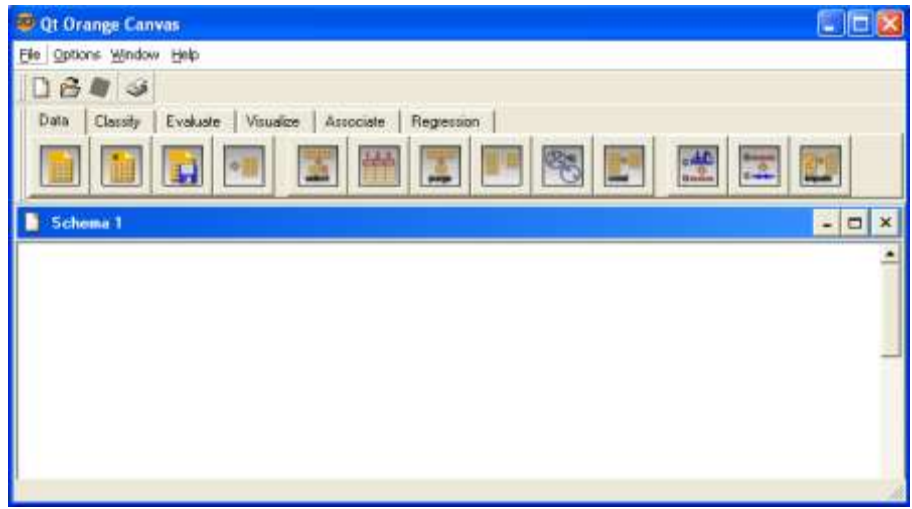

**Figure 4.9 – ORANGE Data Mining Software Canvas**

### 4.4.2. Pre-Processing

For this prototype, the Pre-Processing module was similar to the one constructed for the first prototype (signal acquisition and pre-processing), but we added the amplitude calculation because it would be needed to the feature extraction phase.

### 4.4.3. Feature Extraction

The first step to create a feature-based model is to choose features that characterize each gesture with accuracy. Since this was the second prototype, we already have some prior knowledge about which characteristics better define the body based gestures. We decided to choose 12 different features, considering gesture starting in the chest and finishing in a body point. In the Feature Extraction module we use the maximum and the minimum values from the X, Y and Z axis. These 6 features are essential to determine the direction and position variation of the gesture. Similarly to what was done in the position-based prototype, we added 3 features with the final value of each gesture, corresponding to the final rotation, but in this case there is no division in rotational classes (Figure 4.10). Finally, the signal's amplitude was also considered, since some gestures have different amplitude variation. The maximum and minimum values were added, as well as the amplitude mean value during the whole gesture. The captured signal is usually noisy and not suitable for a correct feature extraction. We used a smooth algorithm based on the Hanning window, which has a better performance compared with a Moving Average approach, applying it in the Feature Extraction module even before the feature extraction process takes place.

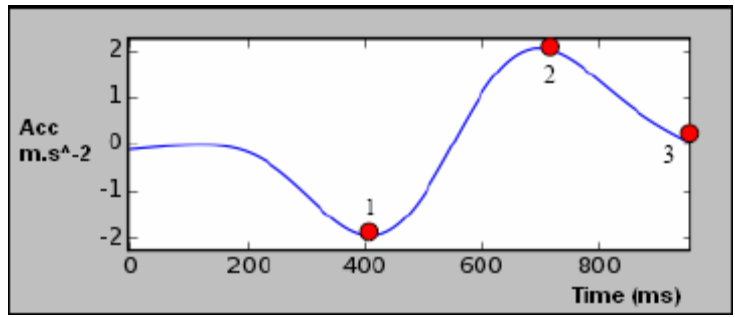

**Figure 4.10 – Features from y axis 1) Minimum Value 2) Maximum Value 3) Final Rotation**

### 4.4.4. Classification

After feature extraction, we were able to generate training sets and fill the feature space with test sets to test the classifier. The classifiers we used were kNN and Naive Bayes classifier (and their specific characteristics are explained in 4.3.4.1 and 4.3.4.2). Training sets were generated using the Classification module, where the data was processed and the output formed a data file, which was read by ORANGE. The test sets were generated with the same module, but they were typically smaller files. In Figure 4.11 we present the construction of both kNN and Naive Bayes classifier using ORANGE visual programming.

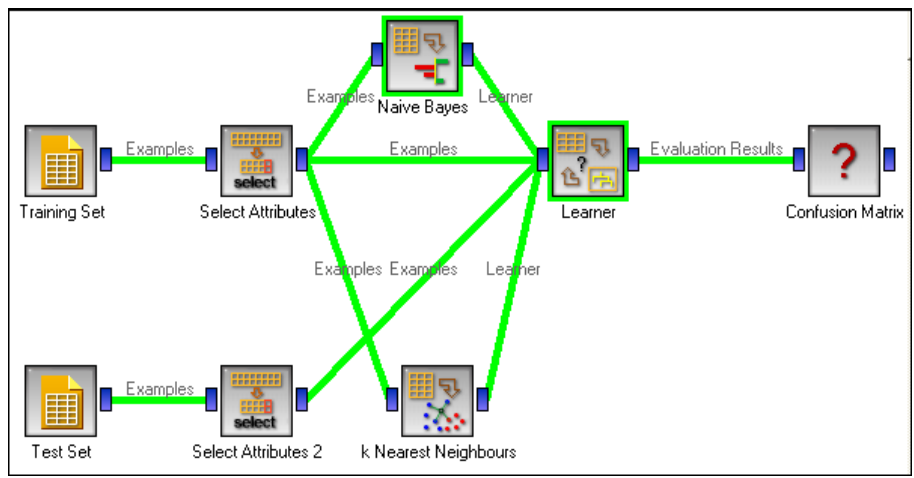

**Figure 4.11 – Construction of kNN and Naive Bayes Classifiers using ORANGE widgets**

Both test and training data pass trough attribute selection, in order to exclude irrelevant attributes or identification attributes (such as the id of user who made the gesture). The training serves as input to kNN and Naive Bayes classifiers, and the test data joins the training data in the test learner, which outputs the needed data to compute a confusion matrix (Table 4.1). The confusion matrix informs about how many sets belonging to class *i* were classified in class *j*. The error free classification would present values only in the matrix diagonal. This confusion matrix shows the final classification results for the test data.  In this case, the horizontal identifier represent the expected result and the vertical identifier the final classification result.

|    | 0  | 1  | 2  | 3  | 4  | 5  | 6  | 7  | 8  | 9  | 10 | 11 |
|----|----|----|----|----|----|----|----|----|----|----|----|----|
| 0  | 95 | 0  | 0  | 0  | 0  | 0  | 0  | 0  | 0  | 0  | 0  | 0  |
| 1  | 1  | 94 | 0  | 0  | 0  | 0  | 0  | 0  | 0  | 0  | 0  | 0  |
| 2  | 0  | 0  | 95 | 0  | 0  | 0  | 0  | 0  | 0  | 0  | 0  | 0  |
| 3  | 0  | 0  | 0  | 95 | 0  | 0  | 0  | 0  | 0  | 0  | 0  | 0  |
| 4  | 0  | 0  | 0  | 0  | 95 | 0  | 0  | 0  | 0  | 0  | 0  | 0  |
| 5  | 0  | 0  | 0  | 0  | 0  | 95 | 0  | 0  | 0  | 0  | 0  | 0  |
| 6  | 0  | 0  | 0  | 0  | 0  | 0  | 95 | 0  | 0  | 0  | 0  | 0  |
| 7  | 0  | 0  | 0  | 0  | 0  | 0  | 0  | 95 | 0  | 0  | 0  | 0  |
| 8  | 0  | 0  | 0  | 0  | 0  | 0  | 0  | 0  | 95 | 0  | 0  | 0  |
| 9  | 0  | 0  | 0  | 0  | 0  | 1  | 0  | 0  | 0  | 92 | 0  | 2  |
| 10 | 0  | 0  | 0  | 0  | 0  | 0  | 0  | 0  | 0  | 0  | 95 | 0  |
| 11 | 0  | 0  | 0  | 0  | 0  | 0  | 0  | 0  | 0  | 0  | 0  | 95 |

Table 4.1 – Example of a confusion matrix generated by ORANGE

#### 4.4.4.1. kNN Classifier

The kNN classifier present on ORANGE software is essentially the standard implementation of the kNN algorithm: it receives the features of the training set, constructs a feature space with those features and classifies new samples calculating the k nearest neighbours on the feature space. However, ORANGE algorithm has a specific characteristic: the k neighbours do not have the same importance for the final decision: a Gaussian formula is used to give more credit to those who are near the new object. ORANGE allows developers to define the k value, change the distance metric to only analyse the neighbours based on the ranking of distances, change the distance formula to Euclidean, Hamming, Manhattan or Maximal, normalize continuous values and ignore unknown values. For this project, we only varied the number of k neighbours for different objectives, and we use kNN with the Euclidian distance, using distances and the Gaussian formula to compute the final classification result.

#### 4.4.4.2. Naive Bayes Classifier

The Naive Bayes classifier can also be adapted using ORANGE widgets: calculation of prior probability may be done using relative frequencies, Laplace estimate or m-estimate. Both discrete and continuous conditional probabilities may also be specified. The classifier was used always with the default values defined by ORANGE (relative frequencies and a method named LOESS for conditional continuous probabilities, which is detailed in ORANGE documentation).

## 4.5. Final prototype

After the evaluation of the fist prototypes, we were able to define which of the two prototypes was able to guarantee a better recognition rate. As will be described in chapter, the feature-based prototype was clearly the most complete, and we decided to use that prototype as basis to the development of a final prototype, able to recognize Mnemonical Body Shortcuts. The main objectives of the construction of the final prototype were:

- Enable real-time recognition of Mnemonical Body Shortcuts, using the feature-based prototype that only supported offline recognition.
- Allow user-testing while mobile.
- Create a user interface for Mnemonical Body Shortcuts, giving more information to the user (more feedback) and allow them to fully control the shortcut triggering.
- Simulate the creation of shortcuts in a mobile device.

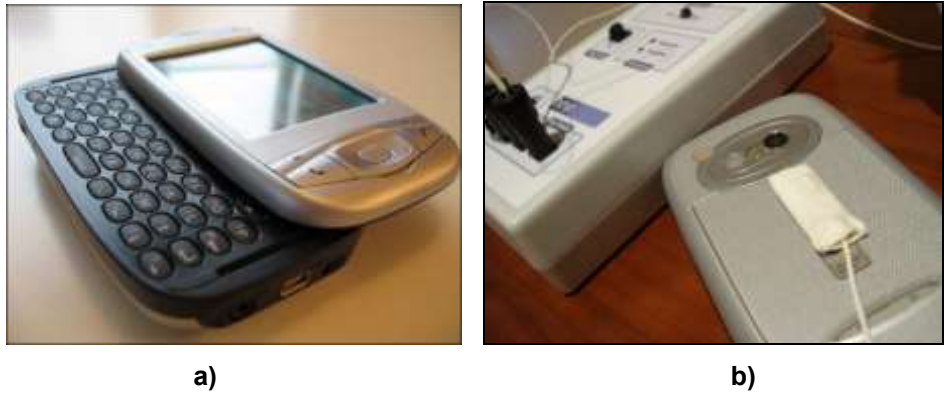

a)                                                              b)

**Figure 4.12 – a) HTC Wizard PDA; b) HTC Wizard with ADXL 330 accelerometer**

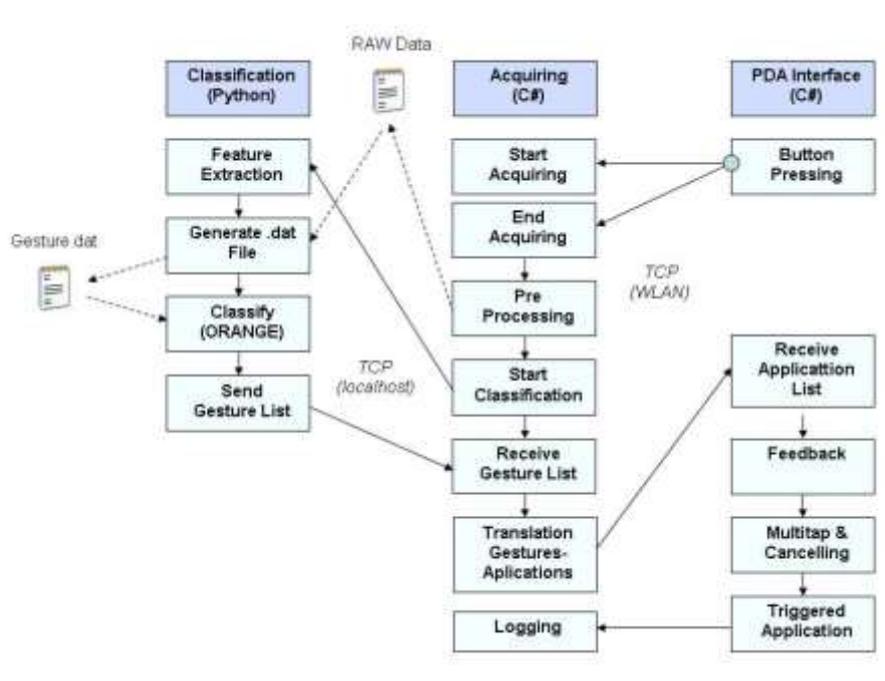

**Figure 4.13 – Final Prototype diagram.**

Since the Pocket Loox 720 was not suitable to perform gestures and simultaneously press an action button, we decided to use a HTC Wizard PDA (Figure 4.12). This PDA is smaller, and it has better located buttons, especially a button on the top left corner of the device that we expected to act well as an action button. In this prototype, instead of pressing the action button during the whole gesture, the user only had to press the button in the start and in the end of the gesture. Our initial thoughts on this prototype was to develop an algorithm to run entirely in the mobile device, but time limitations obliged us to use the available ORANGE libraries, that could not be easily reproduced in C# in the available time. Instead, we used the available library and acquired the signal in an Acquiring module developed in C# running on the laptop. The PDA only receives messages from the laptop with the possible applications to be triggered. In the following sections, we explore in detail the development of this final prototype. The final diagram of the prototype is present on Figure 4.13, where we can see the division on 3 main modules: A Classification module developed in Python, an Acquiring C# module running on a laptop, that also serves as communication pivot and, finally, a PDA Interface module running in .NET Compact Framework in a HTC Wizard, to detect button pressings and give feedback to the user.

### 4.5.1. Classification Module

The Classification module was produced fundamentally to use ORANGE scripting facilities and classify gestures based on pre-processed data. Firstly, it performs a feature extraction algorithm and stores the results on a .dat file, one of the possible file formats readable by ORANGE. The classification algorithm, based on the feature-based prototype, uses the features to classify the gesture. However, we adjusted the algorithm focusing on the conclusions we have made during the feature-based prototype evaluation, which will be discussed in chapter 5. When there is only user training data present, our option was to use Naive Bayes, and when we use the total training set acquired in the feature-based prototype evaluation, we only use the kNN algorithm with k = 50. We decided to use this value because we had more than 50 gestures stored for each class and ORANGE documentation states that a larger k would not significantly change the recognition. Finally, the Classification module sends the information about recognized gestures to the Acquiring module, using a TCP connection to send the information via sockets. This information consists on an array of 3 gestures, ordered by the probability of classification.

### 4.5.2. Acquiring Module

We defined four main objectives to the Acquiring module:
- Acquire and store the signal from the accelerometer between two key presses on the action button of the mobile device.
- Communicate with the Classification module to allow classification and send the recognition results to the PDA.
- Translate the recognized gestures to applications, accordingly to the associations defined in a GUI.
- Produce logs based on the recognition and user interaction.

This module works as a pivot between the Interface module and Classification module. Firstly, it is able to communicate via TCP over WLAN with a PDA, and receive the information about key presses. Then, the signal is acquired, stored, and it communicates with the Classification module to enable the reading on the data file. After recognition, this module receives a list of gestures and translates gestures to applications. The first gesture is translated to all the applications associated with that gesture, and then the first application of the two remaining gestures is also added. The application list is then sent to the Interface module. Finally, the module receives information from the Interface module (cancelling, application triggered and time spent) to log each gesture in a .txt file.

### 4.5.3. PDA Interface Module

The programming developed directly in the PDA was made to recognize key presses, send that information to the laptop, receive the recognized gestures (already in terms of applications to be triggered) and finally, but most importantly, provide users with feedback and tools to control the Mnemonical Body Shortcuts application. The interface gives 3 types of feedback: visual, audio and vibrational. After a gesture and appropriate feedback, the user has the opportunity to cancel or alter his selection. These mechanisms are useful when the user makes a mistake or gives up launching an application. On the other hand, even when the user draws a desirable gesture, the system can trigger the wrong application. This happens when two or more gestures are associated with the same body point or when a gesture is misrecognized (close body points). Thus, the user can navigate through a list of shortcuts, ordered by recognition certainty. Both mechanisms allow users to effectively control shortcut triggering and therefore be confident on its use. In the next sections, we will describe in more detail the feedback and user control facilities provided by this prototype.

#### 4.5.3.1. Audio Feedback

In terms of audio feedback, we recorded audio samples for each one of the 21 applications with available shortcut (Agenda, Internet Browser, Calculator, Camera, Calling Mother, Calling Peter, Calling Andreia, Contacts, Alarm, GPS, Photos, Voice Recorder, Time, Games, Messenger, MP3, Pedometer, SMS, Temperature, Voicemail, MS Word). Those audio samples have direct correspondence with the triggered application, i.e. "Music" or "Calendar". In a full implementation, we would have female or male voices, and users might also be able to record voice samples for each shortcut.

#### 4.5.3.2. Visual Feedback

The implementation of a gestural interface has as one of the main objectives the reduction of visual workload to users. However, it is important to associate visual feedback to other types of feedback, since it might be helpful in some scenarios. We provide a simple visual feedback – a small rectangle that appears on screen when a gesture is detected (Figure 4.14). That rectangle has a blue and green progress bar that runs for 2 seconds. Those 2 seconds represent the time that users have to cancel the shortcut or use *Multichoice* to switch to other application (the *Multichoice* feature is explained in detail in section 4.5.4). The information about the progress of the two-second timer is only available in visual feedback. Furthermore, we also added the name of the shortcut to be triggered on the top of the progress bar. When the shortcut is triggered, we present the regular screen of that function in the mobile device.

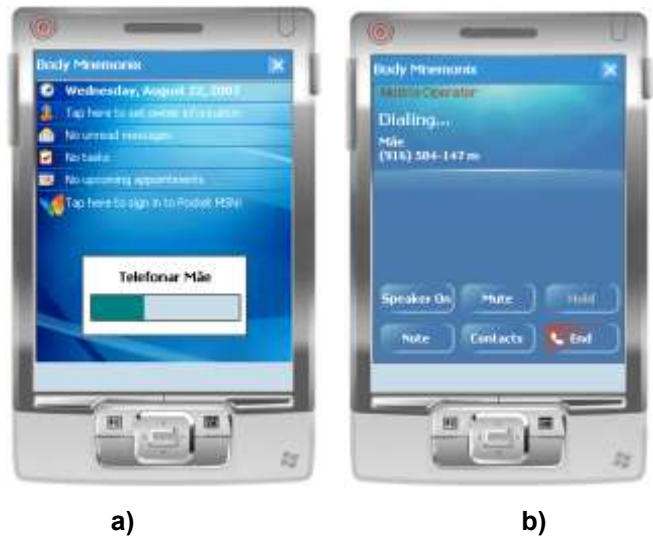

a)          b)

**Figure 4.14 – Visual feedback given by the Final Prototype. a) Progress Bar and shortcut to be triggered; b) The shortcut was triggered.**

#### 4.5.3.3. Vibrational Feedback

The vibrational feedback has great importance because we know that users certainly have the mobile device in the hand when performing a gesture, but they might not ear audio feedback or look to the screen. This type of feedback may be, in some cases, the only feedback users can have when triggering a Mnemonical Body Shortcut. Since it was not possible to identify each recognized gesture through vibrational feedback, we decided to inform users about the certainty of the recognition. The vibration of the mobile device varies from 0,25 seconds when the recognition percentage given by the ORANGE classification algorithm (x) is above 85%, 1 second for 65% < x < 85% and 2 seconds for x > 65%. This enables users to be aware of gestures that are recognized with less certainty, thus more probable to be incorrectly recognized. When facing a long vibration, users can cancel the recognition or confirm the application to be triggered. The vibrational characteristics might also be adjustable in a final deployment of this project.

### 4.5.4. User Control

Audio, visual and vibrational feedback are extremely helpful to inform users on the different variables regarding gestural recognition and shortcut triggering, but they would be useless if not developed together with mechanisms that allow users to control the behaviour of the mobile device and shortcuts. We developed two methods of interaction with Mnemonical Body Shortcuts, cancelling and *Multichoice*. The cancelling feature is easy to understand: users can abort a shortcut whenever they want, but only within the time frame of two seconds, represented by the progress bar. To cancel, users have available a special button, with the only function of cancelling the shortcut (Figure 4.15 b)). The *Multichoice* mechanism has the function of switch between different shortcuts, and is used by clicking the action button (Figure 4.15 a)) during the 2-seconds delay.

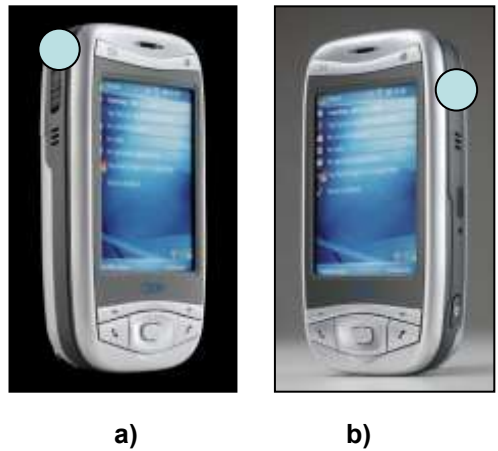

**a)**          **b)**

**Figure 4.15 – a) Action Button; b) Cancelling Button**

Each time the *Multichoice* is used, the progress bar returns to 0, and if there are no more suggestions, it simply cancels the shortcut triggering. There are two situations when using *Multichoice* might be useful. Firstly, if users store two or more gestures in the same body part, it is possible to reach the second or third stored application. In example, some user has stored in the head both the contact list and calculator, in this order. If he makes a gesture towards the head, firstly the mobile device will trigger the contact list application, but if the user presses the action button again during the 2-second delay, it would reset the delay and begin the triggering of the calculator application. Secondly, when a gesture is incorrectly recognized, users might also use *Multichoice* to search for the next gesture suggested by the recognizer. It is important to know that the recognizer will always suggest the applications stored in the body part recognized in first place, but when those end, it also has two shortcuts to trigger by order, consistent with the two other gestures that had more probability of being recognized. In example, a user makes a gesture towards the head, but the system recognized the only application he stored in the ear. If he presses the action button during the 2-second delay, it is likely to begin the triggering the application he stored in the head.

Both cancelling and *Multichoice* mechanism allow users to feel more under control of the shortcut triggering and correct some recognizing errors that may happen during the utilization of Mnemonical Body Shortcuts.

# 5

# Results and Discussion

After the development of each one of the three Inertial Sensing prototypes, we conducted evaluation tests. The objective of each evaluation test varied with the aimed prototype. With the Position-Based and the Feature-Based prototype evaluation, our main goal was to test the efficiency of each algorithm, in order to enhance them and also to compare them both and choose the most complete for the development of a final interface. The efficiency of the algorithm was also tested in the final usability tests, but we also examined the usability of the created interface, namely evaluating the different forms of feedback, gesture cancelling and the *Multichoice* feature.

This chapter follows the order present on the previous chapter, beginning with the evaluation of the Position-Based prototype in section 5.1. In section 5.2 we describe user tests and results on the Feature-Based prototype and finally, in chapter 5.3, we explain how we directed the usability tests on the Final Prototype and its results. Each one of the sections ends with a discussion on each prototype evaluation results.

## 5.1. Position-Based Prototype Evaluation

After the development of the Position-Based prototype, our focus was to test how well would it behaves in terms of efficiency. We would only pass to a next phase of the development if we were certain that this algorithm was able to correctly recognize gestures with a good recognition rate. This prototype had two different modes that were separately tested, the default gestures recognizer and the personalized gesture recognizer. Users tested both prototypes while standing, because we thought that these results would give us already a fast and reliable measurement of the efficiency of the recognizer and also because we did not had available a mobile device sufficiently easy to handle.

### 5.1.1. Test Results

Tests on this prototype were realized in the month of March of 2007 at IST-Taguspark, during 2 days. We selected a range of 10 users, averaging 24 years.

In first place, we tested the efficiency of the recognizer on the 10 default gestures. We began our test with the insertion of user's height on the application, in order to recalculate the distances using the *eight-head* approach. After a brief demonstration of each gesture, users were prompted to perform 5 random gestures out of the available 10 gestures, 4 times each, totalling 20 gestures. It is important to refer that the algorithm was recognizing the whole set of 10 gestures, but only five were prompted to the user. Recognition results for each one of the 10 default gestures are present in Table 5.1 a). The best detected gesture was the one towards the neck, with a 100% recognition rate, while the worst one was the gesture towards the ear, with 60% recognition rate. The general efficiency measurement of the default gestures recognizer was set on 82%. The final recognition result is not simply the mean of the recognition rate of the 10 gestures because some gestures were performed more times.

| Gesture | Recognition Rate |
|---|---|
| Neck | 100% |
| Navel | 90% |
| Mouth | 85% |
| Chest | 85% |
| Head | 85% |
| Wrist | 85% |
| Leg | 80% |
| Eye | 75% |
| Shoulder | 75% |
| Ear | 60% |
| TOTAL | 82% |

| Gesture | Recognition Rate |
|---|---|
| Leg | 85% |
| Mouth | 94% |
| Navel | 64% |
| Neck | 95% |
| Ear | 55% |
| TOTAL | 71% |

a)                                                                                   b)

**Table 5.1 – a) Recognition rates for the 10 default gestures; b) Five most selected body parts and correspondent recognition rates**

After the test on default gestures, we asked users to freely choose 5 body parts, this time to measure the algorithmic efficiency of personalized gestures towards each body part. In the beginning of the test, users repeated the 5 selected gestures, 5 times each, in the training mode. This training set was used to recognize afterward gestures, when users were prompted to perform 4 times each gesture, totalling 20 gestures, as it was done with the default gestures recognizer. Since there were many different chosen body parts, there is not much significance on recognition results for each part, but in Table 5.1 b) we present the 5 most selected body parts and the recognition results for each one. The final recognition rate for personalized gestures was set on 71%.

### 5.1.2. Discussion

Since the objective of this test was to measure how well the algorithm performed, the number of possible conclusions is reduced. In terms of default gestures, the final recognition rate of 82 % was below our expectations. This low result is explainable due to the characteristics of the algorithm: the position calculation when there was rotation involved did not prove to be a good gesture differentiating characteristic. Because of that issue, some different results with similar rotations were misclassified. Personalized gestures revealed an even lower recognition rate of 72%. Users sometimes picked similar body parts or gestures that were misclassified by the recognizer, and the problem of position measurements with rotation also applies to this case. Besides, since we did not include outlier detection, when one training was badly performed, it had a great influenced many misclassified gestures.

With this algorithm, the usage of personalized gestures is impracticable, and that would provoke a major failure regarding the objectives for our interface that we defined in the Design Guidelines. If only using the default gestures, users would be restricted to a limited set of 10 gestures and obliged to learn all of those gestures.

In conclusion, this algorithm was clearly not proper for a full implementation. We cannot forget that this algorithm was tested while standing, so we would expect an even larger error margin for recognition tests with mobile used, since mobility introduces much noise in the received signal. It was not further developed since we embraced the task of creating a feature-based algorithm, but this algorithm could be largely enhanced if the rotation influence on the position calculation was reduced. Besides, we learned that we should ask users to select their personalized gestures after we explain default gestures, because in this test we noticed that users tend to choose gestures similar with those we have explained before in the gestural personalization phase.

## 5.2. Feature-Based Prototype Evaluation

The implementation of the feature-based prototype had the same objectives of the position-base prototype: develop a reliable platform, in terms of gesture recognition, to be later enhanced with appropriate user-control tools and feedback. For that reason, user tests on the feature based prototype were also focused on finding the recognition rate, not only for default gestures but also for personalized gestures.

### 5.2.1. Test Results

Tests were realized in the month of June of 2007 at IST-Taguspark, during 3 days. We selected a range of 20 users, and they were asked to perform a set of 12 gestures, five times each, totalling 60 gestures per user. Gestures were not performed randomly: users sequentially performed 5 times each type. These gestures were similar to those default gestures tested with the position-based prototype, adding a gesture towards the hip and the back, and they were also performed while standing. The offline analysis used data from only 19 users (one was discarded) because one user did not performed gestures we explained in a proper manner, even after a second explanation. Using the 60 gestures from 19 users, we were able to combine them and test different scenarios:

- Recognition rate using the gestures from the other 18 users as training set, aiming both 12 gestures and 5 random gestures.
- Recognition rate with 1,2 and 3 trainings for each gesture.

- Recognition rate using the gestures from other 18 users as training set but also with 1,2 or 3 user trainings, aiming both 12 gestures and 5 random gestures.

After we collected all the data from the 20 users, an offline evaluation was performed, using different training and testing sets and both Naive Bayes and kNN classifiers (Table 5.2 presents the overall results).

| User Training | | |
|---|---|---|
| 12 Gestures | | |
| 1 Training | 79,5% | 88,5% |
| 2 Trainings | 86,8% | 92,4% |
| 3 Trainings | 91,9% | 92,8% |
| 5 gestures | | |
| 1 Training | 88,2% | 90,8% |
| 2 Trainings | 96,1% | 98,2% |
| 3 Trainings | 96,3% | 97,9% |
| Total Training Set | | |
| 12 Gestures | 93,6% | 92,8% |
| 5 Gestures | 97,3% | 96,2% |
| Total Training Set + User Training | | |
| 12 gestures | | |
| 1 Training | 93,8% | 93,2% |
| 2 Trainings | 94,3% | 92,4% |
| 3 Trainings | 95,8% | 95,0% |
| 5 gestures | | |
| 1 Training | 95,1% | 94,7% |
| 2 Trainings | 96,1% | 95,8% |
| 3 Trainings | 96,8% | 97,9% |
| Knn | Bayes | |

Table 5.2 – Results of the Feature-Based prototype evaluation

The test was divided in two phases:

**User Training**

In this first phase, we tested the recognition rate using as training set only the gestures performed by the user. The training set varied between 1, 2 or 3 gestures. This approach was tested using the whole set of 12 gestures but also using 5 random gestures, which was the mean number of key shortcuts a user commonly have available, thus a estimative of how many gesture would be chosen in a daily-usage scenario.

**Total Training Set**

The second phase was based on using the whole set of training from all users. This set of 1080 gestures worked as a training set, and each user's gestures were classified using that training set, adding none, one, two or three user trainings, also with the 12 and 5 gestures set. The final results of these tests are also available in Table 5.2 and the confusion matrix of 12 and 5 gesture test using only the training set (without user training) and kNN classifier are available in Table 5.3 and Table 5.4 respectively. We only show the confusion matrix for kNN because the prototype achieved the best results with this classifier.

| Gestures | Mouth | Shoulder | Chest | Navel | Ear | Back | Head | Wrist | Neck | Leg | Eye | Hip |
|---|---|---|---|---|---|---|---|---|---|---|---|---|
| Mouth | 87,5% | 6,2% | 0,0% | 0,0% | 5,1% | 0,0% | 0,0% | 0,0% | 0,0% | 0,0% | 0,0% | 0,0% |
| Shoulder | 4,2% | 90,7% | 2,0% | 1,0% | 0,0% | 0,0% | 0,0% | 0,0% | 0,0% | 0,0% | 0,0% | 0,0% |
| Chest | 0,0% | 1,0% | 94,9% | 0,0% | 0,0% | 0,0% | 0,0% | 0,0% | 0,0% | 0,0% | 0,0% | 0,0% |
| Navel | 0,0% | 0,0% | 3,0% | 95,8% | 0,0% | 0,0% | 0,0% | 0,0% | 0,0% | 0,0% | 0,0% | 0,0% |
| Ear | 4,2% | 0,0% | 0,0% | 0,0% | 92,9% | 0,0% | 0,0% | 0,0% | 0,0% | 0,0% | 0,0% | 0,0% |
| Back | 3,1% | 0,0% | 0,0% | 1,0% | 1,0% | 97,8% | 0,0% | 0,0% | 0,0% | 0,0% | 0,0% | 2,9% |
| Head | 1,0% | 2,1% | 0,0% | 0,0% | 0,0% | 0,0% | 97,8% | 0,0% | 0,0% | 0,0% | 2,0% | 0,0% |
| Wrist | 0,0% | 0,0% | 0,0% | 0,0% | 0,0% | 0,0% | 2,2% | 94,6% | 0,0% | 0,0% | 1,0% | 4,8% |
| Neck | 0,0% | 0,0% | 0,0% | 0,0% | 0,0% | 0,0% | 0,0% | 3,3% | 100,0% | 0,0% | 2,0% | 0,0% |
| Leg | 0,0% | 0,0% | 0,0% | 0,0% | 0,0% | 1,1% | 0,0% | 0,0% | 0,0% | 95,5% | 0,0% | 9,5% |
| Eye | 0,0% | 0,0% | 0,0% | 0,0% | 1,0% | 0,0% | 0,0% | 1,1% | 0,0% | 0,0% | 94,9% | 0,0% |
| Hip | 0,0% | 0,0% | 0,0% | 2,1% | 0,0% | 1,1% | 0,0% | 1,1% | 0,0% | 4,5% | 0,0% | 82,9% |

Table 5.3 – Confusion Matrix for the Feature-Based prototype with 12 default gestures.
Columns – Expected Result; Lines – Classification Result

| Gestures | Mouth | Shoulder | Chest | Navel | Ear | Back | Head | Wrist | Neck | Leg | Eye | Hip |
|---|---|---|---|---|---|---|---|---|---|---|---|---|
| Mouth | 89,5% | 7,0% | 0,0% | 0,0% | 0,0% | 0,0% | 0,0% | 0,0% | 0,0% | 0,0% | 0,0% | 0,0% |
| Shoulder | 0,0% | 93,0% | 0,0% | 0,0% | 0,0% | 0,0% | 0,0% | 0,0% | 0,0% | 0,0% | 0,0% | 0,0% |
| Chest | 0,0% | 0,0% | 100% | 0,0% | 0,0% | 0,0% | 0,0% | 0,0% | 0,0% | 0,0% | 0,0% | 0,0% |
| Navel | 0,0% | 0,0% | 0,0% | 100% | 0,0% | 0,0% | 0,0% | 0,0% | 0,0% | 0,0% | 0,0% | 0,0% |
| Ear | 0,0% | 0,0% | 0,0% | 0,0% | 100% | 0,0% | 0,0% | 0,0% | 0,0% | 0,0% | 0,0% | 0,0% |
| Back | 5,3% | 0,0% | 0,0% | 0,0% | 0,0% | 100% | 0,0% | 0,0% | 0,0% | 0,0% | 0,0% | 5,7% |
| Head | 0,0% | 0,0% | 0,0% | 0,0% | 0,0% | 0,0% | 100% | 0,0% | 0,0% | 0,0% | 0,0% | 0,0% |
| Wrist | 0,0% | 0,0% | 0,0% | 0,0% | 0,0% | 0,0% | 0,0% | 93,8% | 0,0% | 0,0% | 0,0% | 0,0% |
| Neck | 0,0% | 0,0% | 0,0% | 0,0% | 0,0% | 0,0% | 0,0% | 4,2% | 100% | 0,0% | 0,0% | 0,0% |
| Leg | 0,0% | 0,0% | 0,0% | 0,0% | 0,0% | 0,0% | 0,0% | 0,0% | 0,0% | 100% | 0,0% | 3,8% |
| Eye | 0,0% | 0,0% | 0,0% | 0,0% | 0,0% | 0,0% | 0,0% | 0,0% | 0,0% | 0,0% | 100% | 0,0% |
| Hip | 5,3% | 0,0% | 0,0% | 0,0% | 0,0% | 0,0% | 0,0% | 2,1% | 0,0% | 0,0% | 0,0% | 90,6% |

Table 5.4 – Confusion Matrix for the Feature-Based prototype with 5 default gestures
Columns – Expected Result; Lines – Classification Result

### 5.2.2. Discussion

A feature based approach achieved a high recognition rate in the majority of the tests, both using user training and with the general training set of 1080 gestures. Naive Bayes and kNN algorithms were tested, and Naive Bayes performed better when only user training was present (low number of sample gestures), while kNN achieved better results with a large set of training. Considering the results of isolated user training of the 12 gestures set, the best recognition was achieved with 3 trainings with 92,8%. This recognition rate, although acceptable, is still vulnerable to some possible erroneous classifications. However, we do not believe users would want to use 12 gestures simultaneously. The test using a reduced set of 5 gestures achieved, using Naive Bayes, a recognition rate of 98,2% with only 2 gestures, with no positive impact of a third training. For those default gestures, user training seems to be a good approach, but this test does not guarantee the same recognition rate using free gestures. It is also problematic if users perform training gestures inconsistently, because it would reflect a lower recognition rate.

Results were also positive considering the usage of the training set of 1080 gestures (1140 gestures minus the 60 gestures performed by each user). Using all the 12 gestures, we achieved a recognition rate of 93,6%. This recognition rate is achieved without any user training, which is a crucial point for a good user acceptance. This value reaches 97,3% when considering 5 gestures. When we increasingly introduce the training set of the user, the recognition rate did not increase significantly using kNN algorithm, but it influenced positively Naive Bayes by 2 percent points. Yet, kNN algorithm still has the best performance using the total training set. User training could be added not by explicitly asking the user to train the system, but instead using an adaptive approach: when a user correctly performs a gesture, it should be possible to enrich the training set and successively increase the recognition rate. Such an approach would, based on the results of this prototype, significantly enhance the recognition.

The study on this prototype proved the feature-based approach as the most successful and appropriate. However, there were some untested scenarios: we only tested gestures that were pre-defined by us and we did not gave users the free will to define their own gestures; gestures in a movement scenario were not tested; lastly, users performed each type of gesture repeatedly (we did not introduced a random factor), so it is probable that gestures are performed more equally in this test than they would be performed on a daily basis. In resume, we tend to believe that recognition rates would decrease in a real-life scenario but maintain an acceptable margin, capable to perform as a suitable gestural interaction algorithm

## 5.3. Final Usability Tests

With the final prototype developed, we were finally able to fully test the Mnemonical Body Shortcuts recognition and user interface. In the start of the development of the usability test guide, we had in mind three main objectives:

1) Test the recognition accuracy of the algorithm, but this time focusing also on mobility scenarios and on truly personalized gestures.
2) Test user acceptance on the developed feedback and also their adaptability to the user control tools we made available.
3) Have a deeper look on user's opinion on this new type of interaction.

In fact, we intended to have a good analysis on four different aspects of interface usability: Effectiveness, Learnability, Likeability and Usefulness. These four aspects will be referenced in the discussion section of the usability tests on the final prototype.

Usability tests were conducted at IST-Taguspark, during 3 days in the late August of 2007, in the room 1.4.32. The room was set up as shown in Figure 5.1: Users were able to walk a continuous path, around a table, while interacting with our system. We placed the laptop in a position where it was easy to monitor both the user and the results on the laptop screen. Besides, we installed a Wireless Router in order to provide the communication between the PDA and the Laptop.

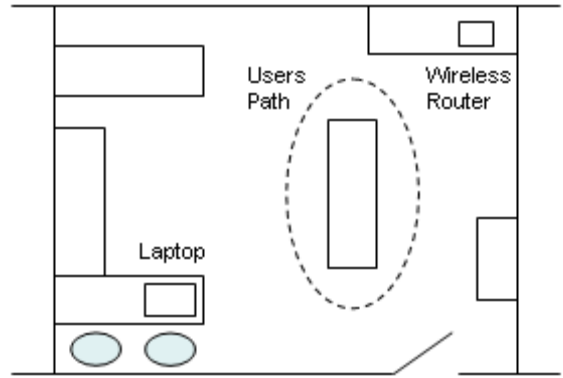

**Figure 5.1 – Room setup for the final usability tests**

Users were able to move around the table and perform gestures because we had a small case where users transported the Bioplux4 device (Figure 5.2). The Usability Test protocol is available in Appendix D, and was divided in 4 main phases:

**Personalized Mnemonical Body Shortcuts**

We began the first phase of usability tests by asking users to make their own associations between body parts and applications, in a total of 5 Mnemonical Body Shortcuts. This is important because they did not have contact with the default gestures, so these associations were really personal and authentic. After this step, users trained the system with one gesture for each Mnemonical Body Shortcut, and then they were randomly prompted to perform those gestures 20 times while standing and other 20 times while moving. The process was repeated for 2 and 3 trainings for each gesture. This phase allow us to know what would be the recognition rate of the system for personalized gestures in a realistic scenario.

**Default Mnemonical Body Shortcuts**

As stated in the fourth chapter, we also made available 12 default Mnemonical Body Shortcuts, which can be used without any training. In this phase, we demonstrated each one of the default gestures to users and then they had to perform 24 gestures considering all the 12 default gestures, one set of 24 while standing and other while mobile. The process was repeated but we randomly selected 5 gestures from the total set, and users had to perform 20 gestures using the new set, also in the two mobility situations. Using these results, we will know how the default gestures perform, with special interest on the recognition rate using a set of 5 gestures, because it is more close to the number of gestures users would want to use in a daily basis.

**Feedback and User control**

The Personalized and Default shortcuts evaluation phase of the usability test focus essentially on discovering the effectiveness of the recognition algorithm. In the Feedback and User control phase, we intended to test the feedback and user control mechanisms present on our prototype. To achieve this goal, we asked users to perform 20 Mnemonical Body Shortcuts while moving, using the same random default gestures selected in the second phase. However, this time we also explained users which applications were in each body part and we asked for the applications instead of the body part. We varied the applications between applications stored as first priority in a body part and others in a second or third priority, in order to fully test the *Multichoice* feature. Using this method, we were able to evaluate user's reaction to the feedback, usefulness of *Multichoice* and know how users will handle recognition errors.

**Questionnaire**

Finally, we wanted to perceive user's opinion on different aspects of the interface, such as the suitability of the feedback, *Multichoice* and cancelling features, their confidence on the gesture recognizer, ease of use while moving, probability of using the system on a real-life scenario and some other suggestions. Essentially, the last phase of the test guide is important to know about the Likeability characteristic of our approach

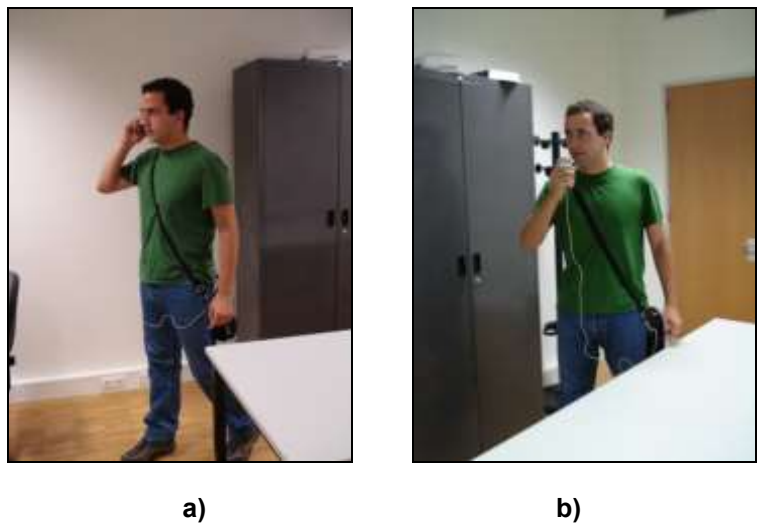

a) b)

**Figure 5.2 – Final Usability Test a) Gesture to the ear b) Gesture to the mouth**

### 5.3.1. Test Results

The following results are divided in the four phases described in the last section (detailed results in Appendix D):

**Personalized Mnemonical Body Shortcuts**

The Personalized Shortcuts phase refers to the recognition results on the personalized gestures, with 1, 2 or 3 trainings. The extracted results give us a perspective on the evolution of the recognizer, both in a standing position or while moving (Figure 5.3). In a static position, the recognition rate evolved from 70% with 1 training, to 81% with two trainings (+15%) and finally 89.5% with three trainings (+10%). While moving, the first rate was 64.5%, raised to 72.5% with two trainings (+12%) and 80.5% with three trainings (+11%).

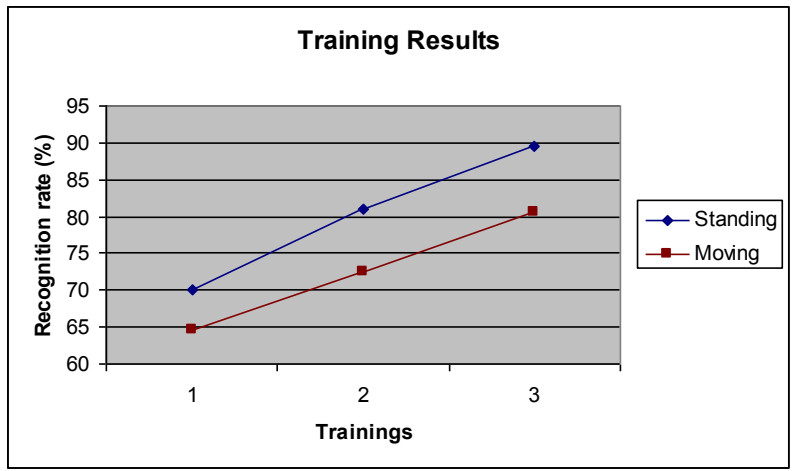

**Figure 5.3 – Results for personalized gestures, with 1, 2 and 3 trainings, while standing (blue with rhombuses) or moving (red with squares)**

**Default Mnemonical Body Shortcuts**

In the second phase, we focused on finding the recognition rate for default gestures, also while standing and moving. When using the whole set of 12 gestures, and randomly prompting users to perform all those gestures twice, users achieved results of 76.6% while moving and 84.2% while standing. When considering only 5 random gestures, results achieved better recognition rates, 90% for the mobility scenario and 92,5% for the static scenario (Figure 5.4). In Table 5.5 we present the confusion matrix where we joined the results for both mobility settings for the 12 gestures, while Table 5.6 represents the same results but for the test with 5 random gestures.

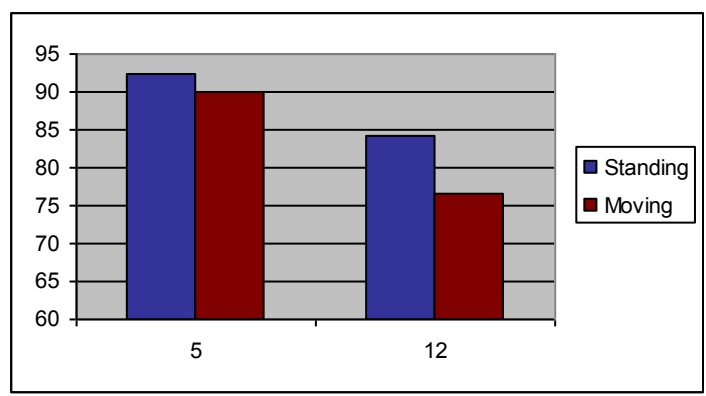

**Figure 5.4 – Results for default gestures, while standing or moving and also with 5 or 12 default gestures**

|       | Mouth | Chest | Elbow | Navel | Neck | Head | Back | Ear | Hip | Leg | Wrist | Eye | TOTAL |
|-------|-------|-------|-------|-------|------|------|------|-----|-----|-----|-------|-----|-------|
| Mouth | 27 | 3 | 0 | 0 | 0 | 0 | 0 | 0 | 0 | 0 | 0 | 0 | 30 |
| Chest | 3 | 34 | 3 | 2 | 0 | 1 |   | 1 | 1 | 1 | 0 | 0 | 46 |
| Elbow | 5 | 1 | 34 | 1 | 0 | 0 | 0 | 0 | 1 | 0 | 0 | 0 | 42 |
| Navel | 0 | 2 | 0 | 35 | 0 | 0 | 2 | 0 | 3 | 1 | 0 | 0 | 43 |
| Neck  | 0 | 0 | 0 | 0 | 40 | 2 | 0 | 2 | 0 | 0 | 0 | 0 | 44 |
| Head  | 2 | 0 | 2 | 0 | 0 | 35 | 0 | 1 | 0 | 0 | 1 | 0 | 41 |
| Back  | 1 | 0 | 0 | 0 | 0 | 0 | 38 | 0 | 0 | 3 | 0 | 0 | 42 |
| Ear   | 2 | 0 | 0 | 0 | 0 | 0 | 0 | 10 | 1 | 1 | 1 | 0 | 15 |
| Hip   | 0 | 0 | 0 | 2 | 0 | 0 | 0 | 0 | 33 | 12 | 0 | 0 | 47 |
| Leg   | 0 | 0 | 0 | 0 | 0 | 0 | 0 | 0 | 1 | 22 | 0 | 0 | 23 |
| Wrist | 0 | 0 | 1 | 0 | 0 | 1 | 0 | 0 | 0 | 0 | 38 | 0 | 40 |
| Eye   | 0 | 0 | 0 | 0 | 0 | 1 | 0 | 26 | 0 | 0 | 0 | 40 | 67 |
| TOTAL | 40 | 40 | 40 | 40 | 40 | 40 | 40 | 40 | 40 | 40 | 40 | 40 | 480 |
| %     | 67,5 | 85 | 85 | 87,5 | 100 | 87,5 | 95 | 25 | 82,5 | 55 | 95 | 100 | 80,4 |

Table 5.5 – Confusion Matrix of joined results for 12 default gestures
Columns – Expected Results; Lines – Classification Results

|       | Mouth | Chest | Elbow | Navel | Neck | Head | Back | Ear | Hip | Leg | Wrist | Eye | TOTAL |
|-------|-------|-------|-------|-------|------|------|------|-----|-----|-----|-------|-----|-------|
| Mouth | 37 | 1 | 0 | 1 | 0 | 0 | 3 | 1 | 1 | 0 | 1 | 0 | 45 |
| Chest | 2 | 38 | 1 | 2 | 0 | 0 | 0 | 0 | 0 | 0 | 0 | 0 | 43 |
| Elbow | 6 | 1 | 46 | 0 | 0 | 0 | 1 | 0 | 1 | 0 | 0 | 0 | 55 |
| Navel | 0 | 0 | 0 | 29 | 0 | 0 | 0 | 0 | 0 | 1 | 0 | 0 | 30 |
| Neck  | 0 | 0 | 0 | 0 | 40 | 1 | 0 | 5 | 0 | 0 | 0 | 0 | 46 |
| Head  | 2 | 0 | 0 | 0 | 0 | 15 | 0 | 0 | 0 | 0 | 0 | 0 | 17 |
| Back  | 0 | 0 | 1 | 0 | 0 | 0 | 28 | 0 | 0 | 0 | 0 | 1 | 30 |
| Ear   | 0 | 0 | 0 | 0 | 0 | 0 | 0 | 18 | 0 | 0 | 0 | 0 | 18 |
| Hip   | 0 | 0 | 0 | 0 | 0 | 0 | 0 | 0 | 29 | 0 | 0 | 0 | 29 |
| Leg   | 0 | 0 | 0 | 0 | 0 | 0 | 0 | 0 | 0 | 31 | 0 | 0 | 31 |
| Wrist | 0 | 0 | 0 | 0 | 0 | 0 | 0 | 0 | 1 | 0 | 23 | 0 | 24 |
| Eye   | 1 | 0 | 0 | 0 | 0 | 0 | 0 | 0 | 0 | 0 | 0 | 31 | 32 |
| TOTAL | 48 | 40 | 48 | 32 | 40 | 16 | 32 | 24 | 32 | 32 | 24 | 32 | 400 |
| %     | 77 | 95 | 95,8 | 90,6 | 100 | 93,8 | 87,5 | 75 | 90,6 | 96,8 | 95,8 | 0 | 91,25 |

Table 5.6 – Confusion Matrix of joined results for 5 default gestures
Columns – Expected Results; Lines – Classification Results

**Feedback and User control**

In the two first phases, the final results were only based on the performance of the classification algorithm. However, in the Feedback and User control phase, when the focus was on the user interface, there are a lot more variables to consider in order to extract useful information that can lead us to pertinent conclusions. To better understand results, we divided the conclusion of each shortcut as one of 5 types:

1) Correct recognition, when users triggered the wanted shortcut with only one gesture and without *Multichoice*.

2) Correct recognition with *Multichoice*, when users triggered a shortcut that is stored in second or third order of preference in the same body part.

3) Error with *Multichoice* correction, when users reached the wanted application using *Multichoice*, even when other gesture was recognized.

4) Error with cancel, when a wrong gesture was recognized but users cancelled the shortcut at least one time. In this case, the user was able to reach the wanted application in the second or third gesture.

5) Error, when an unwanted application was triggered.

|  | Percentage | Time (Average) |
|---|---|---|
| Correct recognition | 62,43% | 3,3s |
| Correct recognition w/ *Multichoice* | 23,3% | 4,5s |
| Error with *Multichoice* correction | 7,1% | 4,6s |
| Error with Cancel | 3,8% | 6s |
| Errors (wrong application triggered) | 3,3% | --- |

**Table 5.7 – Percentage and Average time for each type of shortcut conclusion**

In Table 5.7, we registered the percentage of each type of shortcut conclusion, and also the average time users spent to reach the application we defined. In this phase, users made a total of 210 gestures (10 gestures more than supposed due to errors). When the wrong application was triggered, we did not register the spent time.

We made use of results of Table 5.7 to construct Table 5.8 and analyse the impact of having the *Multichoice* and cancelling mechanism in the interface. If we did not have any of those mechanisms, we estimate that 17.95% of the gestures would end in an unwanted shortcut. If using only the cancelling method, we estimate a error margin of 14,14% and 10,8% if using only the *Multichoice* mechanism. As stated in Table 5.7, we only registered 3,3% of shortcut errors (triggering other application rather that the one we asked for).

|  | Percentage |
|---|---|
| Errors without *Multichoice* or Cancelling | 17,95% |
| Errors using only Cancelling | 14,14% |
| Errors using only *Multichoice* | 10,8% |
| Errors | 3,3% |

**Table 5.8 – Impact of *Multichoice* and cancelling on error margin**

Finally, Table 5.9 presents some miscellaneous results, namely the click and time average for all Mnemonical Body Shortcuts with the correct result, respectively 2.5 clicks and 3.8 seconds. Furthermore, we also took note of stops during the interaction with our system, and users only stopped during a Mnemonical Body Shortcut in 1.4% of the cases.

|  | Percentage |
|---|---|
| Clicks Average | 2,5 clicks |
| Time Average | 3,8 seconds |
| Stops during Shortcuts | 1,4% |

**Table 5.9 – Miscellaneous Results**

**Questionnaire**

In the final phase of the usability studies, we asked users about different aspects of the Mnemonical Body Shortcuts interface, and in some questions they had to classify the interface in a range from 1 to 5. Firstly, we wanted to know their opinion about the interface feedback and control mechanism. In terms of feedback importance (Figure 5.5), the average classifications of each type were: Visual 3.1; Audio 4.9; Vibrational 3.7. When we asked for a general classification on the system feedback, the average was 4.5 (Figure 5.6).

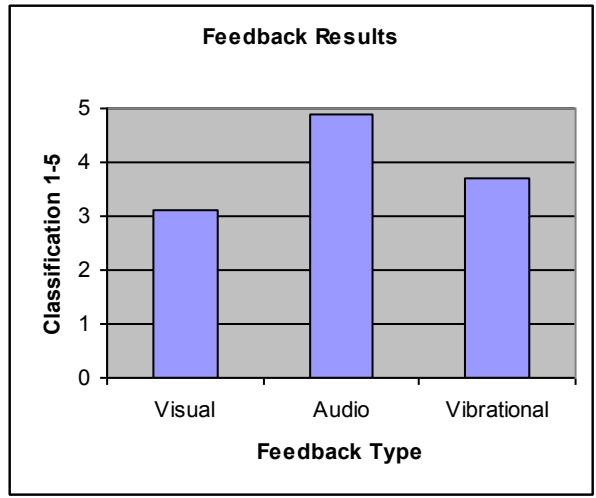

**Figure 5.5 – Feedback classification results from user questioning.**

In Figure 5.6 we also have available the average classification for other questions. Users classified the *Multichoice* function as 4.8/5, the confidence on the gesture recognized as 4/5. The advantage of using Mnemonical Body Shortcuts while walking was also classified as 4/5, and the speed of access to the applications as 4.1/5. Finally, users were asked if they liked the application (4.8/5), and 90% of them stated that they were capable of using it in private and 80% in public (Figure 5.7).

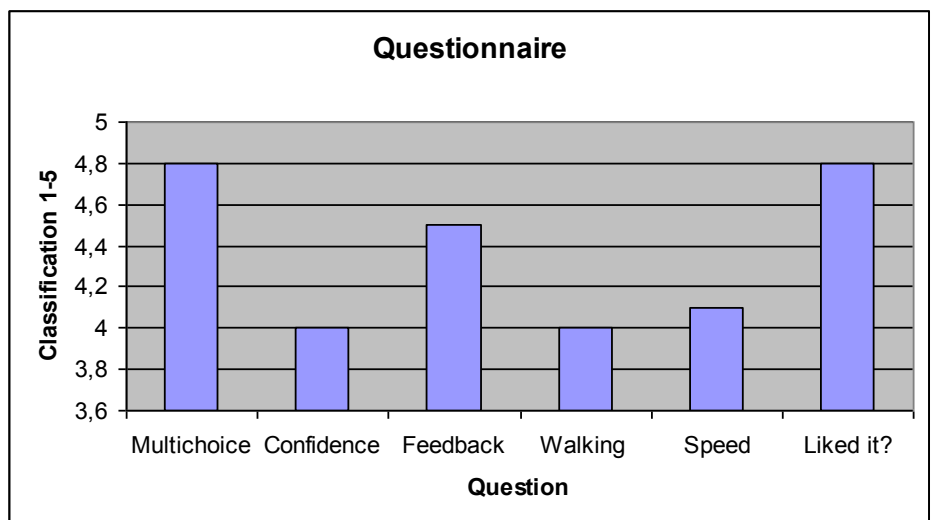

**Figure 5.6 – User classification for 6 different features or questions.**

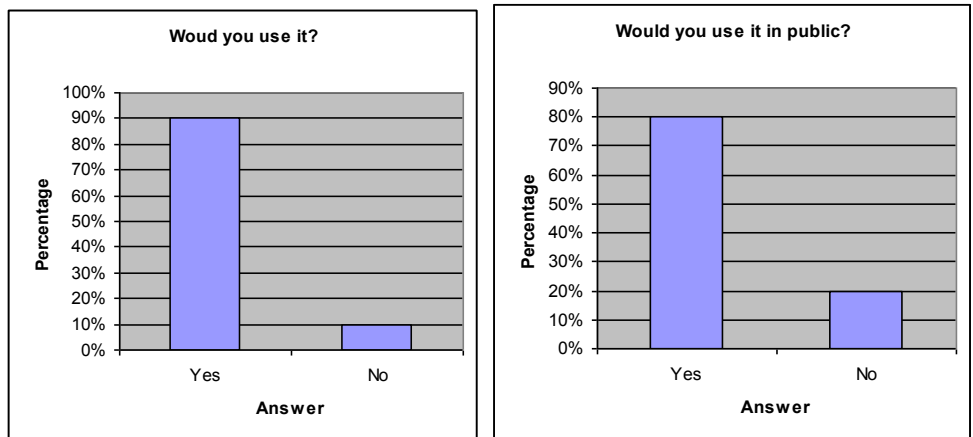

**Figure 5.7 – Results for the questions "Would you use the System" (left) and "Would you use it in public?" (right)**

We also asked users about possible enhancements to the system, and they came with some suggestions, described in Appendix D.

### 5.3.2. Discussion

Usability tests on the final prototype were diverse, so we focused on splitting the results on four different usability aspects, Effectiveness, Usefulness, Learnability and Likeability.

**Effectiveness**

The effectiveness of our approach is measurable in terms of recognition rate of the various scenarios we have tested. In terms of personalized gestures, the system had a final performance of 80.5% recognition while moving and 89.5% while standing with three trainings. While these final results are not totally satisfying, especially while moving, we can observe that they represent a large increase when compared with the results with 1 or 2 trainings (which will be discussed in the Learnability section) and reflect the recognition of totally uncontrolled gestures by the user, because we did not have any influence in the gesture selection. The effectiveness on the recognition of default gestures also has interesting results. As it was expected, when using the whole set of 12 gestures, the recognition rate was not very high (achieving 84.2% while standing), but in a real scenario we do not believe that users would want to have available all the default gestures. For that reason, we also tested default gestures with only 5 available gestures, and results were very positive (90% while moving and 92.5% while standing), especially if keeping in mind that these results do not rely on any user training, and could be largely improved if user training was added. We believe that these results prove that our system has a good effectiveness, suitable to the demanded task. It is also important to justify why recognition results dropped from the second prototype to the final prototype. Both prototypes used almost the same classification algorithm, and the differences are essentially on the test scenario. While users where on a standing position and repeated the each type of gestures 5 times consecutively in the second prototype tests, in the final usability tests they where also moving and were prompted to perform gestures randomly. Besides, in the second prototype we did not perform a study with truly personalized gestures (training results were constructed with gestures we asked users to perform). In fact, usability tests on the final prototype were far more demanding, which reflected results more close to the results it would achieve in real-life utilization. For example, since we had to delimit the space of interaction because of limitations of the Bluetooth Connection of the Bioplux4 system to the Laptop, we had to make users walk continuously in a room. They had to walk around a table, so there were many situations where they were obliged to perform a gesture when curving around the table. This utilization is even more demanding for a recognition algorithm, because in a real life scenario most of the gestures would be performed while walking to the front. This was one of the main causes for having a lower recognition rate for gestures performed when mobile.

**Usefulness**

We are able to measure the usefulness of our system when analysing if it accomplishes to give users what they need. In our work, we identified some requirements to the system that needed to be addressed to reduce user difficulties while using a mobile device: provide interaction while moving, with a low number of key presses, within a small time frame and with a low error rate. Most of the usefulness results of our prototype were given by the third phase of the usability test. In this phase, users triggered many shortcuts while moving, and all their actions were recorded. Final results demonstrated that only 3.3% of the Mnemonical Body Shortcuts ended with the triggering of an application we did not ask. This result shows the importance and the usefulness of the control mechanisms that the system provides to users. Besides, users only stopped 1.3% of the times while performing a gesture, demonstrating that this method is suitable for *on-the-move* interaction. In terms of key-presses and time to trigger a shortcut, users only needed an average of 2.5 clicks (including both clicks to start and end the gesture) and 3.8 seconds to reach their objective, which prove that our approach is able to trigger shortcuts rapidly and also with a reduced number of clicks. These results testify that our system achieves the needed usefulness to fulfil user needs, especially because they fulfil the defined usability goals for this area: triggering error less than 10% (3.3%), unintentional stops in less than 5% of the cases (1.3%), trigger applications in a less than 5 seconds in average (3.8 seconds) and need less than an average of 4 clicks to trigger a shortcut (2.5 clicks).

**Learnability**

The Learnability factor of a system usually refers to how users evolve while performing tasks with the system. However, our usability tests also had to focus on the recognition rate, thus varying the utilization scenarios. Furthermore, a test on user Learnability had to be performed while using a system on a daily basis, but hardware restrictions did not allow us to perform those tests. Due to these facts, we cannot distinct a learning curve of the utilization of Mnemonical Body Shortcuts. We can, however, analyse the Learnability of the system, observing the improvement across training phases. In Figure 5.3 we presented the results of the different training phases, and it is possible to observe that recognition rate was growing at an enormous rate in each phase. It is possible to conclude that the recognition rate would still increase many points if we constantly updated the training set. The system would be enhanced if we used a dynamic training set during time. A dynamic approach would also bring many benefits to the recognition while moving, because this prototype was only trained with standing gestures. In conclusion, we believe that the system has a great potential to learn with users and achieve even better recognition rates. User Learnability has still to be further tested, but during the usability test our perception was that users felt increasingly more comfortable while using the Mnemonical Body Shortcuts.

**Likeability**

In order to know what was user's opinion about the different aspects of the interface, we conducted a questionnaire after the usability tests, rating some characteristics from 1 to 5.In general terms, results were very positive (Table 5.6). There are some important conclusions from the results

- Users felt the feedback as suitable and appropriate (classification averaging 4.5), and prefer the audio feedback when compared with the vibrational and visual feedback.
- They classified the *Multichoice* feature as very important (avg. 4.8), and see the system as a fast mechanism to use while mobile (avg. 4) and to rapidly access applications (4.2)
- Generally, they liked the system (avg. 4.8), and 90% of users stated that they would use the system in private, while 80% would also use it in public.

These results are the clear reflection that users liked Mnemonical Body Shortcuts and, even having some reserves in the first usages, ended the usability tests appreciating the system. The usability goal of 50% of users willing to use our system was clearly surpassed.

**Suggestions**

Users were, in the last phase of usability tests, very interested in giving the opinion about the system and trying to explore their limitation to help us enhancing the Mnemonical Body Shortcuts approach. We considered a couple of suggestions as very interesting:

- Cancelling a shortcut shaking the mobile device is an alternative to the cancelling button, but we should guarantee a low rate of false positives.
- Be able to personalize the delay time, because some users felt that 2 seconds was too short time, especially in the first usages of the system.
- Give more feedback to users when a gesture is cancelled, because when users cancelled the gesture or reached the final application using *Multichoice* there was no confirmation of that cancelling. Users were obliged to look to the mobile device to confirm that they did not trigger any shortcut.

- The *Multichoice* feature could also be used with a tilt mechanism. Further testing should be done to validate the usefulness and feasibility of such interaction

The physical issues regarding the PDA would be easily surpassed if using a lighter mobile device with buttons with a better placement for our approach. Diverse results were discussed, but the essential objectives that we defined in the usability goals were achieved. The comparison between usability goal, metric and result is present in Table 5.10 for all the six usability goals.

| Usability Goal | Metric | Result |
| --- | --- | --- |
| Gesture memorization | Errors < 10% | 6% |
| Clicks to trigger applications | Clicks < 4 | 2.5 clicks |
| Time to trigger applications | Time < 5 seconds | 3,8 sec. |
| Mobility | < 5% unintentional stops | 1,3% |
| Likeability | > 50% willing to use the system | 90% |
| Shortcut triggering | < 10% errors | 3,3% |

**Table 5.10 – Usability goals and final results**

# 6

# Conclusions and Future Work

In the beginning of our work, we defined as main objective the creation of a gestural interface for mobile devices capable of surpassing some issues present on actual interaction with these devices. After the development of our system, we are able to present some conclusions and suggest future activities capable of enhancing the work described in this document.

## 6.1. Conclusions

Our approach to enhance mobile interaction is based on the creation of gestural shortcuts to the most used applications, using the body-space as a repository of meaningful relations with the applications to be triggered, named Mnemonical Body Shortcuts. This interface should be usable in diverse mobile and social environments and provide fast and accurate gesture recognition. In order to persecute the best options, we explored the different available technologies able to provide mobile devices with gestural recognition, and ended up using two of them, Radio Frequency Identification (RFID) and Accelerometers. Our approach was firstly validated with an RFID prototype, where user evaluation showed that, even against some established key shortcuts, Mnemonical Body Shortcuts had better recall results. The accelerometer was used to surpass physical limitations of a RFID prototype. With accelerometers, we explored two alternatives to provide gestural recognition, chose one for the final prototype and enhanced the gestural recognition interface with appropriate feedback and user-control mechanisms, namely with the implementation of gesture cancelling facilities and the *Multichoice* feature, capable of switching between applications. The most important results came from the usability studies on this last prototype:

- The recognition algorithm proved to be suitable to the task of recognizing body-based gestures, not only while standing but in demanding mobility settings. It was also possible to conclude that there is an excellent evolution on the recognition rate when user training is added.
- With our approach, it is possible to use the system without any training if using the default gestures, also with appropriate recognition accuracy.
- When using the full interface, with feedback and user control (cancelling and *Multichoice* features), we were able to significantly reduce the number of possible errors that would happen if applications were triggered immediately after recognition.

- We were also able to conquer user's opinion about the system: in the final test, the questionnaire on the multiple characteristics of the system was very positive, as well as the feedback given by users. Most of them stated that they enjoyed using the interface and they would appreciate using it in a daily basis.

The results that were described throughout this thesis are sufficient for us to state the general objectives we have defined in the start of the work as totally accomplished. Besides, the specific usability goals we defined were also totally completed. We can, in the actuality, provide users with a system that is able to enhance the way we interact with our mobile devices, surpassing actual solutions. Our research and implementation also surpassed previous works because it was able to provide an overall look through the diverse aspects inherent to the creation of a mobile gestural interface, such as concept evaluation, gestural recognition, feedback and user-control mechanisms.

## 6.2. Future Work

It is possible to define many enhancements to our work, as well as suggestions to future work that can use Menmonical Body Shortcuts as basis to explore other areas of mobility interaction using inertial sensing. The future work within the subject of Mnemonical Body Shortcuts can be divided in three main areas:

**Development**

During the development phase and user testing, some ideas came to mind, but were not implemented on the final prototype. It should be interesting to create a prototype that would be able to learn with user gestures not only in a defined training phase but also during regular utilization. We believe that this approach would significantly increase the recognition rate with little utilization. Cancelling shortcuts could be triggered with a gesture and should also be enhanced with more feedback. Users should be able to parameterize aspects of the interface such as feedback or confirmation time. Most importantly, if a mobile device with accelerometer and open API was available, one should develop Mnemonical Body Shortcuts to run only in the mobile device.

**Tests**

Usability tests on the final prototype lacked on analysing users' learning curve while interacting with the system. Further testing should be done, focusing on this specific characteristic. Furthermore, we are aware that our tests were performed in a controlled environment, and these tests lack some relation with the utilization of mobile devices. If a self-contained development could be achieved (using mobile devices with inertial sensing), we would be able to make some field tests, giving test mobile devices during some days, and then retrieve results from that usage.

**Mobile Interaction Enhancement**

A successful implementation of this system in commercial mobile devices is still difficult, but we believe that, if we correctly communicate to the capabilities of gesture-based interaction, many manufactures would give more attention to this type of interaction. First implementations of a gestural interface could start with simpler gesture to perform some actions, such as answering a call when lifting the mobile device to the ear, or either make proper use of the contextual information given by an accelerometer. These simpler actions would be the first step to introduce to users this new method of interaction. We developed a gestural interface, but there is still much work to do if we want to introduce it to a vaster audience.

# A

# Task Analysis

## A.1 Protocol for Actual Panorama Analysis

1. **User Characterization**

    a) Name:_______________________
    b) Age: _____
    c) Sex:
       ☐ M ☐ F

    d) Academic Habilitations:
       ☐ 4ª Classe  ☐ 9º Ano  ☐ 12º Ano  ☐ Ensino Superior

2. **Mobile device usage**

    a) Do you use cell phone or PDA?
       ☐ Cell phone  ☐ PDA

    b) Model :______________________
    c) Frequency of usage of the mobile device:

       ☐ One time per week or less
       ☐ One time per day or less
       ☐ Up to 10 times per day
       ☐ More than 10 times per day

    d) Used Applications:

       | | |
       |---|---|
       | Contact List | ☐ |
       | Calls | ☐ |
       | Send SMS | ☐ |
       | Send MMS | ☐ |
       | Listen to Music | ☐ |
       | Listen to Radio | ☐ |
       | Take Photos | ☐ |
       | See Photos | ☐ |
       | Film | ☐ |
       | See Videos | ☐ |
       | Browse the Internet | ☐ |
       | E-mails | ☐ |
       | Awakening Alarm | ☐ |
       | Clock | ☐ |
       | Games | ☐ |

   Voice Recorder   ☐
   Calendar/Agenda  ☐
   Bluetooth/Infrared  ☐
   Calculator    ☐
   Other_______________________________

e) Number of most frequent contacts:_____

**3. Shortcuts in Mobile Devices**

 **3.1 – Key Shortcuts**

 a) Do you use key shortcuts? ☐ Yes ☐ No
 b) Utilization Frequency:
  ☐ One time per week or less
  ☐ One time per day or less
  ☐ Up to 10 times per day
  ☐ More than 10 times per day
 c) How many key shortcuts you have available? _____
 d) To you remember all relations between shortcuts and applications? ☐ Yes ☐ No
  Example:_______________________________________________

 e) Do you have any difficulty in the utilization of key shortcuts?
______________________________________________________________________
______________________________________________________________________

 **3.2 – Voice Shortcuts**

 b) Do you use voice shortcuts? ☐ Yes ☐ No
 b) Utilization Frequency:
  ☐ One time per week or less
  ☐ One time per day or less
  ☐ Up to 10 times per day
  ☐ More than 10 times per day
 f) How many voice shortcuts you have available? _____
 g) To you remember all relations between shortcuts and applications? ☐ Yes ☐ No
  Example:_______________________________________________

 h) Do you have any difficulty in the utilization of voice shortcuts?
______________________________________________________________________
______________________________________________________________________

## A.2 Protocol for User Observation

**1) Selection of most frequent applications**

Ask users to select the 3 most used applications, from the previous list. Then, ask users to reach those applications with their mobile device, counting the number of necessary clicks for each.

|   | Application | Contact |
|---|---|---|
| **1** | | |
| **2** | | |
| **3** | | |

Notes:
______________________________________________________________________
______________________________________________________________________

**2) Selection of most frequent contacts**

Ask users to select the 3 most used contacts. Then, ask users to use their own mobile device to call those 3 contacts, while we count the number of necessary clicks for each.

| **Contact** | Clicks |
|---|---|
| **1** | |
| **2** | |
| **3** | |

Notes:
______________________________________________________________________
______________________________________________________________________

# A.3 Task Analysis Results

**1. User Characterization**

    **b)** Age: Average of **24,45** years

    **c)** Sex:                                          **d)** Academic Applications:

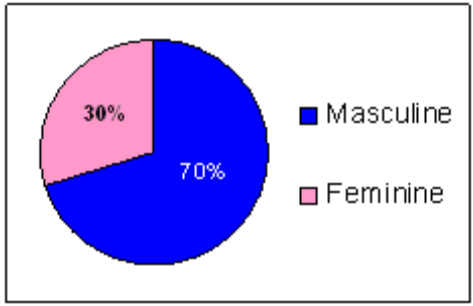 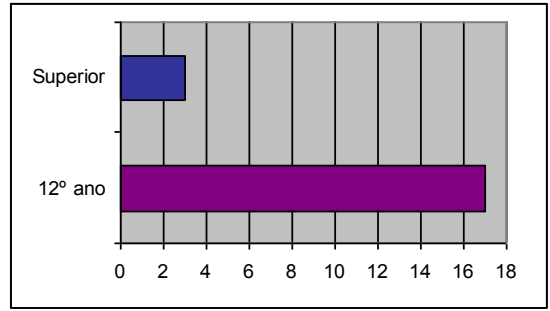

**2. Mobile devices usage**

    **a)** Cell Phone / PDA usage        **c )** Frequency of utilization of the mobile device

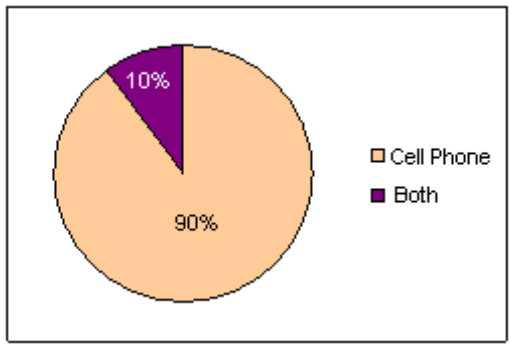 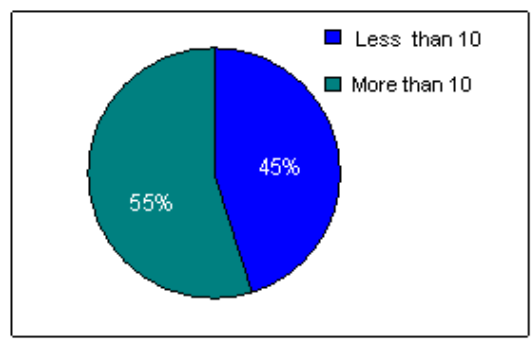

    **d)** Most used applications

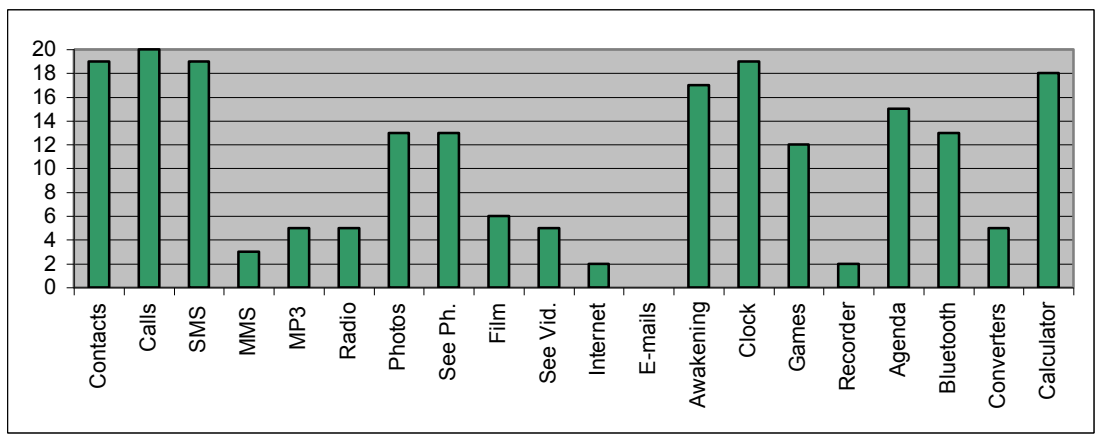

    **e)** Most frequent contacts: averages **5,7 contacts**

## 2. Shortcuts in Mobible devices

### 2.1 – Key Shortcuts

**a)** Uses Key Shortcuts          **b)** Frequency of Utilization:

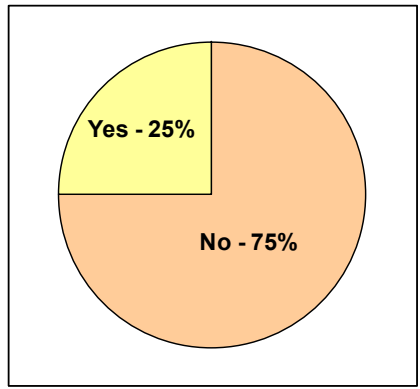 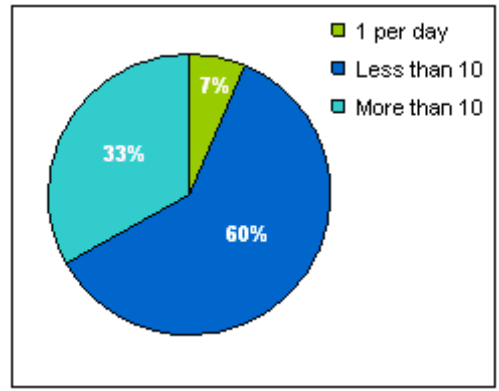

**c)** How Many Key Shortcuts normally use: average of **4,73,** mode e median of **3.**

**d)** Shortcut memorization on key shortcuts: only **one** user stated that he usually forgets where the applications are.

### 2.2 – Voice Shortcuts

**A)** Uses Voice Shortcuts: **100% don't use**

## 3. User Observation:

Click average to reach the 3 most frequent applications and 3 most frequent contacts

Aplications: **3,54**   Contacts: **4,7**

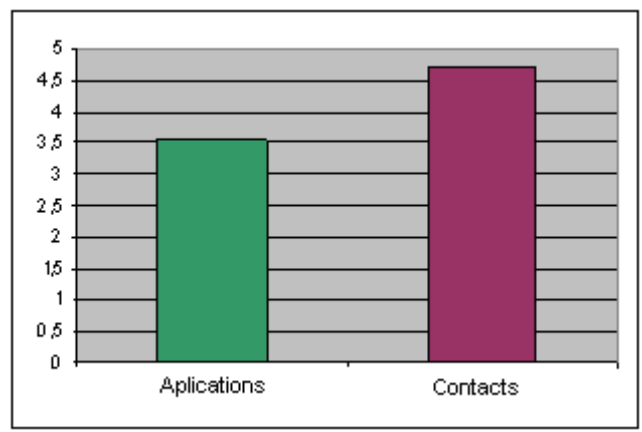

# B

# Answering the 11 Task Analysis Questions

This appendix presents the answers to the 11 Task Analysis questions:

**Who is going to use the system?**

The system is supposed to be used by a general audience that already uses mobile devices, with variant literacy, ages and physical characteristics.

**What tasks do they perform?**

Actually, users interact with mobile devices using graphical user interfaces, based on menu selection to reach the applications. In terms of shortcuts, two solutions are currently used: voice and key shortcuts. While key shortcuts are used by many users, voice shortcuts are not so common. In terms of applications, the most used ones are making calls, sending SMS, contact list, see/take photos, clock, games, agenda and Bluetooth connection.

**What tasks are desired?**

Users expect to be able to reach the most used applications in a faster and more suitable manner, with a less number of clicks.

**How are the tasks learned?**

Firstly, users learn no reach applications using the default menu selection. Menus are usually defined in hierarchical levels that users follow to reach the wanted application. Later, they are able to program voice or key shortcuts and personalize their access.

**Where are tasks performed?**

Tasks are performed in a mobile device, most of the times in cell phones but also in PDA's or Smart Phones. They are performed in diverse mobility settings (while standing, walking or even running), with variable noise, lighting and social constraints.

**What is the relationship between user and data?**

User's information is restricted to their mobile device

**What other tools does the user have?**

Some mobile devices have point-and-click interfaces using touch screens, but generally they are also based on menu selection.

**How do users communicate with each other?**

Not relevant question.

**How often are the tasks performed?**

Users need their mobile device to realize some task at least one time per day (93%), while 60% perform tasks up to 10 times per day and 33% more than 10 times per day.

**What are the time constraints on the tasks?**

The time to access an application should be minimal, because it is the main motivator to use shortcuts and users might be involved in other tasks at the same time and need some information rapidly.

**What happens when things go wrong?**

We define "wrong" as the situation when a user selects an unwanted application. In that case, the solution is to exit the application that was triggered, which is usually done by a cancelling button in all mobile devices.

# C

# RFID Prototype Evaluation

## C.1 Protocol for RFID Prototype Evaluation

Name: _______________________________________________

**Associating applications to keys and body parts (use numbers in the body below)**

|   | Application | Key | Tag |
|---|---|---|---|
| 1 |  |  |  |
| 2 |  |  |  |
| 3 |  |  |  |
| 4 |  |  |  |
| 5 |  |  |  |

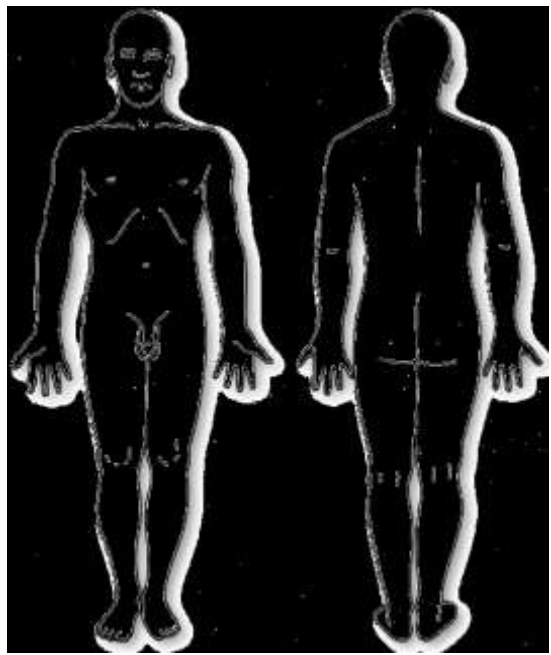

**Shortcut Execution**

a) Equip the user with the RFID tags in the previously chosen places

b) Ask the user to reach some body shortcuts, with intervals of 5 seconds, following this order:

**2-4-1-3-5-4-1-3-5-2-2-5-3-1-4-5-3-1-4-2**

c) Record the logo f the application with the user name.

d) Complete the table with the data present in the log:

Coments: ____________________________________________

| Shortcut | Correct | Mnemonical Error | Reading Error | Key Error |
|---|---|---|---|---|
| 2 | | | | |
| 4 | | | | |
| 1 | | | | |
| 3 | | | | |
| 5 | | | | |
| 4 | | | | |
| 1 | | | | |
| 3 | | | | |
| 5 | | | | |
| 2 | | | | |
| 2 | | | | |
| 5 | | | | |
| 3 | | | | |
| 1 | | | | |
| 4 | | | | |
| 5 | | | | |
| 3 | | | | |
| 1 | | | | |
| 4 | | | | |
| 2 | | | | |
| # | | | | |
| % | | | | |

**User Classification**

1) Is it easy to use the system to trigger the applications?____________________
2) Would you fell comfortable if performing these gestures in public?________________
   ___________________________________________________________________
3) Would you use this method if the tags could be invisibly placed?________________________________________________________
   a. If not, would you use this method if tags were not required?
      ___________________________________________________________
4) Is this system different from other shortcut mechanisms?
   ___________________________________________________________

**Thank You!**

# C.2 Results on RFID Prototype Evaluation

**1) Relational table between Applications and Body Parts (Mnemonics)**

| Application /Body | Finger | Ear | Eye | Head | Wrist | Chest | Hand | Mouth |
|---|---|---|---|---|---|---|---|---|
| **Send SMS** | **6** | | | | | 1 | **10** | |
| **Make Call** | | **12** | | 1 | | | | 3 |
| **Contacts** | | 1 | | 2 | | **5** | 3 | |
| **Clock** | | | 1 | | **10** | | | |
| **Awakening Alarm** | | **3** | 2 | 2 | 2 | | | |
| **Agenda** | | | | 1 | | **3** | 1 | |
| **MP3** | | 2 | | | | | | |
| **Receive SMS** | | | | | | 1 | | |
| **Photos** | | | **8** | 2 | | | | |
| **Calculator** | | | | | | | 3 | |

**2) Some of the most relevant exampeles**

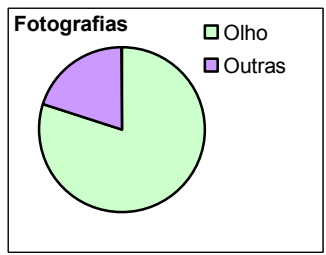
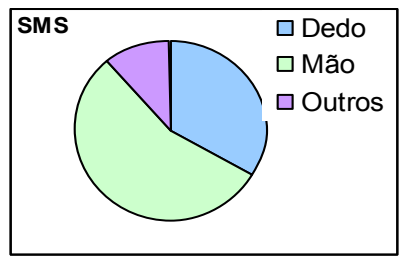
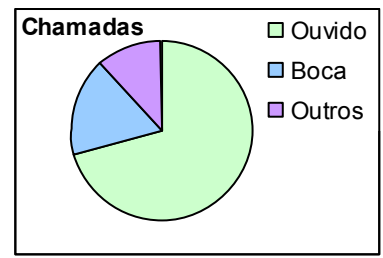
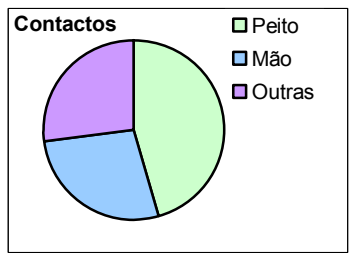
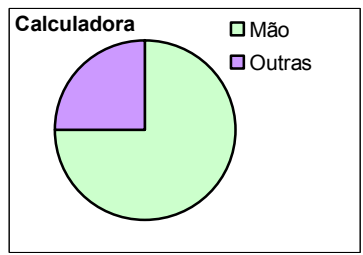
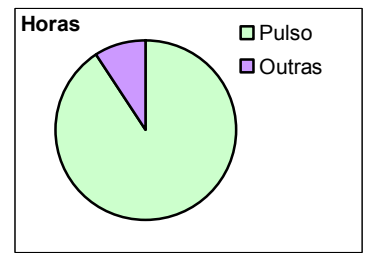

**a)** Error percentage on the utilization of key shortcuts, gestural mnemonics and reading errors, when one hour passed from the shortcut selection.

**Key Errors: 9%**

**Reading Errors: 6%**

**Gestural Mnemonics errors: 0.8%**

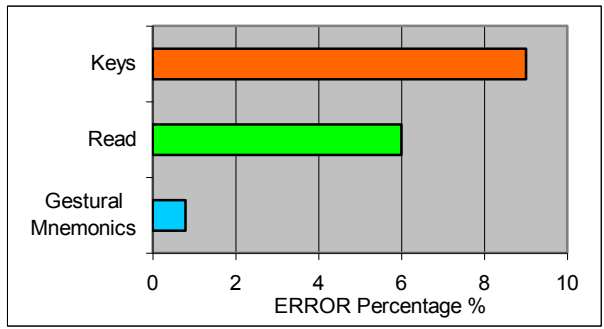

**b)** Error percentage on the utilization of key shortcuts and gestural mnemonics, when one week passed from the shortcut selection.

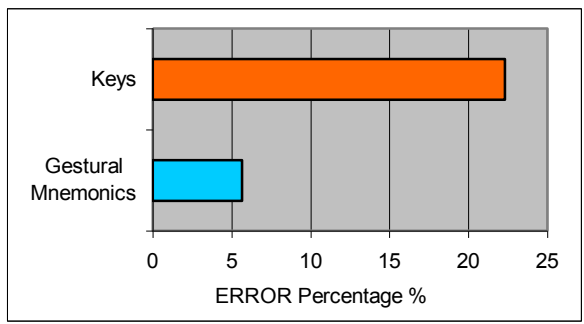

# D

# Final Prototype Evaluation

## D.1 Protocol for Final Prototype Evaluation

## 1 Pre-Questionnaire

### 1.1 User Characterization

    a) Name:_________________________
    b) Age: _____
    c) Sex: ☐ M ☐ F
    d) Academic Habilitations:
        ☐ 4ª Classe ☐ 9º Ano ☐ 12º Ano ☐ Ensino Superior

### 1.2 Mobile device usage

    a) Do you use cell phone or PDA?
        ☐ Cell phone ☐ PDA

    b) Model :____________________________

    c) Frequency of usage of the mobile device:

        ☐ One time per week or less

        ☐ One time per day or less

        ☐ Up to 10 times per day

        ☐ More than 10 times per day

    d) Do you use key shortcuts? ☐ Yes    ☐ No
    e) Do you use voice shortcuts? ☐ Yes    ☐ No

## 2  Introduction to Mnemonical Body Shortcuts

**2.1 Explain to the user the concept of mnemonical Body**

**2.2 Ask the user to chose 5 personal Mnemonical Body Shortcuts:**

| Body Part | Aplication |
|---|---|
|  |  |
|  |  |
|  |  |
|  |  |
|  |  |

**Possible Aplications**: Agenda, Browser, Calculator, Camera, Call Mother, Call Friend, Contacts, Awakening Alarm, Photos, GPS, Voice Recorder, Hours, Games, Messenger, Mp3, SMS, Temperature, Voicemail, Word.

The body parts are chosen freely.

## 3  Personalized Gestures and System Learnability

**3.1 In first place, the user should experiment one of the default gestures, with the objective of a better understand how he should perform gestures.**

**3.2 Personalized Gestures**

The user should train the selected gestures, in the following order:

**1 – Train the 5 gestures one time;**

    Test 20 random gestures while standing

    Test 20 random gestures while moving

**2 - Train the 5 gestures one more time;**

    Test 20 random gestures while standing

    Test 20 random gestures while moving

**3 - Train the 5 gestures for the third time**

    Test 20 random gestures while standing

    Test 20 random gestures while moving

## 4  Default Gestures

**4.1 The user should start by knowing the 12 default gestures that are recognized by the system without training.**

The test monitor should exemplify each one of the 12 gestures one time, repeating if the user has any doubt.

**4.2  Teste de eficiência**

Before the test starts, a random list of 5 default gestures should be generated, to be used during this test.

**1 – Test the set of 12 default gestures, 2 times each**

    Test 24 gestures while standing

    Test 24 gestures while moving

**2 – Test the set of 5 default gestures, 4 times each**

    Test 20 gestures while standing

    Test 20 gestures while moving

## 5 – Interface Test

In the last part of the test, we intend to test the developed user interface.

**5.1 Firstly, the user should experiment some Mnemonical Body Shortcuts using the full interface, with feedback, cancelling and *Multichoice* mechanisms**

**5.2 Realize the Interface Test.**

The Interface Test will be performed using 20 Mnemonical Body Shortcuts, using 5 random default gestures defined before. In this case, we will prompt users to access some applications and not to make some specific gesture. We should guarantee that users are obliged to use *Multichoice* to reach some applications. The user should perform the test while walking.

To get all the needed results, an excel sheet should be completed during tests.

**5.3 Feedback Questionnaire**

**5.3.1 From 1 to 5, how do you classify the advantage of visual feedback? ________**

**5.3.2 From 1 to 5, how do you classify the advantage of audio feedback? ________**

**5.3.3 From 1 to 5, how do you classify the advantage of vibrational feedback? ____**

**5.3.4 Do you distinguish the different vibrational times? _______**

**Are those different vibrational times useful? \_\_\_\_\_\_\_\_**

**5.3.5 From 1 to 5, how do you classify the advantage of using Multichoice? \_\_\_\_\_\_**

# 6 – Final Questionnaire

**6.1 From 1 to 5, how do you classify your confidence on the gesture recognizer? \_\_\_\_\_**

**6.2 From 1 to 5, how do you classify the feedback given after each recognized gesture? \_\_\_\_\_\_\_\_**

**6.3 From 1 to 5, how do you classify the advantages of using Mnemonical Body SHortcuts when you are walking? \_\_\_\_\_\_\_**

**6.4 From 1 to 5, how do you classify the advantages of using Mnemonical Body Shortcuts in terms of fast access to the applications? \_\_\_\_\_\_**

**6.5 Do you like using the application? Classify from 1 to 5 \_\_\_\_\_\_**

**6.6 Would you use Mnemonical Body Shortcuts if they were available in your mobile device?**
\_\_\_\_\_\_\_\_\_\_\_\_\_\_\_\_\_\_\_\_\_\_\_\_\_\_\_\_\_\_\_\_\_\_\_\_\_\_\_\_\_\_\_\_\_\_\_\_\_\_\_\_\_\_\_\_\_\_\_\_\_\_\_\_\_\_\_\_\_\_
\_\_\_\_\_\_\_\_\_\_\_\_\_\_\_\_\_\_\_\_\_\_\_\_\_\_\_\_\_\_\_\_\_\_\_\_\_\_\_\_\_\_\_\_\_\_\_\_\_\_\_\_\_\_\_\_\_\_\_\_\_\_\_\_\_\_\_\_\_\_
\_\_\_\_\_\_\_\_\_\_\_\_\_\_\_\_\_\_\_\_\_\_\_\_\_\_\_\_\_\_\_\_\_\_\_\_\_\_\_\_\_\_\_\_\_\_\_\_\_\_\_\_\_\_\_\_\_\_\_\_\_\_\_\_\_\_\_\_\_\_

**6.7 Would you use them in public?**
\_\_\_\_\_\_\_\_\_\_\_\_\_\_\_\_\_\_\_\_\_\_\_\_\_\_\_\_\_\_\_\_\_\_\_\_\_\_\_\_\_\_\_\_\_\_\_\_\_\_\_\_\_\_\_\_\_\_\_\_\_\_\_\_\_\_\_\_\_\_
\_\_\_\_\_\_\_\_\_\_\_\_\_\_\_\_\_\_\_\_\_\_\_\_\_\_\_\_\_\_\_\_\_\_\_\_\_\_\_\_\_\_\_\_\_\_\_\_\_\_\_\_\_\_\_\_\_\_\_\_\_\_\_\_\_\_\_\_\_\_

**6.8 Do you have any suggestion?**
\_\_\_\_\_\_\_\_\_\_\_\_\_\_\_\_\_\_\_\_\_\_\_\_\_\_\_\_\_\_\_\_\_\_\_\_\_\_\_\_\_\_\_\_\_\_\_\_\_\_\_\_\_\_\_\_\_\_\_\_\_\_\_\_\_\_\_\_\_\_
\_\_\_\_\_\_\_\_\_\_\_\_\_\_\_\_\_\_\_\_\_\_\_\_\_\_\_\_\_\_\_\_\_\_\_\_\_\_\_\_\_\_\_\_\_\_\_\_\_\_\_\_\_\_\_\_\_\_\_\_\_\_\_\_\_\_\_\_\_\_

**Thank You!**

## D.2 Final Prototype Evaluation Results

### 1 - Personalized Gestures and System Learnability

The Table D.1 presents the individualized results (in % of recognition) for each one of the 10 tested users, regarding personalized gestures with different training sets. Figure D.1 resumes the results in a Line Graphic.

|         | 1 Train  |        | 2 Trains |        | 3 Trains |        |
|---------|----------|--------|----------|--------|----------|--------|
|         | Standing | Moving | Standing | Moving | Standing | Moving |
| User1   | 60%      | 60%    | 70%      | 65%    | 85%      | 65%    |
| User2   | 75%      | 85%    | 100%     | 85%    | 90%      | 85%    |
| User3   | 95%      | 75%    | 90%      | 75%    | 95%      | 85%    |
| User4   | 50%      | 45%    | 75%      | 65%    | 85%      | 80%    |
| User5   | 65%      | 50%    | 65%      | 70%    | 95%      | 75%    |
| User 6  | 80%      | 70%    | 95%      | 70%    | 100%     | 85%    |
| User 7  | 80%      | 75%    | 90%      | 80%    | 90%      | 85%    |
| User 8  | 55%      | 60%    | 75%      | 90%    | 95%      | 90%    |
| User 9  | 80%      | 65%    | 80%      | 70%    | 85%      | 75%    |
| User 10 | 60%      | 60%    | 70%      | 55%    | 75%      | 80%    |
| Average | 70%      | 64,5%  | 81%      | 72,5   | 89,5%    | 80,5%  |

**Table D.1**

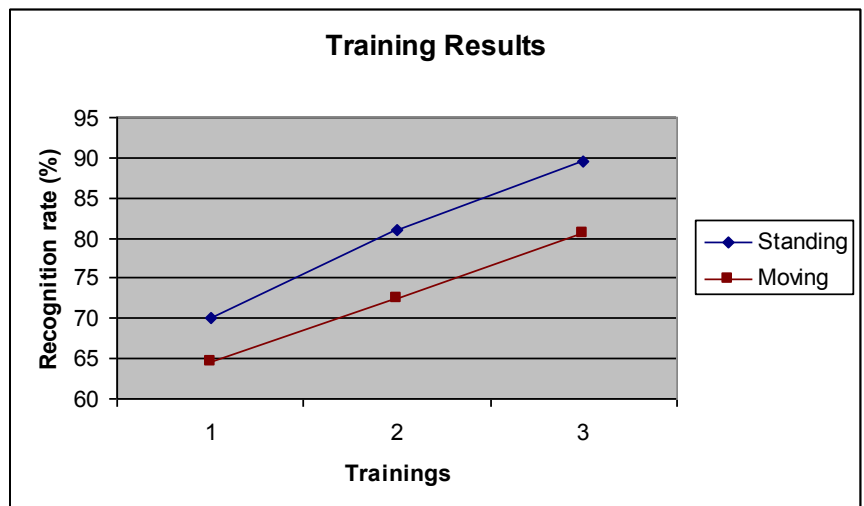

**Figure D.1**

## 2 - Default Gestures

In this section we reveal the detailed results on the recognition of 12 and 5 default gestures.

### 2.1 - 12 Default Gestures

Confusion Matrixes D.2 and D.3 correspond, respectively, to the recognition of 12 default gestures while standing and moving. Confusion Matrix D.4 joins the results of D.2 and D.3.

|       | Mouth | Chest | Elbow | Navel | Neck | Head | Back | Ear | Hip | Leg | Wrist | Eye | TOTAL |
|-------|-------|-------|-------|-------|------|------|------|-----|-----|-----|-------|-----|-------|
| Mouth | 14    | 0     | 0     | 0     | 0    | 0    | 0    | 0   | 0   | 0   | 0     | 0   | 14    |
| Chest | 1     | 20    | 2     | 0     | 0    | 1    | 0    | 1   | 1   | 0   | 0     | 0   | 26    |
| Elbow | 3     | 0     | 18    | 0     | 0    | 0    | 0    | 0   | 0   | 0   | 0     | 0   | 21    |
| Navel | 0     | 0     | 0     | 19    | 0    | 0    | 2    | 0   | 1   | 0   | 0     | 0   | 22    |
| Neck  | 0     | 0     | 0     | 0     | 20   | 0    | 0    | 1   | 0   | 0   | 0     | 0   | 21    |
| Head  | 0     | 0     | 0     | 0     | 0    | 18   | 0    | 0   | 0   | 0   | 0     | 0   | 18    |
| Back  | 1     | 0     | 0     | 0     | 0    | 0    | 18   | 0   | 0   | 1   | 0     | 0   | 20    |
| Ear   | 1     | 0     | 0     | 0     | 0    | 0    | 0    | 6   | 0   | 0   | 1     | 0   | 8     |
| Hip   | 0     | 0     | 0     | 1     | 0    | 0    | 0    | 0   | 17  | 6   | 0     | 0   | 24    |
| Leg   | 0     | 0     | 0     | 0     | 0    | 0    | 0    | 0   | 1   | 13  | 0     | 0   | 14    |
| Wrist | 0     | 0     | 0     | 0     | 0    | 1    | 0    | 0   | 0   | 0   | 19    | 0   | 20    |
| Eye   | 0     | 0     | 0     | 0     | 0    | 0    | 0    | 12  |     | 0   | 0     | 20  | 32    |
| TOTAL | 20    | 20    | 20    | 20    | 20   | 20   | 20   | 20  | 20  | 20  | 20    | 20  | 240   |
| %     | 70    | 100   | 90    | 95    | 100  | 90   | 90   | 30  | 85  | 65  | 95    | 100 | 84,1  |

**Table D.2 - Confusion Matrix of results for 12 default gestures (Standing)**
**Columns – Expected Results; Lines – Classification Results**

|       | Mouth | Chest | Elbow | Navel | Neck | Head | Back | Ear | Hip | Leg | Wrist | Eye | TOTAL |
|-------|-------|-------|-------|-------|------|------|------|-----|-----|-----|-------|-----|-------|
| Mouth | 13    | 3     | 0     | 0     | 0    | 0    | 0    | 0   | 0   | 0   | 0     | 0   | 16    |
| Chest | 2     | 14    | 1     | 2     | 0    | 0    | 0    | 0   | 0   | 1   | 0     | 0   | 20    |
| Elbow | 2     | 1     | 16    | 1     | 0    | 0    | 0    | 0   | 1   | 0   | 0     | 0   | 21    |
| Navel | 0     | 2     | 0     | 16    | 0    | 0    | 0    | 0   | 2   | 1   | 0     | 0   | 21    |
| Neck  | 0     | 0     | 0     | 0     | 20   | 2    | 0    | 1   | 0   | 0   | 0     | 0   | 23    |
| Head  | 2     | 0     | 2     | 0     | 0    | 17   | 0    | 1   | 0   | 0   | 1     | 0   | 23    |
| Back  | 0     | 0     | 0     | 0     | 0    | 0    | 20   | 0   | 0   | 2   | 0     | 0   | 22    |
| Ear   | 1     | 0     | 0     | 0     | 0    | 0    | 0    | 4   | 1   | 1   | 0     | 0   | 7     |
| Hip   | 0     | 0     | 0     | 1     | 0    | 0    | 0    | 0   | 16  | 6   | 0     | 0   | 23    |
| Leg   | 0     | 0     | 0     | 0     | 0    | 0    | 0    | 0   | 0   | 9   | 0     | 0   | 9     |
| Wrist | 0     | 0     | 1     | 0     | 0    | 0    | 0    | 0   | 0   | 0   | 19    | 0   | 20    |
| Eye   | 0     | 0     | 0     | 0     | 0    | 1    | 0    | 14  | 0   | 0   | 0     | 20  | 35    |
| TOTAL | 20    | 20    | 20    | 20    | 20   | 20   | 20   | 20  | 20  | 20  | 20    | 20  | 240   |
| %     | 65    | 70    | 80    | 80    | 100  | 85   | 100  | 20  | 80  | 45  | 95    | 100 | 76,6  |

**Table D.3 - Confusion Matrix of results for 12 default gestures (Moving)**
**Columns – Expected Results; Lines – Classification Results**

|       | Mouth | Chest | Elbow | Navel | Neck | Head | Back | Ear | Hip | Leg | Wrist | Eye | TOTAL |
|-------|-------|-------|-------|-------|------|------|------|-----|-----|-----|-------|-----|-------|
| Mouth | 27    | 3     | 0     | 0     | 0    | 0    | 0    | 0   | 0   | 0   | 0     | 0   | 30    |
| Chest | 3     | 34    | 3     | 2     | 0    | 1    | 0    | 1   | 1   | 1   | 0     | 0   | 46    |
| Elbow | 5     | 1     | 34    | 1     | 0    | 0    | 0    | 0   | 1   | 0   | 0     | 0   | 42    |
| Navel | 0     | 2     | 0     | 35    | 0    | 0    | 2    | 0   | 3   | 1   | 0     | 0   | 43    |
| Neck  | 0     | 0     | 0     | 0     | 40   | 2    | 0    | 2   | 0   | 0   | 0     | 0   | 44    |
| Head  | 2     | 0     | 2     | 0     | 0    | 35   | 0    | 1   | 0   | 0   | 1     | 0   | 41    |
| Back  | 1     | 0     | 0     | 0     | 0    | 0    | 38   | 0   | 0   | 3   | 0     | 0   | 42    |
| Ear   | 2     | 0     | 0     | 0     | 0    | 0    | 0    | 10  | 1   | 1   | 1     | 0   | 15    |
| Hip   | 0     | 0     | 0     | 2     | 0    | 0    | 0    | 0   | 33  | 12  | 0     | 0   | 47    |
| Leg   | 0     | 0     | 0     | 0     | 0    | 0    | 0    | 0   | 1   | 22  | 0     | 0   | 23    |
| Wrist | 0     | 0     | 1     | 0     | 0    | 1    | 0    | 0   | 0   | 0   | 38    | 0   | 40    |
| Eye   | 0     | 0     | 0     | 0     | 0    | 1    | 0    | 26  | 0   | 0   | 0     | 40  | 67    |
| TOTAL | 40    | 40    | 40    | 40    | 40   | 40   | 40   | 40  | 40  | 40  | 40    | 40  | 240   |
| %     | 67,5  | 85    | 85    | 87,5  | 100  | 87,5 | 95   | 25  | 82,5| 55  | 95    | 100 | 80,4  |

**Table D.4 - Confusion Matrix of joined results for 12 default gestures**
**Columns – Expected Results; Lines – Classification Results**

## 2.2 - 5 Default Gestures

Confusion Matrixes D.5 and D.6 correspond, respectively, to the recognition of 12 default gestures while standing and moving. Confusion Matrix D.7 joins the results of D.6 and D.5.

|       | Mouth | Chest | Elbow | Navel | Neck | Head | Back | Ear | Hip | Leg | Wrist | Eye | TOTAL |
|-------|-------|-------|-------|-------|------|------|------|-----|-----|-----|-------|-----|-------|
| Mouth | 22    | 0     | 0     | 0     | 0    | 0    | 2    | 0   | 0   | 0   | 1     | 0   | 25    |
| Chest | 1     | 20    | 1     | 1     | 0    | 0    | 0    | 0   | 0   | 0   | 0     | 0   | 23    |
| Elbow | 1     | 0     | 22    | 0     | 0    | 0    | 1    | 0   | 0   | 0   | 0     | 0   | 24    |
| Navel | 0     | 0     | 0     | 15    | 0    | 0    | 0    | 0   | 0   | 1   | 0     | 0   | 16    |
| Neck  | 0     | 0     | 0     | 0     | 20   | 0    | 0    | 3   | 0   | 0   | 0     | 0   | 23    |
| Head  | 0     | 0     | 0     | 0     | 0    | 8    | 0    | 0   | 0   | 0   | 0     | 0   | 8     |
| Back  | 0     | 0     | 1     | 0     | 0    | 0    | 13   | 0   | 0   | 0   | 0     | 1   | 15    |
| Ear   | 0     | 0     | 0     | 0     | 0    | 0    | 0    | 9   | 0   | 0   | 0     | 0   | 9     |
| Hip   | 0     | 0     | 0     | 0     | 0    | 0    | 0    | 0   | 15  | 0   | 0     | 0   | 15    |
| Leg   | 0     | 0     | 0     | 0     | 0    | 0    | 0    | 0   | 0   | 15  | 0     | 0   | 15    |
| Wrist | 0     | 0     | 0     | 0     | 0    | 0    | 0    | 0   | 1   | 0   | 11    | 0   | 12    |
| Eye   | 0     | 0     | 0     | 0     | 0    | 0    | 0    | 0   | 0   | 0   | 0     | 15  | 15    |
| TOTAL | 24    | 20    | 24    | 16    | 20   | 8    | 16   | 12  | 16  | 16  | 12    | 16  | 200   |
| %     | 92    | 100   | 92    | 94    | 100  | 100  | 81   | 75  | 94  | 94  | 92    | 94  | 92,5  |

**Table D.5 - Confusion Matrix of results for 5 default gestures (Standing)**
**Columns – Expected Results; Lines – Classification Results**

|       | Mouth | Chest | Elbow | Navel | Neck | Head | Back | Ear | Hip | Leg | Wrist | Eye | TOTAL |
|-------|-------|-------|-------|-------|------|------|------|-----|-----|-----|-------|-----|-------|
| Mouth | 15    | 1     | 0     | 1     | 0    | 0    | 1    | 1   | 1   | 0   | 0     | 0   | 20    |
| Chest | 1     | 18    | 0     | 1     | 0    | 0    | 0    | 0   | 0   | 0   | 0     | 0   | 20    |
| Elbow | 5     | 1     | 24    | 0     | 0    | 0    | 0    | 0   | 1   | 0   | 0     | 0   | 31    |
| Navel | 0     | 0     | 0     | 14    | 0    | 0    | 0    | 0   | 0   | 0   | 0     | 0   | 14    |
| Neck  | 0     | 0     | 0     | 0     | 20   | 1    | 0    | 2   | 0   | 0   | 0     | 0   | 23    |
| Head  | 2     | 0     | 0     | 0     | 0    | 7    | 0    | 0   | 0   | 0   | 0     | 0   | 9     |
| Back  | 0     | 0     | 0     | 0     | 0    | 0    | 15   | 0   | 0   | 0   | 0     | 0   | 15    |
| Ear   | 0     | 0     | 0     | 0     | 0    | 0    | 0    | 9   | 0   | 0   | 0     | 0   | 9     |
| Hip   | 0     | 0     | 0     | 0     | 0    | 0    | 0    | 0   | 14  | 0   | 0     | 0   | 14    |
| Leg   | 0     | 0     | 0     | 0     | 0    | 0    | 0    | 0   | 0   | 16  | 0     | 0   | 16    |
| Wrist | 0     | 0     | 0     | 0     | 0    | 0    | 0    | 0   | 0   | 0   | 12    | 0   | 12    |
| Eye   | 1     | 0     | 0     | 0     | 0    | 0    | 0    | 0   | 0   | 0   | 0     | 16  | 17    |
| TOTAL | 24    | 20    | 24    | 16    | 20   | 8    | 16   | 12  | 16  | 16  | 12    | 16  | 200   |
| %     | 63    | 90    | 100   | 88    | 100  | 88   | 94   | 75  | 88  | 100 | 100   | 100 | 90    |

**Table D.6 - Confusion Matrix of results for 5 default gestures (Moving)**
**Columns – Expected Results; Lines – Classification Results**

|       | Mouth | Chest | Elbow | Navel | Neck | Head | Back | Ear | Hip | Leg | Wrist | Eye | TOTAL |
|-------|-------|-------|-------|-------|------|------|------|-----|-----|-----|-------|-----|-------|
| Mouth | 37    | 1     | 0     | 1     | 0    | 0    | 3    | 1   | 1   | 0   | 1     | 0   | 45    |
| Chest | 2     | 38    | 1     | 2     | 0    | 0    | 0    | 0   | 0   | 0   | 0     | 0   | 43    |
| Elbow | 6     | 1     | 46    | 0     | 0    | 0    | 1    | 0   | 1   | 0   | 0     | 0   | 55    |
| Navel | 0     | 0     | 0     | 29    | 0    | 0    | 0    | 0   | 0   | 1   | 0     | 0   | 30    |
| Neck  | 0     | 0     | 0     | 0     | 40   | 1    | 0    | 5   | 0   | 0   | 0     | 0   | 46    |
| Head  | 2     | 0     | 0     | 0     | 0    | 15   | 0    | 0   | 0   | 0   | 0     | 0   | 17    |
| Back  | 0     | 0     | 1     | 0     | 0    | 0    | 28   | 0   | 0   | 0   | 0     | 1   | 30    |
| Ear   | 0     | 0     | 0     | 0     | 0    | 0    | 0    | 18  | 0   | 0   | 0     | 0   | 18    |
| Hip   | 0     | 0     | 0     | 0     | 0    | 0    | 0    | 0   | 29  | 0   | 0     | 0   | 29    |
| Leg   | 0     | 0     | 0     | 0     | 0    | 0    | 0    | 0   | 0   | 31  | 0     | 0   | 31    |
| Wrist | 0     | 0     | 0     | 0     | 0    | 0    | 0    | 0   | 1   | 0   | 23    | 0   | 24    |
| Eye   | 1     | 0     | 0     | 0     | 0    | 0    | 0    | 0   | 0   | 0   | 0     | 31  | 32    |
| TOTAL | 48    | 40    | 48    | 32    | 40   | 16   | 32   | 24  | 32  | 32  | 24    | 32  | 400   |
| %     | 77    | 95    | 95,8  | 90,6  | 100  | 93,8 | 87,5 | 75  | 90,6| 96,8| 95,8  | 0   | 91,25 |

**Table D.7 - Confusion Matrix of joined results for 5 default gestures**
**Columns – Expected Results; Lines – Classification Results**

## 2.3 - Comparison

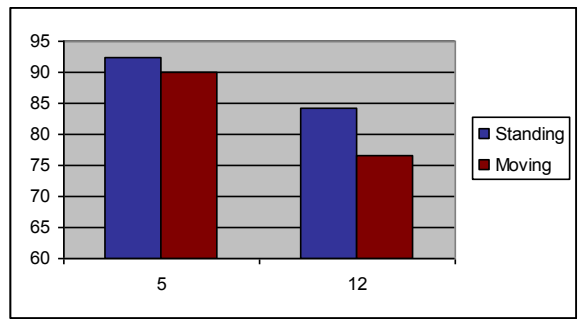

**Figure D.2 – Results for default gestures, while standing or moving and also with 5 or 12 default gestures**

## 3 - Interface Test

In this section we present the detailed results for the Interface test. From Table D.8 we extracted the results for all other graphics and tables on this subject (such as D.9, D.10 and D.11).

|  | User1 | User2 | User3 | User4 | User5 | User6 | User7 | User8 | User9 | User10 |  | Metric |
|---|---|---|---|---|---|---|---|---|---|---|---|---|
| Number of Gestures | 24 | 22 | 22 | 20 | 20 | 20 | 20 | 21 | 21 | 20 | 210 | Total |
| (1)Correct | 9 | 15 | 13 | 11 | 13 | 12 | 14 | 14 | 13 | 15 | 61,4 | % |
| (2)Correct with multitap | 4 | 4 | 4 | 6 | 5 | 4 | 6 | 6 | 6 | 4 | 23,3 | % |
| (3)Error with correction | 7 | 0 | 3 | 3 | 0 | 2 | 0 | 0 | 0 | 0 | 7,1 | % |
| (4)Cancelled at least one time | 2 | 2 | 2 | 0 | 0 | 0 | 0 | 1 | 1 | 0 | 3,8 | % |
| General Clicks Average | 3,3 | 2,6 | 2,7 | 2,7 | 2,3 | 2,3 | 2,4 | 2,4 | 2,5 | 2,2 | 2,5 | Mean |
| Average Time for (1) | 3,4 | 3,4 | 3,2 | 3,2 | 3,5 | 3,3 | 3,4 | 3,4 | 3,3 | 3,4 | 3,3 | Mean |
| Average Time for (2) | 4,5 | 3,9 | 4,6 | 4,7 | 4,1 | 4,4 | 5 | 4,1 | 4,4 | 5,0 | 4,5 | Mean |
| Average Time for (3) | 4,4 | - | 4,7 | 5 | - | 4,8 | 0 | - | - | - | 3,78 | Mean |
| Average Time for (4) | 7,1 | 5,2 | 6,5 | - | - | - | - | 5,5 | 5 | - | 5,85 | Mean |
| Errors | 0 | 1 | 0 | 0 | 2 | 2 | 0 | 0 | 1 | 1 | 3,3 | % |
| Number of Stops | 0 | 1 | 0 | 0 | 0 | 0 | 2 | 0 | 0 | 0 | 1,4 | % |
| Number of Cancellings | 4 | 2 | 2 | 0 | 0 | 0 | 0 | 1 | 1 | 0 | 4,7 | % |

**Table D.8 – Complete Results Table for Interface Test**

|  | Percentage | Time (Average) |
|---|---|---|
| Correct recognition | 62,43% | 3,3s |
| Correct recognition w/ *Multichoice* | 23,3% | 4,5s |
| Error with *Multichoice* correction | 7,1% | 4,6s |
| Error with Cancel | 3,8% | 6s |
| Errors (wrong application triggered) | 3,3% | --- |

**Table D.9 – Percentage and Average time for each type of shortcut conclusion**

|  | Percentage |
|---|---|
| Errors without *Multichoice* or Cancelling | 17,95% |
| Errors using only Cancelling | 14,14% |
| Errors using only *Multichoice* | 10,8% |
| Errors | 3,3% |

**Table D.10 – Impact of *Multichoice* and cancelling on error margin**

|  | Percentage |
|---|---|
| Clicks Average | 2,5 clicks |
| Time Average | 3,8 seconds |
| Stops during Shortcuts | 1,4% |

Table D.11 – Miscellaneous Results

## 4 - Final Questionnaire

In the final questionnaire, we were able to retrieve results with two different points of interest. Firstly, users were asked about feedback and user control mechanism, and the results are present in table D.12. Lastly, they were prompted to classify some system characteristics, such as their confidence on the recognizer or if they use it in public. Those results are reported in table D.13. We recommend to overview the protocol to know what the questions for each line were.

| Questions | User1 | User2 | User3 | User4 | User5 | User6 | User7 | User8 | User9 | User10 | Average |
|---|---|---|---|---|---|---|---|---|---|---|---|
| Visual Feedback | 5 | 3 | 3 | 3 | 1 | 2 | 3 | 3 | 4 | 4 | 3,1 |
| Audio Feedback | 5 | 5 | 5 | 5 | 5 | 4 | 5 | 5 | 5 | 5 | 4,9 |
| Vibrational Feedback | 4 | 4 | 3 | 5 | 1 | 3 | 4 | 4 | 4 | 5 | 3,7 |
| Distinct Vibrations? | Yes | Yes | Yes,2 | Yes | Yes | Yes,2 | Yes | Yes | Yes | Yes | |
| Useful? | Yes | Yes | Yes | Yes | Yes | Yes | Yes | Yes | Yes | Yes | |
| Is Multitap Useful? | 5 | 5 | 5 | 5 | 5 | 5 | 5 | 5 | 4 | 4 | 4,8 |

Table D.12 – Feedback and User Control Results

| Questions \ User | User1 | User2 | User3 | User4 | User5 | User6 | User7 | User8 | User9 | User10 | Average |
|---|---|---|---|---|---|---|---|---|---|---|---|
| Confidence | 4 | 4 | 4 | 4 | 4 | 4 | 4 | 4 | 4 | 4 | 4 |
| Feedback | 5 | 4 | 5 | 5 | 5 | 4 | 4 | 4 | 4 | 5 | 4,5 |
| Walking | 4 | 4 | 4 | 5 | 4 | 4 | 5 | 5 | 4 | 5 | 4,4 |
| Speed | 4 | 4 | 4 | 5 | 5 | 3 | 4 | 4 | 3 | 5 | 4,1 |
| Liked | 5 | 5 | 5 | 5 | 5 | 4 | 4 | 5 | 5 | 5 | 4,8 |
| Would Use it | Yes | Yes | No | Yes | Yes | Yes | Yes | Yes | Yes | Yes | 90% |
| Use in Public | Yes | Yes | No | Yes | Yes | No | Yes | Yes | Yes | Yes | 80% |

Table D.13 – Miscellaneous Results

**Suggestions:**

- Have a more accessible cancelling button, with the possibility of having a finger on the action button and other in the cancel button.
- Have all the possible shortcuts in the *Multichoice* feature.
- Have a cancelling gesture.
- Have a different start movement, such as the pocket.
- Have a personalized delay time (more than 2 seconds).
- Give more feedback when a gesture is cancelled (both with *Multichoice* and cancel button).
- Use tilt for *Multichoice*.
- Substitute the probability-based vibration to a more subtle vibration for all the delay time.
- Use a smaller PDA.

# Bibliography

[1] J. Landt, "*The history of RFID*", IEEE Potentials, Vol. 24, Issue 4, pp 8- 11, 2005.

[2] RFID Journal – The History of RFID Technology. Available in:
http://www.rfidjournal.com/article/articleview/1338/1/129 (9/2007).

[3] J. A. Landay and T. R. Kaufmann, "*User Interface Issues in Mobile Computing*", Proceedings of the Fourth Workshop on Workstation Operating Systems, Napa, CA, IEEE Computer Society Press (1993), pp. 40–47, 1993.

[4] S. Brewster, "*Overcoming the Lack of Screen Space on Mobile Computers*", Personal and Ubiquitous Computing, Vol. 6 Number 3, p.188-205, May 2002

[5] A. Oulasvirta, S. Tamminen, V. Roto, J. Kuorelahti, "*Interaction in 4-second bursts: the fragmented nature of attentional resources in mobile HCI*", Proceedings of the SIGCHI conference on Human factors in computing systems, USA, 2007

[6] G. H. Forman , J. Zahorjan, "*The Challenges of Mobile Computing*", Computer, v.27 n.4, p.38-47, 1994

[7] RFID Journal - Nokia Unveils RFID Phone Reader. Available in:
http://www.rfidjournal.com/article/articleview/834/1/13/ (9/2007)

[8] S. Kristoffersen , F. Ljungberg, ""*Making place" to make IT work: empirical explorations of HCI for mobile CSCW*", Proceedings of the international ACM SIGGROUP conference on Supporting group work, p.276-285, United States, 1999

[9] R. Want, K. Fishkin, A. Gujar and B.Harrison, "*Bridging physical and virtual worlds with electronic tags*". Proceedings of the SIGCHI conference on Human factors in computing systems: the CHI is the limit, p.370-377, United States, 1999.

[10] J. Demsar, B. Zupan, G. Leban  "*Orange: From Experimental Machine Learning to Interactive Data Mining*", White Paper (www.ailab.si/orange), Faculty of Computer and Information Science, University of Ljubljana, 2004.

[11] H. Silva, "*Feature selection in pattern recognition systems*". Master's thesis, Universidade Técnica De Lisboa Instituto Superior Técnico, 2007.

[12] R. Tenmoku, M. Kanbara, and N. Yokoya: "*A Wearable Augmented Reality System Using Positioning Infrastructures and a Pedometer*", Proceedings of International Symposium on Wearable Computers, pp. 110-117, 2003

[13] Available in Hitachi Metals Website: http://www.hitachimetals.com/ (9/2007)

[14] S. Willis, S. Helal, "*RFID Information Grid and Wearable Computing Solution to the Problem of Wayfinding for the Blind User in a Campus Environment*", Proceedings of the Ninth IEEE International Symposium on Wearable Computers, pp. 34 – 37, 2005.

[15] A. Madhavapeddy, D. Scott, R. Sharp, E. Upton, "*Using camera-phones to enhance human-computer interaction*". In 6th International Conference on Ubiquitous Computing. 2004.

[16] A. Schmidt, H.W. Gellersen, and C. Merz. "*Enabling implicit human computer interaction: A wearable RFID-tag reader*". Proceedings of the 4th IEEE International Symposium on Wearable Computers, p.193, October 18-21, 2000

[17] K. Fishkin, M. Philipose, and A. Rea, "*Hands-On RFID: Wireless Wearables for Detecting Use of Objects*", Ninth International Symposium on Wearable Computers (ISWC' 05), pp. 38-43, 2005.

[18] M. Rohs, "*Real-world interaction with camera-phones*", In: International Symposium on Ubiquitous Computing Systems, 2004.

[19] O. Ozer, O. Ozun, C. Tuzel, V. Atalay and A. Etin, "*Vision-Based Single-Stroke Character Recognition for Wearable Computing*", IEEE Intelligent Systems, Vol. 16, pp. 33-37, 2001.

[20] R. Headon, G. Coulouris, "*Supporting Gestural Input for Users on the Move*". Procedings of IEE Eurowearable '03, pp 107—112 2003.

[21] A. Feldman, E. Tapia, S. Sadi, P. Maes, C. Schmandt, C. "*ReachMedia: on-the-move interaction with everyday objects*", Ambient Intelligence Group, MIT Media Lab, 2005

[22] J. Sandsjö, "*Movement Thinking as a Way to Approach Computational Device Design*", In Workshop Proceedings of Approaches to Movement-Based Interaction, 2005.

[23] T. Pering, R. Ballagas, and R. Want. "*Spontaneous marriages of mobile devices and interactive spaces*". Communications of the ACM, Vol. 48, Issue 9, pp. 53-59, 2005

[24] C. Baber P. Smith, J.Cross, D. Zasikowski, J. Hunter, "*Wearable technology for crime scene investigation, Wearable Computers*", Ninth IEEE International Symposium on Wearable Computers (ISWC'05)  pp. 138-143, 2005

[25] T. Kurata, T. Kato, M. Kourogi, J. Keechul, and K. Endo. "*A functionally-distributed hand tracking method for wearable visual interfaces and its applications*". In IAPR MVA'02, pp. 84–89, 2002.

[26] T. Kurata, T. Okuma, M. Kourogi, and K. Sakaue, "*The Hand Mouse: GMM Hand-color Classification and Mean Shift Tracking*", Proceedings. IEEE ICCV Workshop on Recognition, Analysis, and Tracking of Faces and Gestures in Real-Time Systems, pp. 119-124, 2001

[27] M. Petersen, O. Iversen, P. Krogh, M. Ludvigsen, "*Aesthetic Interaction. A pragmatist's aesthetics of interactive systems*", In Proceedings of the 2004 conference on Designing Interactive Systems, pp 269-275, 2004.

[28] H. Sasaki, T. Kuroda, Y. Manabe and K. Chihara, "*HIT-Wear: A Menu System Superimposing on a Human Hand for Wearable Computers*", in Proc. of International Conference on Artificial Reality and Teleexistence (ICAT 99), pp.146-153, 1999.

[29] E. Toye, R. Sharp, A. Madhavapeddy, D. Scott, E. Upton, A. Blackwell, "*Interacting with mobile services: an evaluation of camera-phones and visual tags*", Personal and Ubiquitous Computing, pp97 106, 2006

[30] P. Pirhonen, S. A. Brewster, and C. Holguin, "*Gestural and Audio Metaphors as a Means of Control in Mobile Devices*," Proceedings of the SIGCHI conference on Human factors in computing systems, pp. 291 - 298, 2002.

[31] N. Fiedlander, K. Schlueter, and M. Mantei, "*Bullseye! When Fitt's Law Doesn't Fit*" Proceedings of the SIGCHI conference on Human factors in computing systems, pp. 257 - 264, 1998

[32] V. Kostakos, E. O'Neil, "*A directional stroke recognition technique for mobile interaction in a pervasive computing world*". In People and Computer XVII, Proc. HCI 2003: Designing for Society. pp. 197–206, 2003.

[33] S. Brewster, J. Lumsden, M. Bell, M. Hall, S. Tasker, "*Multimodal 'eyes-free' interaction techniques for wearable devices*", Proceedings of the SIGCHI conference on Human factors in computing systems, 2003.

[34] Graffiti - Available in: http://www.palm.com/us/products/input/ (9/2007)

[35] J. Farringdon, A. Moore, N. Tilbury, J. Church and P. Biemond "*Wearable Sensor Badge & Sensor Jacket for Context Awareness*", Third International Symposium on Wearable Computers (ISWC'99), pp.107-113, 1999.

[36] M. Mathie, B. Celler, "*A system for monitoring posture and physical activity using accelerometers*", in: Proceedings of the 23rd Annual International Conference of the IEEE Engineering in Medicine and Biology Society, pp. 3654–3657, 2001.

[37] S. Perrin, A. Cassinelli, and M. Ishikawa, "*Gesture recognition using laser-based tracking system*", Proceedings of the Sixth IEEE International Conference on Automatic Face and Gesture recognition, pp. 541-546, 2004.

[38] C. Metzger, M. Anderson, and T. Starner. "*Freedigiter: A contact-free device for gesture control*". Eighth IEEE International Symposium on Wearable Computers (ISWC'04), pp 18–21, 2004.

[39] E. Costanza, S. A.Inverso, R. Allen. "*Toward Subtle Intimate Interfaces for Mobile Devices Using an EMG Controller*". Proceedings of the SIGCHI conference on Human factors in computing systems, pp.481 486, April 2005.

[40] A. Chamberlain, and R. Kalawsky. "*Comparative Investigation into Two Pointing Systems for Use with Wearable Computers While Mobile*", Eighth IEEE International Symposium on Wearable Computers (ISWC'04), pp. 110-117, 2004.

[41] M. Raghunath, C. Narayanaswami, "*User Interfaces for Applications on a Wrist Watch*". In Personal and Ubiquitous Computing, pp 17-30, 2002.

[42] R. Duta et.al, "*Pattern Classification*", Chichester: John Wiley & Sons, 2001.

[43] K. Hinckley, M. Sinclair, "*Touch-Sensing Input Devices*", Proceedings of the SIGCHI conference on Human factors in computing systems, 223-230, 1999

[44] U. Maurer, A. Smailagic, D. Siewiorek, M. Deisher, "*Activity Recognition and Monitoring Using Multiple Sensors on Different Body Positions*," International Workshop on Wearable and Implantable Body Sensor Networks (BSN'06), pp. 113-116, 2006.

[45] J. Fistre, A. Tanaka, "*Real Time EMG Gesture Recognition for Consumer Electronics Device Control*", Sony CSL Paris Open House Poster, 2002.

[46] A Haro, K Mori, T Capin, S. Wilkinson, "*Mobile Camera-based User Interaction*," Lecture notes in computer science, 2005

[47] T. R. Hansen, E. Eriksson, A. Lykke-Olesen, "*Mixed interaction space: designing for camera based interaction with mobile devices*," CHI '05 extended abstracts on Human factors in computing systems, 2005

[48] J.Mantyjarvi, M. Lindholm, E. Vildjiounaite, S.-M.Makela, and H. Ailisto. "*Identifying users of portable devices from gait pattern with accelerometers*". IEEE International Conference on Acoustics, Speech and Signal Processing (ICASSP'05), pp. 973–976, 2005.

[49] A. Crossan, R. Murray-Smith, S. Brewster, J. Kelly, and B, Musizza, "*Gait Phase Effects in Mobile Interaction*," CHI '05 extended abstracts on Human factors in computing systems, pp. 1312-1315, 2005.

[50] S. Strachan, R. Murray-Smith. "*Muscle Tremor as an Input Mechanism*". In Annual ACM Symposium on User Interface Software and Technology, 2004.

[51] M. Hachet, J. Pouderoux, P. Guitton, "*A camera-based interface for interaction with mobile handheld computers*", Proceedings of the 2005 symposium on Interactive 3D graphics and games, pp. 65-72, 2005.

[52] O. Cakmakci, J. Coutaz, K. V. Laerhoven, and H. Gellersen. "*Context awareness in systems with limited resources*". In Proceedings of AIMS-2002, Artificial Intelligence in Mobile Systems, 2002.

[53] K. Wheeler, C. Jorgensen, "*Gestures as Input: Neuroelectric Joysticks and Keyboards*". IEEE Pervasive Computing, Vol. 2, No. 2, pp. 56-61, 2003.

[54] E. Costanza, S. A.Inverso, R. Allen. "*EMG as a Subtle Input Interface for Mobile Computing*". Mobile Human-Computer Interaction – MobileHCI 2004, pp. 426-430, 2004

[55] E. Choi, W. Bang, S. Cho, J. Yang, D. Kim, S. Kim. "*Beatbox music phone: gesture based interactive mobile phone using a tri-axis accelerometer*", IEEE International Conference on Industrial Technology ICIT 2005, pp. 97-102, 2005.

[56] A. Benbasat, J. Paradiso, "*An Inertial Measurement Framework for Gesture Recognition and Applications*", Revised Papers from the International Gesture Workshop on Gesture and Sign Languages in Human-Computer Interaction, pp. 9-20, 2001

[57] M. Petersen, O. Iversen, P. Krogh, M. Ludvigsen, "*Aesthetic interaction: a pragmatist's aesthetics of interactive systems*", Proceedings of the 2004 conference on Designing interactive systems: processes, practices, methods, and techniques, pp. 269-276, 2004.

[58] J. Kela, P. Korpipää, J. Mäntyjärvi, S. Kallio, G. Savino, L. Jozzo, S. Di Marca "*Accelerometer-based gesture control for a design environment*". Personal and Ubiquitous Computing, Springer-Verlag, vol. 10, pp. 285-299, 2005.

[59] J. Mäntyjärvi, J. Kela, P. Korpipää, S. Kallio. "*Enabling fast and effortless customisation in accelerometer based gesture interaction*". Proceedings of the 3rd international conference on Mobile and ubiquitous multimedia, pp. 25-31, 2004.

[60] S. Kallio, J. Kela, J. Mäntyjärvi "*Online Gesture Recognition System for Mobile Interaction*". IEEE International Conference on Systems, Man and Cybernetics, pp. 2070-2076, 2003.

[61] J. Rekimoto, "*GestureWrist and GesturePad: Unobtrusive Wearable Interaction Devices*", Fifth International Symposium on Wearable Computers (ISWC'01), pp. 21, 2001.

[62] T. Fuhrmann, M. Klein, and M. Odendahl. "*The Bluewand as interface for ubiquitous and wearable computing environments*". In Proceedings of the 5th European Personal Mobile Communications Conference (EPMCC'03), 2003.

[63] A. Wilson, S. Shafer, "*XWand: UI for intelligent spaces*", Proceedings of the SIGCHI conference on Human factors in computing systems, pp. 545-552, 2003.

[64] J. Ängeslevä, I. Oakley, S. Hughes, S. and S. O'Modhrain, "*Body Mnemonics: Portable Device. Interaction Design Concept*", Proceedings of UIST, 2003.

[65] I. Jang, W. Park, "*Signal processing of the accelerometer for gesture awareness on handheld devices*". In The 12th IEEE Int. Workshop on Robot and Human Interactive Communication, 2003.

[66] J. Linjama, T. Kaaresoja, "*Novel, minimalist haptic gesture interaction for mobile devices*". In Proceedings of the 3rd Nordic Conference on Human-Computer interaction, pp. 457-458, 2004

[67] J. Rekimoto, "*Tilting operations for small screen interfaces*", Proceedings of the 9th annual ACM symposium on User interface software and technology, pp.167-168, 1996.

[68] B. Harrison, K. Fishkin, A. Gujar, C. Mochon, R. Want, "*Squeeze me, hold me, tilt me! An exploration of manipulative user interfaces*", Proceedings of the SIGCHI conference on Human factors in computing systems, pp.17-24, 1998.

[69] K. Hinckley, J. Pierce, M. Sinclair, E. Horvitz, "*Sensing techniques for mobile interaction*", Proceedings of the 13th ACM UIST, pp .91-100, 2000

[70] J. Bartlett, "*Rock 'n' Scroll Is Here to Stay*", IEEE Computer Graphics and Applications, v.20 n.3, pp.40-45, 2000

[71] D. Wigdor, R. Balakrishnan, "*TiltText: using tilt for text input to mobile phones*", Proceedings of the 16th annual ACM symposium on User interface software and technology, pp. 81-90, 2003.

[72] K. Partridge, S. Chatterjee, V. Sazawal, G. Borriello, R. Want, "*TiltType: accelerometer supported text entry for very small devices*", Proceedings of the 15th annual ACM symposium on User interface software and technology, pp. 201-204, 2002.

[73] C. H. Chen and P. Wang, "Handbook of Pattern Recognition and Computer Vision". World Scientific Publishing Company, 2005.